\documentclass[reprint,amsmath,amssymb,aps,prx,nofootinbib]{revtex4-2}
\usepackage{graphicx}
\usepackage{color}
\usepackage{dcolumn}
\usepackage{latexsym}

\usepackage[normalem]{ulem}
\usepackage{hyperref,amssymb}
\usepackage{url}
\usepackage{color,soul}
\usepackage{graphicx}
\usepackage{verbatim}
\usepackage{multirow}
\usepackage{amsmath}
\newcommand{\beq}{\begin{eqnarray}}
\newcommand{\eeq}{\end{eqnarray}}
\usepackage{mathrsfs}
\usepackage{float,soul}
\usepackage[dvipsnames]{xcolor}
\usepackage{mathtools}
\usepackage{slashed}
\usepackage{physics}
\usepackage{graphicx}
\usepackage{epstopdf}
\usepackage{subfigure}
\usepackage[font=small]{caption}
\usepackage{hyperref}
\usepackage{bbold}
\usepackage{wasysym}
\usepackage{feynmp}
\usepackage{hyperref}
\hypersetup{colorlinks}
\usepackage{ragged2e}
\usepackage{siunitx}
\usepackage[utf8]{inputenc}
\usepackage{enumitem}

\makeatletter
\newcommand{\printfontsize}{\f@size pt}
\makeatother

\begin{document}
\title{An asymptotically solvable model of many-body critical phases: mobility edges, scars, and inverted scars}

\author{Yi-Ting Tu}
\author{Zi-Jian Li}
\author{Sankar Das Sarma}
\affiliation{Condensed Matter Theory Center and Joint Quantum Institute, Department of Physics, University of Maryland, College Park, Maryland 20742, USA}

\begin{abstract}
While the prethermal regime of random many-body localized (MBL) systems is dominated by accidental many-body resonances, another class of resonances, originating from the underlying potential structure, is expected in large-size deterministic systems.
It is known that this class of resonances can lead to single-particle critical phases that are neither localized nor extended, but the consequences in interacting systems remain unclear.
In this work, we construct an asymptotically solvable model of a one-dimensional nearest-neighbor interacting spin chain, whose spatial structure induces a hierarchy of mirror-like many-body resonances.
We derive two phases in the thermodynamic limit, characterized by the satisfaction and violation of a version of the weak eigenstate thermalization hypothesis (ETH).
While these two phases are similar to the usual MBL and ETH phases, there exist rare eigenstates that behave like the opposite phase, interpreted as many-body scars and inverted scars.
Surprisingly, the two phases can be separated by a finite-temperature phase transition, corresponding to a thermodynamic many-body mobility edge, which was often believed to be impossible.
Our results also suggest the existence of delocalized rare regions in an otherwise-localized interacting Aubry-Andr\'e model, even if there are no low-disorder regions like those in random systems.
This challenges the common belief that there is no avalanche instability in quasiperiodic MBL.
\end{abstract}

\maketitle

\section{Introduction}

Statistical mechanics is based on the expectation that a generic isolated system reaches thermal equilibrium by itself, washing away all microscopic information in the initial condition.
Yet it is proposed that for quantum systems with many-body localization (MBL)~\cite{Anderson1958,nandkishore2015many,abanin2019manybody,sierant2025many}, such an assumption fails, with every bit of local information in the initial condition remaining local indefinitely~\cite{Huse2014,Imbrie2017}.
This phenomenon is most widely studied in spin chains with strong random disorder~\cite{basko2006metal, oganesyan2007localization, Znidaric2008, pal2010manybody, Devakul2015, Imbrie2016, Imbrie2016a}, but it also occurs in deterministic systems with quasiperiodic modulation~\cite{Vidal2002, Iyer2013, Mastropietro2015, Khemani2017_PRL, Xu2019_PRR, Vu2022}.

While the stability of MBL in the thermodynamic limit remains an unsolved problem, much of the recent work has instead focused on the \emph{prethermal} regime of such systems~\cite{Long2023,Long2023a,Tu2024, Colbois2024,Hahn2025,Vanoni2026,Padhan2026,Faulend2026}, where the dynamics are dominated by rare many-body resonances~\cite{Gopalakrishnan2015,Khemani2017,Garratt2021,Crowley2022}.
More specifically, a system that behaves like MBL at short timescales delocalizes slowly at longer timescales due to accidental closeness in energy levels between different localized configurations, referred to as 
\emph{accidental} resonances.%

For systems with a deterministic long-range structure, such as quasiperiodic MBL, while accidental resonances may still exist, another type of resonance induced by the underlying structure should be amplified compared to random systems.
In the paradigmatic Aubry-Andr\'e (AA) model, a recent numerical study~\cite{Faulend2026} points to the many-body resonances due to the approximately mirror-symmetric structure of the potential, which may explain the numerical observation~\cite{Padhan2026} of unexpected long-range resonances in such a system.
Many-body resonances in exact mirror-symmetric MBL systems have also been recently studied in Ref.~\cite{Li2025}, where an analytical theory of isolated resonances is constructed.
The possible proliferation of this class of \emph{structural} many-body resonances, as well as the thermodynamic consequences, is the motivation for this paper.

At the single-particle level, the proliferation of structural resonances often leads to single-particle \emph{critical} phases, which are neither localized nor extended, and are characterized by the \emph{singular continuous} spectrum in the mathematical literature
~\cite{gordon1976point,Avron1982SingularCS,Jitomirskaya1994,hof1995singular,koslover2005jacobi,jitomirskaya2019critical,Jitomirskaya2021}.
There are various types of models where this happens for an entire region in the parameter space: models with exact repetition structure (e.g.\ the Fibonacci quasicrystal~\cite{Kohmoto1983,Suto1987,sutHo1989singular,Damanik2016,Jagannathan2021}), models with arbitrarily weak bonds (e.g.\ the extended Aubry-Andr\'e-Harper model~\cite{Hatsugai1990,Han1994,Takada2004,Liu2015,Avila2017}), and models with a parent flat band (e.g.\ Ref.~\cite{Kumar2026}).
There are also models where this happens for a set of rare parameter values in an otherwise localized phase, such as the \emph{rare initial phases} for the Aubry-Andr\'e (AA) model~\cite{Aubry1980, Harper1955} above the critical point~\cite{Jitomirskaya1994,Jitomirskaya2021}, which is due to the proliferation of mirror-like resonances called \emph{phase resonances} in the literature.
Most of these are rigorous mathematical results in infinite-size single-particle quasiperiodic systems with no interaction whatsoever.

The consequences of particle-particle interactions in such single-particle critical phases, however, are much less studied.
Rigorously proving results on such interacting many-body systems is extremely difficult. Existing results in the literature are mostly based on mean-field and related approximations~\cite{Hiramoto1990,Hiramoto1992,Settino2020} and small-size numerics~\cite{mace2019manybody,varma2019diffusive,wang2021manybody}.
In particular, Refs.~\cite{Hiramoto1990,mace2019manybody,varma2019diffusive} study the interacting Fibonacci model, with inconsistent findings on whether the system becomes MBL at large potential strengths. Refs.~\cite{Settino2020,wang2021manybody} propose that single-particle continuous spectra may lead to a many-body phase which is neither MBL nor thermal.
However, small-size numerics are expected to be dominated by finite-size effects and do not directly probe the thermodynamic consequences of structural resonances.
Roughly speaking, the proliferation of structural resonances is expected to be hierarchical in real space, and to see the resonances of $n$ levels in the hierarchy, we need a system size at least exponential in $n$, leading to a Hilbert space dimension at least doubly exponential in $n$.
Therefore, numerically observing the thermodynamic consequences is extremely difficult, if not impossible.

In this paper, we take a different approach that neither relies on numerical data nor dives into general mathematical proofs.
Instead, we construct an asymptotically solvable model to capture the essential behaviors of structural resonances.
Our model is a one-dimensional (1D) nearest-neighbor mixed-field Ising chain with a spatial structure that we call the \emph{hierarchical mirror structure} (HMS), which induces a hierarchy of mirror-like many-body resonances.
This structure can either be thought of as a simplification of the quasiperiodic structure in the single-particle study in Ref.~\cite{Jitomirskaya1994} or as a generalization of the single mirror-like many-body resonance in Ref.~\cite{Li2025}.
The many-body eigenstates of our model can be written down exactly up to a controllable error bound (which can be made arbitrarily small), and the corresponding energies can be controlled in a hierarchical way.
From the rigorous solution, we derive various properties of the model in the thermodynamic limit.
In particular, we show that there are two types of many-body critical phases that generalize the single-particle phase: the \emph{many-body critically extended} (MBC-E) phase and the \emph{many-body critically localized} (MBC-L) phase.
As the MBC-E (MBC-L) phase satisfies (violates) a form of the weak eigenstate thermalization hypothesis (ETH)~\cite{Deutsch1991, Srednicki1994,Mori2018thermalization,Deutsch2018eigenstate}, as well as some other diagnostics,
they are similar to the usual ETH (MBL) phase in some sense.

The main distinction of MBC-E and MBC-L from the usual ETH and MBL phases is the existence of rare eigenstates that behave like those from the opposite phase.
For the MBC-E phase, while most eigenstates have local observable expectation values being thermal, there is a set of eigenstates that are dense (exists near every energy density whenever it is well-defined) but rare (a vanishing fraction in the thermodynamic limit) which are non-thermal.
This can be regarded as a form of quantum many-body scars, which are, in general, isolated non-thermal energy eigenstates in an otherwise thermal system~\cite{Bernien2017,shiraishi2017systematic,Serbyn2021,Chandran2023}.
For the MBC-L phase, the opposite situation occurs: the existence of rare delocalized states in an otherwise MBL spectrum.
Such phenomena, in general, are mentioned less frequently in the literature but have been termed \emph{inverted many-body scars} by some papers~\cite{srivatsa2020manybody,Iversen2022escaping,chen2024inverting,srivatsa2023mobility}.
Curiously, our work involves both concepts.

Surprisingly, the MBC-E and MBC-L phases are separated by a finite-temperature phase transition in some parameter regime, implying the existence of a thermodynamic many-body mobility edge (MBME) under a closed system interpretation, which was previously claimed to be impossible in the literature~\cite{deroeck2016absence, Huang2023b}.
MBMEs are, in general, points on the energy spectrum that separate (typically) localized and extended many-body eigenstates.
While thermodynamic MBMEs have been analytically argued to exist in some models containing long-range interactions~\cite{Pawlik2024}, their existence in short-range spin chains has been controversial.
In small-size numerics on quasiperiodic spin chains whose single-particle orbitals contain a single-particle mobility edge (SPME), one does observe an MBME~\cite{li2015manybody, Setiawan2017, Hsu2018_PRL, Li2016, Ghosh2020transport, Tu2023}.
However, there are arguments that such an MBME is only a finite-size effect~\cite{deroeck2016absence, Huang2023b, Tu2024}.
In Ref.~\cite{Huang2023b}, it was argued that the finite-size MBME due to SPME will be pushed to the edge of the spectrum in the thermodynamic limit, and in Ref.~\cite{deroeck2016absence}, it was argued that the localized phase can be destroyed by delocalized bubbles due to local energy density fluctuations.
Our model therefore shows the limitations of these arguments for short-range hierarchical systems.
In particular, it circumvents the argument of Ref.~\cite{Huang2023b} as it is not based on SPME (the single-particle subspaces are all critical), and it escapes that of Ref.~\cite{deroeck2016absence} due to the lack of a well-defined ``delocalized bubble'' (as a delocalized state can locally look like a localized state but globally become ETH-like).
Another interesting fact is that the localized phase appears at high temperatures in our model, which is an example of the \emph{inverse freezing} phenomenon described in Ref.~\cite{Iadecola2018}, also related to mirror symmetry.

Our results may also have implications for rare-region effects in quasiperiodic systems.
In a random strongly disordered system, as long as the system size is large enough, there must be some rare regions that accidentally have low disorder and consequently locally thermalize.
It is proposed that such a region may expand indefinitely by thermalizing nearby spins, causing an \emph{avalanche} that destroys an otherwise MBL phase~\cite{thiery2018many,morningstar2022avalanches, sels2022bath}.
In a quasiperiodic system, however, there are no rare low-disorder regions.
Although it is shown that an avalanche may occur in the AA model in the presence of a planted thermal inclusion~\cite{tu2023avalanche}, intuitively it is unlikely that such a thermal region can exist naturally in the AA model.
However, our results show that there is a different type of rare region in the AA model, characterized by the local proliferation of structural many-body resonances (which is motivated by the proof of rare initial phases of the single-particle AA model~\cite{Jitomirskaya1994}).
We discuss the possibility that the interplay between structural and accidental resonances can lead to the thermalization of such regions and, therefore, lead to avalanches.

The rest of this paper is organized as follows. In Sec.~\ref{sec:summary}, we present the idea behind our construction, beginning with the background on single-particle critical phases and the simplification of HMS.
We then provide the intuition on the effect of interactions as well as a summary of how such effects lead to the main properties of the MBC phases.
In Sec.~\ref{sec:SPSolvable}, we construct the single-particle version of our solvable model, which serves both as a warmup for the reader and as a conceptual precursor for our many-body model.
In particular, we prove that our simplified real-space structure does produce a single-particle critical phase in the mathematical sense.
In Sec.~\ref{sec:MBSolvable}, we construct our main many-body solvable model for MBC phases, rigorously derive the asymptotic solution, and prove its key properties.
In Sec.~\ref{sec:nonideal}, we go beyond the solvable limit and discuss the possible many-body consequences of some previously studied single-particle models: the Fibonacci quasicrystal, the EAAH model, and the rare-region effect of the AA model (with some detailed analytical constructions provided in Appendics~\ref{sec:symmetricRandom} and \ref{sec:rareConstruction}).
We conclude in Sec.~\ref{sec:conclusion} with a discussion on possible extensions to higher dimensional quasicrystals.

\section{Idea behind the construction}\label{sec:summary}

In this section, we present the idea behind our construction of the asymptotically solvable model.
We begin with some background on single-particle critical phases, from which we extract a simplified real-space structure that we call the \emph{hierarchical mirror structure} (HMS).
Then we discuss the intuition behind the effects of particle-particle interaction in such a single-particle critical phase due to HMS.
Finally, we summarize how such effects lead to the key properties of the two MBC phases that we construct rigorously later in this paper.

\begin{figure*}
    \centering
    \includegraphics[scale=1]{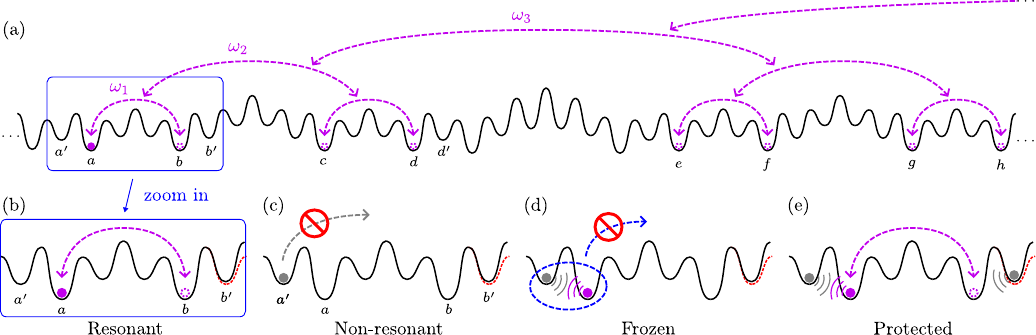}
    \caption{\justifying Illustration of the effects of interaction on a system with hierarchical mirror structure. (a) The transport of such a system can be thought of as hierarchical oscillations between distant regions with frequencies $\omega_1\gg\omega_2\gg \omega_3 \gg \cdots$. (b) For the lowest level of the hierarchy, if a particle is near the mirror center, it is \emph{resonant} between two mirrored sites ($a$ and $b$). (c) Further away from the mirror center, the particle becomes \emph{non-resonant} due to a small asymmetry between sites $a'$ and $b'$ (red dashed line). (d) When the particles in (b) and (c) are both present, the non-resonant particle causes the resonant particle to become \emph{frozen} due to interaction. (e) When two mirror-symmetric non-resonant particles are both present, the resonance of the (purple) particle is \emph{protected}. }
    \label{fig:intro}
\end{figure*}

\subsection{Singular continuous spectra}
Consider a finite single-particle system with Hilbert space $\mathcal{H}$ and Hamiltonian $H$. To study the time evolution of a local state vector $|\psi\rangle\in\mathcal{H},\langle\psi|\psi\rangle=1$, one shall decompose the Hamiltonian into the energy eigenstate basis $|E\rangle$. However, the same eigenstate decomposition cannot be directly carried to an infinite system, as one cannot always find a complete set of $|E\rangle \in \mathcal H$ that is normalizable, i.e., the equation $H|E\rangle = E|E\rangle$ may not have solutions. Such a problem appears when we introduce the concept of the ``scattering states" (which are non-normalizable) to describe some ``eigenstates" in an infinite system.
Luckily, mathematicians have solved this issue by introducing the spectral measure to ``decompose" the Hamiltonian,
\begin{equation}
    \langle\psi|e^{-iHt}|\psi\rangle=\int e^{-iEt} d\mu(E)
\end{equation}
Roughly speaking, the spectral measures describe the overlap between a state vector and an energy eigenstate $d\mu(E) = f(E)dE \sim |\langle\psi|E\rangle|^2 dE$. 

As we know, the single-particle eigenstate can be classified by whether it is localized or extended; the spectral measure can also be decomposed into such parts.
However, in general, there are three parts instead of two:
\begin{equation}
    \mu = \mu_{pp} + \mu_{ac} + \mu_{sc}
\end{equation}
with subscripts standing for \emph{pure point}, \emph{absolutely continuous}, and \emph{singular continuous}, respectively. The pure point part corresponds to localized eigenstates (or bound states), where the energy distribution function $f_{pp}(E)$ is a weighted sum of delta peaks, $f_{pp}(E) = \sum_{i} w_i\delta(E-E_i)$. The absolutely continuous part corresponds to extended states (or scattering states), where $f_{ac}(E)$ is a regular function. The singular continuous part, perhaps less familiar, corresponds to the state where it is neither extended nor localized, where $f_{sc}(E)$ is distributed on a measure-zero uncountable set (such as the Cantor set) without concentration on any particular point (which excludes the pure point case), and zero elsewhere (which excludes the absolutely continuous case). Thus, the singular continuous spectrum usually has a fractal-like structure. One famous and physically relevant example of this case is Hofstadter's butterfly (at irrational magnetic flux ratios)~\cite{hofstader1976energy}. 

We first discuss the consequence of the singular continuous spectrum on transport. Roughly speaking, the distribution of $f_{sc}$ implies that the scales of the
energy gaps $\delta$ can be arbitrarily small and discontinuous. Specifically, if we sort the scales of $\delta$ in descending order, we get a discrete sequence $\{\delta_n\}$ with $\delta_n \rightarrow 0$ when $n\rightarrow \infty$. (We require $\delta_n \gg \delta_{n+1}$ for theoretical convenience later on.)
Thus, the transport of a local state (particle) at a certain timescale $t$ is only governed by a series of the energy eigenstates whose gaps are around $\delta_n\sim 1/t$. 
The particle can be considered ``localized" since no other eigenstates are involved in the transport at a short time interval (which should be larger than the diffusion time) related to $t$. However, in a longer time interval, the particle becomes ``extended" as other eigenstates with a smaller gap $\delta_{n+1}$ get involved. This phenomenon is usually referred to as anomalous diffusion or transport~\cite{fractal1992artuso,anomalous1996piechon,anomalous1998schulz}.

In interacting many-body systems, the above classification no longer works.
Nevertheless, the localized and extended spectra roughly correspond to MBL and ETH phases in many-body systems.
Therefore, it is reasonable to guess that there is a many-body counterpart of the singular continuous spectrum as well, which is the main motivation of our paper. 
Note that we would not try to give a precise and rigorous definition of the many-body singular continuous spectrum, but rather study the consequences of interactions.

\subsection{Hierarchical mirror structures} 
The transport property we discussed in the previous section motivates the construction of a class of models with a singular continuous spectrum. In particular, one can consider the situation in which each energy gap $\delta_n$ is related to a structural resonance; thus, the hierarchy of $\delta_n$ implies a hierarchy of the system structure. In this manuscript, we will consider the system structure as the mirror-symmetric type in 1D lattice systems, which we call the \emph{hierarchical mirror structure} (HMS).

The HMS of a system refers to the case where the underlying system parameters (potential, hopping strength, and so on) themselves have HMS.
Specifically, our model is constructed with a series of self-similar subchains. Each of them is approximately mirror-symmetric with a different length scale. The longer subchains contain the shorter ones on one side of the mirror symmetry point. A cartoon picture of an HMS system and the dynamics of a local particle in it are shown in Fig.~\ref{fig:intro}(a) (where lattice sites are visualized as small potential wells).
Initially, a particle is placed at site $a$. Due to the (approximate) mirror symmetry point nearby, it will first oscillate back and forth between sites $a$ and $b$ at time $t\sim \omega_1^{-1}$. When $t\sim \omega_2^{-1}$, the previous particle configuration (oscillating between sites $a$ and $b$) will oscillate between the pair $a,b$ and its mirror pair $d,c$, and so on.

Note that the ``approximate mirror symmetry'' inside HMS we mentioned above is not a rigorous definition.
Even though there have been some concrete bounds in the mathematical literature~\cite{Jitomirskaya1994,hof1995singular} for some special cases, it is difficult to articulate general conditions/requirements of what counts as an approximate mirror symmetry.
Even ``good enough to induce resonances'' cannot be easily rewritten as a rigorous definition, especially when we consider the generalization to many-body systems.
Instead, our strategy for both single-particle (Sec.~\ref{sec:SPSolvable}) and many-body (Sec.~\ref{sec:MBSolvable}) cases will be to show rigorously that what is definitely going to happen if we assume the approximate mirror structure is good enough in some particular limit.
Then, we will discuss in Sec.~\ref{sec:nonideal} what we can say about some commonly studied critical models beyond this solvable limit (part of it involves a more general but non-solvable model of the freezing/protection condition presented in Appendix~\ref{sec:symmetricRandom}).

Note that the HMS is not the only structure that can lead to singular continuous spectra.
For example, the \emph{frequency resonance} discussed in Ref.~\cite{jitomirskaya2019critical} (based on Refs.~\cite{gordon1976point,Avron1982SingularCS}) instead has hierarchies of Bloch-wave-like resonances over a large number of regions.
In most of the well-studied models, HMS may ``mix'' with other types of structures.
For example, in the Fibonacci quasicrystal, there are both mirror-like and other kinds of structural resonances (see Sec.~\ref{sec:fibonacci}).
We believe that the results of this paper can be generalized to more complicated types of resonances, although the description is expected to be much more complex.

\subsection{Effects of interaction}

In this subsection, we provide an intuitive explanation of the role of interaction in a many-body system with HMS, based on the cartoon picture in Fig.~\ref{fig:intro}(a).

Supposing the separation of the transport timescale still holds in the many-body cases (which is realized by HMS in our cases), one can study the interaction effects on the whole system by studying each level of the hierarchy separately. Let us first study the lowest level, which refers to the subchain from site $a'$ to $b'$ in Fig.~\ref{fig:intro}(a). In other words, we focus on the system behavior at $t_1\sim \omega_1^{-1}$ and ignore the long-time behavior ($t\gg t_1$) for now. Suppose that half of this subchain is disordered enough to have a finite-size MBL description, and the coupling is local in the middle of the chain; the many-body eigenstates of the whole subchain can therefore be theoretically constructed~\cite{Li2025}. As long as the interaction is strong enough and we are not too close to the mirror center, the effect of interaction is to \emph{synchronize} the resonances (oscillations) of the individual particles initially sitting at different potential wells (as depicted in Fig.~1 of Ref.~\cite{Li2025}).

It is important to emphasize that the existence of an HMS forbids the mirror symmetry point from being exact in the whole system, since an exact symmetry point should, by definition, belong to the ``highest" level of the hierarchy, and we cannot find a higher one, which makes the theory self-inconsistent as we require a hierarchy of mirror centers approaching the exact limit.  Thus, the symmetry point can at most be locally exact.
At the lowest level, this explicit symmetry breaking is visualized in Fig.~\ref{fig:intro}(b)--(e) by the red dashed line on site $b'$, stressing that $a'$ and $b'$ do not have exactly the same potential.
Due to the energy detuning of the potential, the (gray) particle in Fig.~\ref{fig:intro}(b) becomes non-resonant between $a'$ and $b'$ (but it may still oscillate in higher levels, such as between sites $a'$ and $d'$).
Note that this non-resonant behavior should be generic for all symmetry-breaking terms, as the tunneling amplitude between mirror sites becomes smaller as we move further away from the mirror center.

Now, interesting things happen when the resonant (purple) and the non-resonant (gray) particles are both present and interact with each other [Fig.~\ref{fig:intro}(d)]: they become collectively \emph{frozen}, where the purple particle cannot resonate as it does in the non-interacting scenario.
There are two ways to understand this phenomenon.
One is via the synchronization picture in Ref.~\cite{Li2025}, which states that when these two particles oscillate, they must oscillate together. Thus, once the potential for these particles on the other side becomes too detuned, the oscillation stops.
Another way is to consider the interaction between two particles as an effective potential. In this case, the purple particle also senses the asymmetry of the original potential (as an effective potential) transferred by the interaction from the gray particle, and becomes non-resonant.

One may bring back the resonance of the purple particle, however, if we put an additional particle at site $b'$ [resulting in two gray particles in Fig.~\ref{fig:intro}(e)].
This can be easily understood through the effective potential picture: now the gray particles on both sides provide an effective potential to the purple particle, and it is largely mirror symmetric if we assume the detuning does not cause much distortion in the orbitals.
In this case, we say that the resonance of the purple particle is \emph{protected} by the two gray particles.
In general, mirror-symmetric configurations protect resonances, so here the resonance between sites $a$ and $b$ is protected by either having no particles at $a'$ and $b'$ [the trivial ``protection'' in Fig.~\ref{fig:intro}(b)] or having both particles at $a'$ and $b'$ [Fig.~\ref{fig:intro}(e)].

Note that the above picture is highly simplified and is only for the illustration of the core idea. The general conditions for freezing and protection are much more complicated and cannot be deduced or described from single-particle dynamics alone. For example, two resonant particles may become frozen when both are present, and the configuration in Fig.~\ref{fig:intro}(e) does not always lead to protected resonance.
We will discuss asymptotically solvable models in Secs.~\ref{sec:SPSolvable} and \ref{sec:MBSolvable} in which a version of this simplified picture is constructed to be exact.
A more realistic (but non-solvable) model of the freezing/protection condition in the still-simplified symmetric random MBL is presented in Appendix~\ref{sec:symmetricRandom}.

\subsection{The many-body critical phases}

Now we are ready to summarize the consequences of the interaction on an infinite system with HMS.
The most natural question is whether the behavior of the infinite system is similar to an MBL phase, an ETH phase, or neither.

To answer the question, we analyze the resonance structure of the particles in the system. Recall from the last subsection that, in the presence of multiple particles with interactions, whether a resonant (purple) particle can participate in the oscillation of the first level ($\sim\omega_1$) depends on whether the particle configuration of non-resonant sites (such as $a'$ and $b'$ in Fig.~\ref{fig:intro}) is mirror symmetric.
The same story can also be applied to all levels of the hierarchy.
For example, whether the purple particle can oscillate between the pair $a,b$ and the mirror pair $d,c$ may depend on whether the configurations of the sites at the left of $a'$ and those at the right of $d'$ are mirror symmetric.
And even if the particle is frozen at a lower level, it may still oscillate in some higher level, depending on the same mirror-symmetric condition of the configurations.

This mirror-symmetric condition is, in fact, connected to the thermodynamic property of the system. To see this, assume that we choose a random initial product state, and ask the question: what is the probability $p_n$ that the $n$th level has a resonating region (in which the particles can resonate from the left half to the right half of the chain)?
The answer can be provided by counting the number of mirror-symmetric configurations inside the non-resonant region.
For example, for the cases in Fig.~\ref{fig:intro}(d) and~\ref{fig:intro}(e), the probability of finding a resonating region (where the purple particle can oscillate) is $p_1=2/4=1/2$, which corresponds to having none or both gray particles.

Next, we ask the question: how likely is it to find \emph{arbitrarily} long resonating regions when $n$ goes to infinity? Assume the resonating region at $(n+1)$th level fully covers the whole region of the $n$th level, then it is equivalent to saying that if a particle is initially placed in a resonating region, how likely it is to be transported to an arbitrarily far distance from its origin in the thermodynamic limit, or how likely a particle can be involved in oscillations in infinitely many levels of the hierarchy.
As the answer to all these equivalent questions, the probability is (assuming the protection conditions can be treated as independent events)
\begin{equation}
    \begin{cases}
        0,\text{ if }\sum_{n=1}^\infty p_n\text{ converges}\\
        1,\text{ if }\sum_{n=1}^\infty p_n\text{ diverges}
    \end{cases}
\end{equation}
Such a result is almost a direct corollary, as $\sum_{n=1}^\infty p_n$ can be interpreted as the expectation value of the number of resonating levels. The first case, in which particles almost certainly do not participate in arbitrarily long-range resonances, is similar to the MBL case and will be considered a \emph{many-body critically localized} (MBC-L) phase; the second case, in which particles almost certainly do, is similar to a many-body extended phase and will be considered a \emph{many-body critically extended} (MBC-E) phase. 
We will construct explicit examples of both phases in Sec.~\ref{sec:MBSolvable}.

The MBC phases have significant differences from the conventional MBL and thermal phases. For a site $j$ in a usual MBL chain, there is a length scale $L_{\rm MBL}$ determined by the primary support of the longest local integral of motion (LIOM) around $j$.
The correlation between site $j$ and another site on any given energy eigenstate must decay exponentially beyond $[j-L_{\rm MBL},j+L_{\rm MBL}]$.
In MBC-L, however, for any $j$ and $L_{\rm MBL}$, there is always a finite fraction (as a function decaying with $L_{\rm MBL}$) of energy eigenstates with $O(1)$ correlation between site $j$ and some site outside $[j-L_{\rm MBL},j+L_{\rm MBL}]$. 
For MBC-E, although we will show that it satisfies a weak form of ETH in section~\ref{sec:MBSolvable} (i.e., a vanishing fraction of eigenstates escape thermalization), the system behaves far from thermal. For example, there is a finite probability of having an arbitrarily long time towards thermalization. The details on the difference between the MBC and the conventional ETH/MBL phases will be discussed in Sec.~\ref{sec:mobilityedge}.

The argument above can also be applied to the picture of energy eigenstates instead of the transport. In particular, the oscillation is replaced by the formation of (many-body) cat states, which are superpositions of the particles being in different positions.
From this point of view, we are able to extract $p_n$ for a thermodynamic ensemble and study the finite-temperature behavior. 
In Sec.~\ref{sec:MBSolvable}, we will construct an explicit model that has a finite-temperature phase transition between MBC-L and MBC-E. In other words, we construct a thermodynamic many-body mobility edge (MBME).
Note that the argument of Ref.~\cite{deroeck2016absence} against the existence of a thermodynamic MBME does not apply due to the lack of well-defined local energy densities, which can be related to the lack of the length scale $L_{\rm MBL}$.

\section{The single-particle solvable model}\label{sec:SPSolvable}

In this and the next section, we construct concrete models with HMS that are asymptotically solvable and satisfy some of the intuitions introduced in Sec.~\ref{sec:summary} in a mathematically rigorous manner.
We will construct the single-particle version in this section, which can be viewed as a simplification of the pictures in Fig.~\ref{fig:intro}(a)--(c).
By ``asymptotically solvable'', we mean that every eigenstate can be labeled and written down explicitly, up to a controllable error term that can be made arbitrarily small, and that the energy spectrum has a clear hierarchical description where the hierarchical energy scales can be arbitrarily separated.

Although several models with a condition similar to HMS have been proven to be singular continuous based on the transfer matrix approach~\cite{Jitomirskaya1994,hof1995singular,koslover2005jacobi}, a many-body generalization is extremely difficult.
As our goal is to construct a many-body generalization, we will instead use the techniques that can be easily generalized to interacting many-body systems. Specifically, we will use a strictly controlled first-order degenerate perturbation theory to allow all eigenstates to be written down explicitly with controllable errors.
We will first prove the singular continuity of the constructed single-particle model in this section and then generalize it to the many-body case in the next section.

Note that the goal of these asymptotically solvable constructions in this and the next sections is to demonstrate that our intuition, summarized in Sec.~\ref{sec:summary}, can indeed be rigorously true, at least in some extreme situations, and to derive their consequences.
What we need here is mathematical existence rather than numerical or experimental feasibility.
Therefore, in the construction below, we will mainly state that the conclusion works as long as some parameters are ``small enough'', rather than providing concrete bounds on the scale of the parameters (which is often a much more difficult task).
``Small enough" here should be construed in the sense of mathematical abstraction, and its significance will be clearer in the details of our various proofs below.

\subsection{Warmup: the all-resonant model}\label{sec:SPAllRes}

\begin{figure*}
    \centering
    \includegraphics[scale=1]{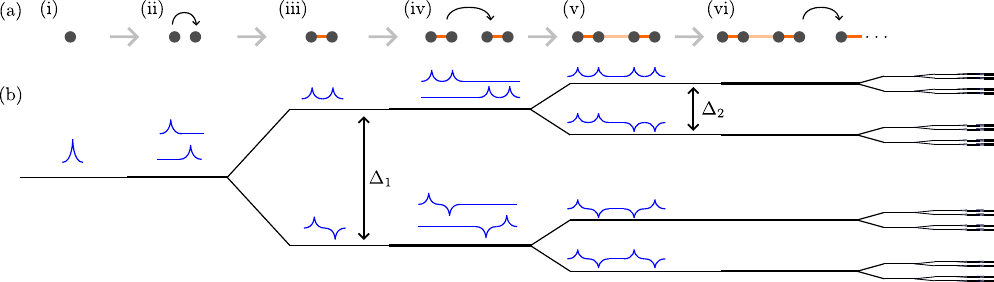}
    \caption{\justifying (a) Iterative construction of the single-particle all-resonant HMS model in Sec.~\ref{sec:SPAllRes}, modeling the purple particle in Fig.~\ref{fig:intro}(a). (b) The spectrum (black) and energy eigenstates (blue), corresponding stepwise with (a).}
    \label{fig:cantor}
\end{figure*}

Before constructing the complete single-particle HMS model, we first consider the simplest case, in which a particle on every site is resonant in all levels.
This case can be thought of as modeling a subsystem of a general HMS chain and demonstrates the basic idea behind the iterative construction and singular continuity.

Let us again illustrate the idea using the cartoon picture in Fig.~\ref{fig:intro}(a). There is a set of sites (more accurately, localized approximate orbitals) to which the purple particle can tunnel (8 are visible in the figure, namely $a$, $b$, $c$, $d$, $e$, $f$, $g$, $h$).
When the sites are more spatially separated (such as $d$ and $e$), the weaker the tunneling amplitude is.
The ranking of the tunneling strength of the 7 ``bonds'' between $a,b$; $b,c$;...; $g,h$ is then marked as $1,2,1,3,1,2,1$, respectively, and so on, where the $j$th term in the sequence is one plus the number of trailing zeros in the binary representation of $j$.
Focusing on the properties above, we consider a 1D chain consisting only of those sites, with the hopping parameters satisfying a similar ranking of tunneling strength. 

This effective model can be constructed iteratively as shown in Fig.~\ref{fig:cantor}(a) with the spectrum and eigenstates visualized stepwise [Fig.~\ref{fig:cantor}(b)].
Specifically, we start with (i) a single site, (ii) make a mirrored copy to become two decoupled sites, and (iii) couple the two sites.
Now we reach the first level of the hierarchy, with eigenstates being even/odd superpositions of two sites with energy splitting $\Delta_1$.
To go to the second level, we again (iv) reflectively copy the chain and (v) couple them with a weaker bond.
Now the spectrum comprises two pairs, each with a splitting $\Delta_2\ll\Delta_1$ (the figure is not to scale), and the eigenstates are approximately equal-weight superpositions of the four sites.
This procedure can be repeated again and again, and the spectrum will converge to a Cantor-like fractal which is of measure zero, and the eigenstates will still be approximately equal-weight superpositions of all sites.

To see the singular continuity of such a model, note that since the energy eigenstates are all approximately equal-weight superpositions, the energy distribution $|\langle j|E\rangle|^2$ can be roughly treated as zero outside the Cantor set and constant inside.
As the weights are evenly divided, there cannot be discrete points where the weights accumulate, leading to the absence of a discrete part.
As the Cantor spectrum is of measure zero, there cannot be a continuous part in the distribution either.
This is the idea behind the proof we will present below for a more general model, which explicitly includes non-resonant sites.

The non-resonant sites will not influence the global spectrum properties until we study the many-body scenario. For completeness, in the full version of our solvable model below, we will include some non-resonant sites as well, although the conclusions remain unchanged if we ignore them in the single-particle case.

\subsection{The solvable HMS Hamiltonian}

\begin{figure*}
    \centering
    \includegraphics[scale=1]{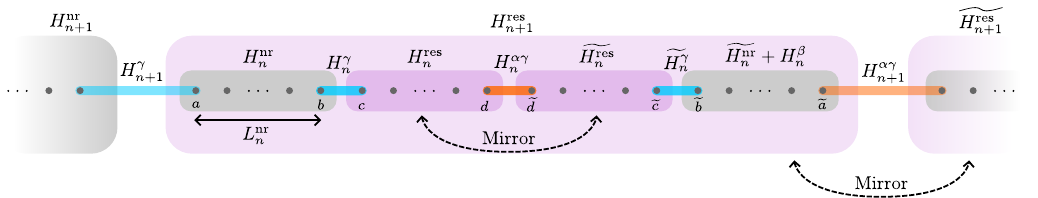}
    \caption{\justifying Recursive construction of the solvable models, which works for both single-particle and many-body versions. At each step, $H^\text{res}_{n+1}$ is formed by making a mirrored copy of $H^\text{res}_n$ (the \emph{resonant region}), adding a pair of new segments $H^\text{nr}_n,\widetilde{H^\text{nr}_n}$ with small symmetry-breaking terms $H^\beta_n$ (the \emph{non-resonant region}), and finally coupling the four segments by bonds $H^\gamma_n,H^{\alpha\gamma}_n,\widetilde{H^{\gamma}_n}$ which are weaker and weaker as $n\to\infty$.}
    \label{fig:solvable}
\end{figure*}

We consider a 1D tight-binding Hamiltonian on the infinite lattice with the form
\begin{equation}\label{eq:SPHinf}
    H_\infty=\sum_{j=-\infty}^\infty t_j(|j\rangle\langle j+1|+|j+1\rangle\langle j|) + \sum_{j=-\infty}^\infty V_j |j\rangle\langle j|
\end{equation}
where the hopping strength $t_j$ and on-site potential $V_j$ are constructed by iteratively connecting finite chains, with the steps described below.

\subsubsection{Initialization of the building blocks}\label{sec:SPInit}

The chain is built by perturbing and connecting a sequence of initial subchains, which we call the \emph{building blocks} of the chain.
They are the resonant region of the first level $H^\text{res}_1$, the non-resonant regions $H^\text{nr}_n,n=1,2,\ldots$ for all levels, as well as their mirror counterparts $\widetilde{H^\text{res}_1}$ and $\widetilde{H^\text{nr}_n}$.
Their lengths are denoted by $L^\text{res}_1$ and $L^\text{nr}_n$, respectively, and are treated as input parameters of our construction.
For convenience, the non-tilde (tilde) ones will be used as the left (right) one in each mirror pair.

Our construction itself only requires that each of the initial building blocks has a non-degenerate energy spectrum, and that all of its eigenstates have nonzero amplitudes at the boundaries.
So almost any choice of $t_j$ and $V_j$ in the block would work.
However, to have a clean asymptotic solution of the model, we will make a specific choice regarding the form of the Hamiltonian.
Let each of these initial building blocks $H^\text{res}_1$ and $H^\text{nr}_n,n=1,2,\ldots$ take the form of
\begin{equation}\label{eq:building-block}
    H=\sum_{j=1}^{L} V_j |j\rangle\langle j|+\delta\cdot\sum_{j=1}^{L-1} t_j(|j\rangle\langle j+1|+|j+1\rangle\langle j|).
\end{equation}
where $L$ is the corresponding size of the blocks.
We will construct the parameters $V_j$ and $t_j$ to be independent uniform random numbers in $[-1,1]$, and $\delta>0$ to be a small number controlled by the error bound, which is elaborated below.
The random numbers are used to avoid accidental degeneracies and symmetries (and will be used to make the thermodynamic limit well-defined in the many-body generalization).
Note that each block is constructed with a different (independent) set of random numbers, and after this initialization step, all the parameters in the blocks are fixed, and when we later reflectively clone the block, each copy will inherit the same fixed choice of numbers.

While $\delta$ itself is to provide the non-zero amplitudes of the eigenstates at the boundary, its smallness is to make the eigenstates of $H$ close to the position eigenstates, which enables a clean analytical description of the final eigenstates up to a controllable error bound.
In order to bound the error from the position eigenstates, note that as $\delta\to 0$, the eigenstates $|\theta_j\rangle$ of $H$ converge to the position eigenstates $|j\rangle$.
So for any choice of an error bound $\epsilon_0>0$, we can fix a small enough $\delta$ such that
\begin{equation}\label{eq:SPInitError}
    |\langle j|\theta_j\rangle|^2>1-\epsilon_0.
\end{equation}
Here we treat $\epsilon_0$ as a block-independent error bound given as a parameter of our system, while $\delta$ is block-dependent and is expected to be smaller for larger blocks.

This approach of making the error bound uniform (regardless of the size of the blocks) can be treated as an idealization of localized systems so that they are strictly devoid of accidental resonances.
This is especially important when we generalize our construction to the many-body case, where controlling accidental resonances in general is still an open problem.
The rest of our construction will follow this philosophy of using artificial limits to avoid accidental resonances, so all the resonances will be of the mirror type.

The notation $|\theta_j\rangle$ will be used to denote the eigenstates of the initial building block that approximate $|j\rangle$, which are distinguished from the eigenstates of other Hamiltonians denoted by $|\psi_j\rangle$, sometimes with additional subscripts.
During the construction, those initial building blocks will be copied, mirrored, and sometimes perturbed, but every site $j$ can be traced back to exactly one of those initial building blocks, and hence the notation $|\theta_j\rangle$ will unambiguously denote its (possibly mirrored) eigenstate that approximates $|j\rangle$.

\subsubsection{The iteration step}\label{sec:SPiter}

Next, we construct the iteration step from $H^\text{res}_n$ to $H^\text{res}_{n+1}$ (for $n=1,2,\ldots$) as illustrated in Fig.~\ref{fig:solvable}.
The site labels ($a,b,c,d$) and limiting parameters ($\alpha,\beta,\gamma$) will be treated as temporary variables within the iteration step. We will not put a subscript $n$ for brevity, although they are conceptually dependent on the level.

First, we make a mirror copy of $H^\text{res}_n$ in the recursion to become $\widetilde{H^\text{res}_n}$.
Specifically, $\widetilde{H^\text{res}_n}$ is formed by replacing the position basis with the mirrored one $|c\rangle\to|\widetilde{c}\rangle$, \dots, $|d\rangle\to|\widetilde{d}\rangle$ in $H^\text{res}_n$.
The eigenstate labeling will also be mirrored.
That is, for an $H^\text{res}_n$ eigenstate $|\psi_j\rangle$ (the label will also be constructed recursively), the corresponding $\widetilde{H^\text{res}_n}$ eigenstate with the same energy is denoted by $|\psi_{\widetilde{j}}\rangle$, where $\widetilde{j}=\widetilde{d}+d-j$ is the mirrored site of $j$.

Next, we place the non-resonant building block $H^\text{nr}_n$ at the left of $H^\text{res}_n$ by replacing $|1\rangle\to|a\rangle,\ldots,|L^\text{nr}_n\rangle\to|b\rangle$ and similarly for the eigenstate labels.
We similarly make a mirrored copy to become $\widetilde{H^\text{nr}_n}$ by $|j\rangle\to|\widetilde{j}\rangle$.
However, since we want it to be non-resonant, we add a symmetry-breaking perturbation
\begin{equation}
    H^\beta_n=\beta\cdot\left[\sum_{j=\widetilde{b}}^{\widetilde{a}-1} t_j'(|j\rangle\langle j+1|+|j+1\rangle\langle j|) + \sum_{j=\widetilde{b}}^{\widetilde{a}}V_j' |j\rangle\langle j|\right]
\end{equation}
so that the right non-resonant block becomes $\widetilde{H^\text{nr}_n}+H^\beta_n$.
Here $t'_j$ and $V'_j$ are, again, independent uniform random numbers in $[-1,1]$, and $\beta>0$ is a number small enough so that the eigenstates of the two blocks still look similar by mirroring (in particular, the eigenstates of $\widetilde{H^\text{nr}_n}+H^\beta_n$ can be labeled consistently with those of $H^\text{nr}_n$), but large enough so that the two blocks do not resonate after we couple all blocks together.
Note that $\beta$ will not be treated as part of the perturbation in the degenerate perturbation step below, but as a parameter generating the unperturbed eigenstates. We will discuss how to determine $\beta$ later.

The four subchains are then coupled by
\begin{equation}
\begin{aligned}
    H^\gamma_n&=\gamma\,(|b\rangle\langle c|+|c\rangle\langle b|),\\
    H^{\alpha\gamma}_n&=\alpha\gamma\,(|d\rangle\langle \widetilde{d}|+|\widetilde{d}\rangle\langle d|),\\
    \widetilde{H^\gamma_n}&=\gamma\,(|\widetilde{c}\rangle\langle \widetilde{b}|+|\widetilde{b}\rangle\langle \widetilde{c}|)
\end{aligned}
\end{equation}
to form the final Hamiltonian
\begin{equation}
    H^\text{res}_{n+1}=H^\text{nr}_n+H^\gamma_n+H^\text{res}_n+H^{\alpha\gamma}_n+\widetilde{H^\text{res}_n}+\widetilde{H^\gamma_n}+\widetilde{H^\text{nr}_n}+H^\beta_n,
\end{equation}
where $\gamma>0$ is a small number to make the degenerate perturbation theory as used below work within some error bound, and will be specified later.
The number $\alpha>0$ is also intended to be small, and will be important in the many-body version of our model.
Here in the single-particle version, however, it does not play an important role, and we may just set $\alpha=1$.

We treat $\gamma$ as a perturbation while $\alpha$ and $\beta$ are not.
That is, the unperturbed and perturbation parts for $H^\text{res}_{n+1}$ are
\begin{align}
    H^\text{unp}_{n+1} &= H^\text{nr}_n+H^\text{res}_n+\widetilde{H^\text{res}_n}+\widetilde{H^\text{nr}_n}+H^\beta_n,\\
    H^\text{pert}_{n+1} &= H^\gamma_n+H^{\alpha\gamma}_n+\widetilde{H^\gamma_n}. \label{eq:SPpert}
\end{align}
The unperturbed eigenstates $|\psi_j\rangle_\text{unp}$ of $H^\text{unp}_{n+1}$ are straightforwardly labeled by site indices between $a$ and $\widetilde{a}$. That is, if a site $j\in[a,b]$, $[c,d]$, $[\widetilde{d},\widetilde{c}]$, or $[\widetilde{b},\widetilde{a}]$, then $|\psi_j\rangle_\text{unp}$ is the eigenstate of $H^\text{nr}_n$, $H^\text{res}_n$, $\widetilde{H^\text{res}_n}$, or $\widetilde{H^\text{nr}_n}+H^\beta_n$, respectively, with the corresponding label of that Hamiltonian defined above.
Now the energy levels of $H^\text{unp}_{n+1}$ are either nondegenerate (if $j\in[a,b]$ or $[\widetilde{b},\widetilde{a}]$) or twofold degenerate (if $j\in[c,\widetilde{c}]$) with subspace $\{|\psi_j\rangle_\text{unp},|\psi_{\widetilde{j}}\rangle_\text{unp}\}$ and subspace Hamiltonian (assume without loss of generality that $j\in[c,d]$)
\begin{equation}
    \gamma\,\begin{pmatrix}
        0 & _\text{unp}\langle\psi_j|d\rangle\langle \widetilde{d}|\psi_{\widetilde{j}}\rangle_\text{unp} \\
        _\text{unp}\langle \psi_{\widetilde{j}}|\widetilde{d}\rangle\langle d|\psi_j\rangle_\text{unp} & 0
    \end{pmatrix},
\end{equation}
which leads to the new eigenstates being even/odd superpositions of them.
Note that here and below, we directly assume the normal situation with probability one. That is, the random numbers we start with do not produce accidental degeneracies and/or accidental symmetries that would make the matrix element vanish.

\begin{figure}
    \centering
    \includegraphics[scale=1]{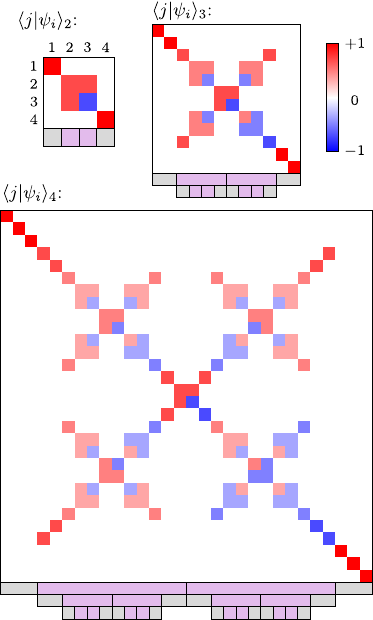}
    \caption{\justifying Visualization of the eigenstates and their labeling of $H^\text{res}_n$ with $L^\text{res}_1=1$, $L^\text{nr}_n=n$. The $(i,j)$ entry of the matrix is $\langle j|\psi_i\rangle_n$, assuming all error bounds are negligible. The figure is for illustrative purposes and not plotted from numerical data. The hierarchical structure is indicated at the bottom of the matrix, with magenta (gray) being (non-) resonance regions. }
    \label{fig:eigenstates}
\end{figure}

By making $\gamma$ small enough, one can make first-order degenerate perturbation theory work as accurately as possible.
The perturbed eigenstates, which are also the eigenstates of $H^\text{res}_{n+1}$ to be used for the next iteration, are then
\begin{equation}\label{eq:SPPert}
    |\psi_j\rangle_{n+1}\approx\begin{cases}
        |\psi_j\rangle_\text{unp} & \text{if }j\in[a,b]\cup[\widetilde{b},\widetilde{a}]\\
        \frac{1}{\sqrt2}(|\psi_j\rangle_\text{unp}+|\psi_{\widetilde{j}}\rangle_\text{unp}) & \text{if }j\in[c,d]\\
        \frac{1}{\sqrt2} (|\psi_{\widetilde{j}}\rangle_\text{unp}-|\psi_j\rangle_\text{unp}) & \text{if }j\in[\widetilde{d},\widetilde{c}] 
    \end{cases}
\end{equation}
Here, the choice of which one is $+$ and which one is $-$ is completely arbitrary, and our specific choice is solely for the convenience of labeling. We have added a subscript $n+1$ so that we can later distinguish the eigenstates for different iteration steps.
For an example of such eigenstates and labeling, see Fig.~\ref{fig:eigenstates}.

\subsubsection{Error bounds}\label{sec:SPerr}

Now we are ready to determine the small numbers $\beta$ and $\gamma$ for the iteration step from $n$ to $n+1$ (note that they implicitly depend on $n$ as we mentioned).
From the discussion above, we have
\begin{multline}\label{eq:SPPertIdeal}
    |\psi_j\rangle_\text{lim}:=\lim_{\beta\to0}\lim_{\gamma\to0}|\psi_j\rangle_{n+1}\\=
    \begin{cases}
        |\psi_j\rangle_n & \text{if }j\in[a,b]\cup[\widetilde{b},\widetilde{a}]\\
        \frac{1}{\sqrt2}(|\psi_j\rangle_n+|\psi_{\widetilde{j}}\rangle_n) & \text{if }j\in[c,d]\\
        \frac{1}{\sqrt2} (|\psi_{\widetilde{j}}\rangle_n-|\psi_j\rangle_n) & \text{if }j\in[\widetilde{d},\widetilde{c}] 
    \end{cases}
\end{multline}
Here $|\psi_j\rangle_n$ are the eigenstates of the Hamiltonian in the same limit
\begin{equation}
    H^\text{lim}_{n+1}:=\lim_{\beta\to0}\lim_{\gamma\to0}H^\text{res}_{n+1}=H^\text{nr}_n+H^\text{res}_n+\widetilde{H^\text{res}_n}+\widetilde{H^\text{nr}_n}
\end{equation}
with consistent labeling.
Note that this does not cause a notational inconsistency between $|\psi_j\rangle_n$ and $|\psi_j\rangle_{n+1}$, as when $j$ is in the block of $H^\text{res}_n$, it is indeed the $|\psi_j\rangle_n$ from the previous iteration.
Also note that the order of limits reflects the conceptual requirements of $\beta$ to be both ``large enough'' to suppress resonances and ``small enough'' to have similar eigenstates from its mirror block.

The above limits imply that as long as $\beta$ and $\gamma$ are chosen to be small enough, we can make $|\psi_j\rangle_{n+1}$ as close to $|\psi_j\rangle_\text{lim}$, and $H^\text{res}_{n+1}$ as close to $H^\text{lim}_{n+1}$ as we want.
For the technical requirements in all the subsequent derivations, we need an error bound parameter $0<\epsilon_n<1$ associated with this iteration, chosen to be smaller than half of any energy difference between subspaces in $H^\text{lim}_{n+1}$.
Then we choose $\beta$ and $\gamma$ to make all of the following rigorous conditions true:
\begin{equation}
    2L^\text{nr}_{n}|\langle \theta_j|\psi_j\rangle_{n+1}|^2<\epsilon_n
\end{equation}
if $\langle \theta_j|\psi_j\rangle_\text{lim}=0$, and
\begin{equation}
    \left|\frac{|\langle \theta_j|\psi_j\rangle_{n+1}|^2-|\langle \theta_j|\psi_j\rangle_\text{lim}|^2}{|\langle \theta_j|\psi_j\rangle_\text{lim}|^2}\right|<\epsilon_n,
\end{equation} 
if $\langle \theta_j|\psi_j\rangle_\text{lim}\neq0$, where $|\theta_j\rangle\approx|j\rangle$ is the eigenstate of (the copy of) the initial building block where $j$ is in.
For energies:
\begin{equation}
    \left\|H^\text{res}_{n+1}-H^\text{lim}_{n+1}\right\|<\epsilon_n,\quad 2|\Delta E^\text{max}_n|<\epsilon_n \label{eq:ineqshift}
\end{equation}
where $\|\cdot\|$ denotes the operator norm, $|\Delta E^\text{max}_n|$ is the maximal energy shift between corresponding eigenstate indices from $H^\text{lim}_{n+1}$ to $H^\text{res}_{n+1}$.
Iterative applications of Eq.~(\ref{eq:SPPertIdeal}), along with the error bounds, lead to the asymptotic solution of our model: each eigenstate can be labeled and expressed exactly up to a controllable error bound (Fig.~\ref{fig:eigenstates}), and the energy shifts at each level are also controlled.

To have a controllable accumulated error as we take $n\to\infty$, we will require $2^n\epsilon_n\to 0$, which implies various notions of convergence we need.
In addition, we require that the accumulated error from $n$ to $\infty$, denoted by $\epsilon_{\geq n}$ (which can be defined by $\prod_{n'=n}^\infty(1+\epsilon_n)-1$ for what we need), satisfies $\epsilon_{\geq n}<2\epsilon_n$.
For physical intuition, we define $\Delta^\text{max(min)}_n$ for the maximum (minimum) subspace degeneracy lifting from $H^\text{lim}_{n+1}$ to $H^\text{res}_{n+1}$, and we will assume informally that $\Delta^\text{max}_{n+1}\ll \Delta^\text{min}_{n}$, so that we have a sequence of timescales $\{t_n\}$ with $1/\Delta^\text{min}_n\ll t_{n}\ll 1/\Delta^\text{max}_{n+1}$ corresponding to the levels.
Some notion of this can be derived as a consequence of the strict error bounds, but we never use it in formal proofs.

As we already mentioned, the purpose of such error-bound requirements is only for existence proofs, and is expected to be far from optimal if we only want the same qualitative behaviors.

\subsubsection{Infinite system}\label{sec:SPinf}

Having defined the initialization and the iteration step, we can now iterate to any finite $n$ given the input parameters $L^\text{res}_1$, $\{L^\text{nr}_n\}$, and the error bounds. However, defining the infinite system Hamiltonian $H_\infty$ requires some special care.
Usually, when we think of taking a system to infinity, we are somehow ``staying in the bulk of the system'' and adding more and more faraway sites.
As our system is highly non-translationally invariant, for ``staying in the bulk'' to be well-defined, we need to see how to embed $H^\text{res}_n$ into $H^\text{res}_{n'}$ for higher $n'$.
To do this, we note that every $H^\text{res}_n$ appears twice inside $H^\text{res}_{n+2}$ (the left one directly, and the right one after two mirrored copying), and in general it appears $2^k$ times in $H^\text{res}_{n+k+1}$, so we have infinitly many ways to iterate to infinity while staying in the middle of the system with a consistent neighborhood. 
We will treat the choice of this embedding as a random sequence and use it to define $H_\infty$, similar to how we use random numbers to initialize the potentials.

Equivalently, we may think of the ways of embedding as a slightly generalized version of the iteration step, in which, when we make a mirrored copy of $H^\text{res}_n$, we can either copy to the right or to the left. Now, all the ways of taking the $n\to\infty$ limit can be described by an infinite binary sequence of ``left'' or ``right,'' corresponding to the direction of copying at the $n$th step.

The physical way to think about this $n\to\infty$ limit is that we first prepare a finite chain $H^\text{res}_n$ and a particle concentrated near the center of the chain.
As time evolves, we can measure the particle to obtain some probability distribution, but there is a maximum timescale $t_n$ above which finite-size effects become important.
In this case, if we want to probe longer timescales, we need to extend the boundary of the chain by embedding the original $H^\text{res}_n$ into that of a higher $n$.
Thus, as long as $n$ is large enough, we can probe up to arbitrarily long timescales $t_n$ with a finite system.

\subsection{Progressive approximations on infinite lattice}\label{sec:SPapprox}

\begin{figure*}
    \centering
    \includegraphics[scale=1]{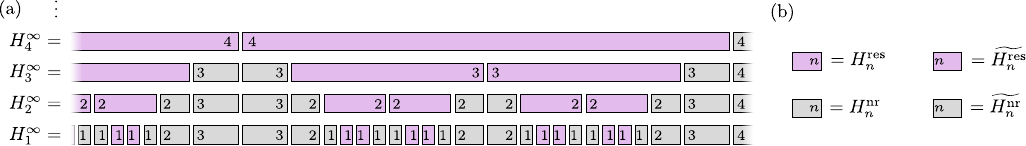}
    \caption{\justifying Progressive approximations of the solvable model on an infinite lattice, which works for both single-particle and many-body versions. The Hamiltonian $H^\infty_n$ at each step is a sum of decoupled segments. In the limit of $n\to\infty$, $H^\infty_n$ converges to the final model $H_\infty$ (in the sense of operator norm in the single-particle case, and term-wise in the many-body case).}
    \label{fig:approx}
\end{figure*}

The above construction defines $H_\infty$ as a limit of the sequence $H^\text{res}_n$, each on a finite chain with size growing in $n$.
However, to study the properties of $H_\infty$, it is better to make it a limit directly on the infinite lattice, so that the limit is taken within a fixed Hilbert space.
That is, we want $H^\infty_n\to H_\infty$ with each $H^\infty_n$ being defined already on the infinite lattice rather than in a finite system.
In this framework, we first make infinite copies of decoupled finite subchains, and then progressively couple these copies with the coupling Hamiltonians ($\alpha$ and $\gamma$ terms at all levels), as well as the symmetry-breaking ($\beta$) terms, to build the $n$th level. 

First, we define the uncoupled Hamiltonian $H^\infty_1$ by placing all initial building blocks ($H^\text{res}_1$ and $H^\text{nr}_n$) in their proper positions, so that they can be progressively coupled (defined below) to $H_\infty$ for a fixed choice of embedding sequence.
The resulting Hamiltonian is a sum of decoupled chains, as shown in Fig.~\ref{fig:approx}.

Now for each $n$, going from $H^\infty_n$ to $H^\infty_{n+1}$ is exactly by our original iteration step that merges $H^\text{nr}_n$, $H^\text{res}_n$, $\widetilde{H^\text{res}_n}$, and $\widetilde{H^\text{nr}_n}$ into $H^\text{res}_{n+1}$.
But now it is happening for an infinite number of copies on the infinite lattice simultaneously.
Note that we have $\|H^\infty_{n+1}-H^\infty_n\|<\epsilon_n$ by the error bound, so that we have the desired limit of $H^\infty_n\to H_\infty$ in the sense of the operator norm.

This construction also allows for a unified labeling of eigenstates for each $n$ (although there are infinitely many choices of eigenbases, as the energy subspaces are highly degenerate, we will use the choice with the most local eigenstates).
Recall that we have defined the labeling of eigenstates for each segment of chains using the position index.
Now, for each $n$, $H^\infty_n$ is just a decoupled collection of them, so we have an unambiguous indexing system by shifting that of each block to its respective range of sites.
We denote the eigenstate of $H^\infty_n$ corresponding to site $j$ as $|\psi_j\rangle_n$, which is localized and normalized as $\langle\psi_j|\psi_j\rangle_n=1$.
In particular, we have $|\psi_j\rangle_1=|\theta_j\rangle$, and for all $j$ in the gray area in Fig.~\ref{fig:approx}, we have $|\psi_j\rangle_n=|\theta_j\rangle$ as well.
That is, $|\psi_j\rangle_n$ only differs from $|\theta_j\rangle$ in the purple areas for $n=2$ and above.
The energy eigenvalues corresponding to $|\psi_j\rangle_n$ are denoted by $E_{j,n}$. However, unlike $H^\infty_n$, the eigenstates $|\psi_j\rangle_n$ do not converge when $n\to\infty$ in the usual sense. For example, ``$|\psi_j\rangle_\infty$'' may not be normalizable. 
The quantity that corresponds to the eigenstates in a finite Hilbert space, and yet has a converging limit to infinity, is the spectral measure~\cite{reed1980methods}, which we will use to formulate our proof in the next subsection.

\subsection{Singular continuity}

Now we are ready to prove that $H_\infty$ is purely singular continuous.
For some background on the rigorous treatment of spectral measure in mathematical physics, see, for example, Ref.~\cite{reed1980methods}. 
We will fix the initial local state as $|\psi\rangle=|\theta_{i}\rangle$ for a general $i\in\mathbb{Z}$.
When $H^\infty_n\to H_\infty$, the spectral measure $\mu_n$ (energy distribution) of $|\theta_{i}\rangle$ for $H^\infty_n$ weakly converges to the spectral measure $\mu$ for $H_\infty$.
That is, for every bounded continuous function $g(E)$ of energy, we have
\begin{equation}
    \lim_{n\to\infty}\int g\,d\mu_n=\int g\,d\mu
\end{equation}
This allows us to discuss the singular continuity of $\mu$ by limiting the properties of $\mu_n$, which is much simpler to calculate.

Since the energy eigenstates of $H^\infty_n$ are purely discrete and labeled by the sites, the spectral measure of the local state $|\theta_i\rangle$ can be described by the weights $w_{j,n}=|\langle \theta_i|\psi_j\rangle_n|^2$, with the corresponding energy $E_{j,n}$. 
That is, the energy integral of $\mu_n$ becomes a discrete sum
\begin{equation}
    \int g\,d\mu_n=\sum_j g(E_{j,n})w_{j,n}
\end{equation}
Moreover, as $H^\infty_n$ is a collection of decoupled chains, it is actually a finite sum.
Therefore, the problem of proving the spectral property of $H_\infty$ reduces to the property of finite sums involving $w_{j,n}$ and $E_{j,n}$.

We proceed by first calculating how the weights $w_{j,n}$ and energies $E_{j,n}$ evolve with $n$, and then show the absence of atomic (pure point) and absolutely continuous parts in $\mu$ based on the properties of $w_{j,n}$ and $E_{j,n}$ to complete the proof.
The core idea of the evolution of the weights in $n$ will have a direct counterpart in the many-body setting, which leads to a weak ETH-like behavior.

\subsubsection{Splitting of the weights on the orbit}\label{sec:splittingweights}

Let us first calculate how $w_{j,n}$ evolves with $n$ for a fixed $j$.
We will see that after some $n_0$, $w_{j,n}$ keeps being divided by halves.
Intuitively, as we increase $n$, a particle initially at site $j$ can oscillate to more and more sites (named as the \emph{orbit}) due to resonances, so the weight initially at site $j$ is evenly divided among those sites in its orbit.

First note that there exists a finite $n_0$ as an upper bound, below which $|\psi_j\rangle_n=|\theta_j\rangle$, so that if $j\neq i$, $|\psi_j\rangle_n$ does not overlap with $|\theta_i\rangle$.
More specifically, $n_0$ is the lowest $n$ such that sites $i$ and $j$ appear in the same $H^\text{res}_n$ or $\widetilde{H^\text{res}_n}$ block in $H^\infty_n$. 
So if $n<n_0$, we have $w_{j,n}=\delta_{i,j}$, and the value of $E_{j,n}$ is irrelevant for $j\neq i$.

Next, for $n=n_0$, due to the perturbative coupling that forms $H^\text{res}_{n}$, $w_{j,n}$ first deviates from $\delta_{i,j}$ (again, we assume that the random numbers do not produce accidental symmetries).

Finally, for $n>n_0$, every time a pair of $H^\text{res}_n$ and $\widetilde{H^\text{res}_n}$ blocks are coupled, the weight $w_{j,n}$ splits into $w_{j,n+1}$ and $w_{\widetilde{j},n+1}$ evenly up to a relative error of $\epsilon_n$.
Therefore, when we go from $n_0$ to $n$, there become $2^{n-n_0}$ numbers of $j'$ among which $w_{j',n}$ is evenly split, up to a relative error of $\epsilon_{\geq n_0}$.
Dynamically, this set of $j'$ forms the orbit that a particle initially at the $j$th eigenstate of $H^\infty_{n_0}$ can oscillate to by the mirror operations of the levels from $n_0$ to $n$.

\subsubsection{Evolution of the energy tree}

Next, we discuss how the energy $E_{j,n}$ evolves corresponding to the evolution of the $w_{j,n}$ for a fixed $j$.
This evolution is basically captured in Fig.~\ref{fig:cantor}, where the step-wise energy splitting forms a tree-like structure that we call an \emph{energy tree}, with layers corresponding to $n$ and branches at that layer corresponding to the energy level at that $n$.
Unlike the simple case of Fig.~\ref{fig:cantor}, however, as we now include non-resonant sites for each level and define the Hamiltonian on an infinite lattice, there will be an infinite number of trees, which makes the description slightly more complicated.

Again, for $n<n_0$, we need to consider the cases of $i=j$ and $i\neq j$ separately.
For $i\neq j$, we have $w_{j,n}=0$ when $n$ is below some $n_0$, so the corresponding $E_{j,n}$ is not relevant; the energy line corresponding to $j$ first appears at $n=n_0$.
For $i=j$, we have instead that $w_{j,n}=1$ below $n_0$, so the line is relevant from the beginning. However, in this case, $j$ is always in the same non-resonant block for all $n<n_0$, so $E_{j,n}$ is simply a constant there.
It first gets shifted (bounded by $\epsilon_{n_0}$) at $n=n_0$.

Now for $n>n_0$, every time a pair of $H^\text{res}_n$ and $\widetilde{H^\text{res}_n}$ blocks are coupled, and the weight $w_{j,n}$ splits into $w_{j,n+1}$ and $w_{\widetilde{j},n+1}$, the energy line $E_{j,n}$ splits into two lines, $E_{j,n+1}$ and $E_{\widetilde{j},n+1}$, whose differences from the original line is by the second inequality in Eq.~(\ref{eq:ineqshift}) $|E_{j,n+1}-E_{j,n}|,|E_{\widetilde{j},n+1}-E_{j,n}|<\frac{1}{2}\epsilon_n$ (so that the splitting between them $|E_{\widetilde{j},n+1}-E_{j,n+1}|<\epsilon_n$). 
Note that these shifts never cause crossing between energy lines with positive weights, as all of those lines at stage $n$ are the spectral lines in $H^\text{lim}_{n+1}$, and $\epsilon_n$ is assumed to be smaller than half of any energy difference of its energy eigenspaces.

Now we can organize all of the $E_{j,n}$ lines with $w_{j,n}\neq 0$ into a set of trees, where each branch at layer $n$ is represented by a label $(j,n)$ and has child branches $(j,n+1)$ and $(\widetilde{j},n+1)$ at the next layer (except when $j=i$, where it only has a single child branch $(j,n+1)$ below some $n_0$).
If a branch $(j,n)$ is not a child of another branch, we call it a \emph{root line}, so that the set of root lines is in one-to-one correspondence with the set of trees.
For a branch $(j,n)$, the energy range $\sup E_{j',n'}-\inf E_{j',n'}$ among all its children $(j',n')$ is called the \emph{width} of the branch $(j,n)$.
The supremum of the total weight $\sup_{n'} \sum_{j'} w_{n',j'}$ among its children $(j',n')$ is called the \emph{final weight} of the branch $(j,n)$, which is bounded by $w_{n,j}(1\pm\epsilon_{\geq n})$.

Although there are an infinite number of trees, most of them will be irrelevant to us.
To see this, note that for a given $n_1$, there is only a finite number of branches below $n_1$, and therefore only a finite number of root lines.
So the rest of the infinite number of trees all have root lines above $n_1$.
Suppose that $n_1$ is large enough so that $i$ is in a resonant block.
Then, for each $n>n_1$, the root lines forming at this stage are due to the formation of $H^\text{nr}_n$. Thus, the number of root lines at $n$ is $2L^\text{nr}_{n-1}$.
By the error bound, we have $w_{n,j}<\epsilon_{n-1}/2L^\text{nr}_{n-1}$ for each root line $(n,j)$, and thus the total final weights of all of these root lines appearing at $n$ is bounded above by $\epsilon_{n-1}(1+\epsilon_{\geq n})$.
By the convergence of accumulated errors from $n_1$ to infinity, we see that as long as $n_1$ is large enough, we can make the final weights of all the root lines above $n_1$ arbitrarily small, and therefore we only need to consider the trees with root lines below $n_1$, a finite number of them.

We will see that the singular continuity of $H_\infty$ is a direct consequence of the asymptotic properties of the width and final weights of the branches of these finite number of trees.

\subsubsection{Absence of pure point spectrum}
In this subsection, we prove that $\mu$ has no pure point part, that is, there is no delta peak in the energy distribution of $|\theta_i\rangle$.
The idea is that having a delta peak in $\mu$ means that the weights in $\mu_n$ must be more and more concentrated around its final position as $n\to\infty$.
So if we show that a small energy window will eventually cover vanishing weight as $n\to\infty$, the desired result will follow.

Take $g_{E_0,\delta}(E)$ to be a family of continuous functions that is zero outside a small energy window $[E_0-\delta,E_0+\delta]$ and has a single peak of size $1$ at $E_0$. The condition that $\mu$ has no delta peak at energy $E_0$ is equivalent to:
\begin{equation}
    \lim_{\delta\to 0}\int g_{E_0,\delta}\, d\mu=0
\end{equation}
By approximating $\mu$ using $\mu_n$, it is equivalent to
\begin{equation}
    \lim_{\delta\to 0}\lim_{n\to\infty}\sum_j g_{E_0,\delta}(E_{j,n})w_{j,n}=0
\end{equation}
So, given $\epsilon>0$, we want to find $\delta_0>0$ such that $0<\delta\leq\delta_0$ implies
\begin{equation}
    \lim_{n\to\infty}\sum_j g_{E_0,\delta}(E_{j,n})w_{j,n}<\epsilon
\end{equation}
Hence, we want to find $\delta_0$ such that as we evolve $n$ to a large enough number, no matter how the new energy lines appear and split, the weights falling into the energy window $[E_0-\delta,E_0+\delta]$ will stay below $\epsilon$.

To find this $\delta_0$, we will use the fact that most trees are negligible. First, choose $n_1$ such that the total final weights of the trees with root lines beyond $n_1$ are below $\epsilon/2$.
This way, we only need to show that a finite number $m$ of trees (those with root lines below $n_1$) will not have a weight greater than $\epsilon/2$ falling into $[E_0-\delta,E_0+\delta]$.
This can be done by showing that each of the $m$ trees will contribute at most $\epsilon/2m$ to such weight.

Now, for each tree with root line $(j,n_0)$, as $n\geq n_0$ grows, its initial weight $w_{j,n_0}$ splits evenly into the orbit of $j$ up to a relative error of $\epsilon_{\geq n}$.
In order to avoid large weights from going into the energy window, we can pick an $n$ such that each branch $(j,n)$ of this tree at this $n$ has a final weight $<\epsilon/4m$, and require $\delta_0$ to be smaller than the width of each of these $(j,n)$.
As different branches do not overlap with each other, $[E_0-\delta,E_0+\delta]$ will at most overlap with two of these branches at the same time, and hence the weight falling into $[E_0-\delta,E_0+\delta]$ due to this tree is below $2\cdot \epsilon/4m=\epsilon/2m$.

With the choice of $\delta_0$ that works for all of the $m$ trees, the total weight that can go into the energy window $[E_0-\delta,E_0+\delta]$ is now bounded by $\epsilon/2+m\cdot \epsilon/2m=\epsilon$, thus completing the proof that $\mu$ has no pure point part.

\subsubsection{Absence of absolutely continuous spectrum}

In this subsection, we prove that $\mu$ has no absolutely continuous part. That is, all the energy distribution is concentrated on a measure-zero set. Physically, this means that there is no ``continuous'' region on the energy distribution function.
The idea is that the energy being distributed on a measure-zero set in $\mu$ means that the weights in $\mu_n$ must be mostly distributed in smaller and smaller regions as $n\to\infty$.
That is, most of the weights can be covered by a set of intervals whose total length is arbitrarily small.
As the weights are distributed in a tree-like structure, we can cover them by having one interval covering each branch.
By showing that the total width goes to zero as $n\to\infty$, the desired result will follow.

To write the condition that $\mu$ has no absolutely continuous part in terms of integrals, we need to construct a family of continuous functions $g_\delta(E)$ that equals 1 except on a set with measure $\delta$, such that
\begin{equation}
    \lim_{\delta\to 0}\int g_\delta\,d\mu=0
\end{equation}
Again, by approximating $\mu$ using $\mu_n$, it is equivalent to
\begin{equation}
    \lim_{\delta\to 0}\lim_{n\to\infty}\sum_j g_{\delta}(E_{j,n})w_{j,n}=0
\end{equation}
So we want to construct such $g_\delta(E)$ so that for all $\epsilon>0$, we can find $\delta_0>0$ such that $0<\delta\leq\delta_0$ implies
\begin{equation}
    \lim_{n\to\infty}\sum_j g_{\delta}(E_{j,n})w_{j,n}<\epsilon
\end{equation}
It is enough to have $g_\delta(E)$ being associated with a finite set of disjoint intervals $\{[E_a-2\delta_a,E_a+2\delta_a]\}$ indexed by $a$, so that $g_\delta(E)=1$ outside these intervals, and  $g_\delta(E)=0$ inside any of $[E_a-\delta_a,E_a+\delta_a]$, and between 0 and 1 otherwise.
By a simple inequality argument, it is enough to show that
given any $\epsilon>0$, we can construct such $\{[E_a-\delta_a,E_a+\delta_a]\}$ with $2\sum_a\delta_a<\epsilon$ so that the sum of weights $w_{j,n}$ with $E_{j,n}$ outside any of these intervals is less than $\epsilon$ for all large enough $n$.

To construct such intervals, we again choose $n_1$ such that the total final weights of the trees with root lines beyond $n_1$ are below $\epsilon$.
So we only need to cover a finite number $m$ of trees, those with root lines below $n_1$.
Our goal can be achieved by covering all branches of each tree (at large $n$) using intervals with total lengths less than $\epsilon/m$.

Now, for each tree with root line $(j,n_0)$, for $n\geq n_0$, there are $2^{n-n_0}$ branches at this $n$.
By the error bound, the width of each of the branches is bounded by $\epsilon_{\geq n}<2\epsilon_n$.
Hence, the total width of all branches at $n$ is less than $2^{n-n_0+1}\epsilon_n$, which goes to zero as $n\to\infty$ by the assumption on the error bounds.
Now pick a large enough $n$ so that $2^{n-n_0+1}\epsilon_n<\epsilon/m$, and construct $2^{n-n_0+1}$ intervals $[E_a-\delta_a,E_a+\delta_a]$, each covering the width of one branch.
Then we successfully cover this tree (at large $n$) using a total length $<\epsilon/m$.

Now, by merging the intervals from all of the $m$ trees (which may result in a smaller total length if they are not disjoint), we get the final set of intervals $[E_a-\delta_a,E_a+\delta_a]$ with a total length less than $m\cdot\epsilon/m=\epsilon$.
For large enough $n$, the only source of weights that they fail to cover is from the trees with root lines beyond $n_1$, which is less than $\epsilon$, thus completing the proof.

\subsection{Conclusion for the single-particle model}

We have constructed the single-particle model of a 1D chain with HMS based on successive applications of first-order degenerate perturbation theory, in which all the energy eigenstates can be solved exactly up to controllable errors that can be made arbitrarily small, and the energy splittings are hierarchically separated.
From these simply expressed properties of the states and spectrum, and based on the idea of splitting weights as we evolve the index $n$ of the level, we rigorously prove that the spectrum of the infinite system has neither a pure point part nor an absolutely continuous part.
That is, our single-particle model is purely singular continuous, which confirms that our solvable model is really a special case of what our physical intuition leads us to, and belongs to the broader class of models we intend to study, as explained in Sec.~\ref{sec:summary}.

The techniques we used in our construction and proofs have been deliberately chosen to be very general and not dependent on the particular properties of a 1D nearest-neighbor tight-binding model.
In addition to the many-body model we are going to construct, our construction above can also be generalized to single-particle models with long-range hopping and/or in higher dimensions.
On the other hand, it also comes with some drawbacks compared to more widely used techniques for 1D nearest-neighbor tight-binding models (such as the transfer matrix approach).
In particular, to avoid the issue of accidental resonance, all the controlled error bounds of the perturbation are required to be extremely small so that accidental resonance is simply impossible.
As we know, single-particle systems can withstand accidental resonances to some extent (as demonstrated by the localized nature of the Anderson and Aubry-Andr\'e models), we expect our error bound to be far from optimal.
Although other techniques can avoid such issues in single-particle systems, there is no direct generalization to many-body cases.
In particular, whether many-body systems can withstand accidental resonances in general is still an open problem.
Therefore, to have a rigorous generalization of HMS to many-body systems, the only mathematical tool available to us will be this perturbative approach with controlled small error bounds.

\section{The many-body solvable model}\label{sec:MBSolvable}

In this section, we will construct our many-body asymptotically solvable model with HMS, analogous to the single-particle construction in Sec.~\ref{sec:SPSolvable}, and derive its properties.
As we already mentioned in Sec.~\ref{sec:SPSolvable}, we have used the construction and proof technique that can be easily generalized to many-body systems.
Therefore, most of the construction steps will be carried over directly from those of Sec.~\ref{sec:SPSolvable}, as well as some of the key steps in proving its properties.
Recall that our single-particle model can be treated as a simplified model of a general chain with HMS, where we only consider a subset of localized approximate orbitals, each corresponding to one site in the modeled chain.
Intuitively, since a weak coupling between two consecutive (approximate) LIOMs can be treated as the two LIOMs being far away with all the LIOMs between them ignored, this many-body model can be similarly understood as modeling a subset of approximate LIOMs in a more general class of many-body HMS chains (see Appendix~\ref{sec:symmetricRandom} for a more detailed model on this). 

Despite the similarities, there is a key ingredient in the many-body case that has no analogy in the single-particle case.
As we summarized in Sec.~\ref{sec:summary} and visualized in Fig.~\ref{fig:intro}, the interaction between particles can produce non-trivial effects on the originally resonant particles, so that originally resonant particles can, in some cases, become non-resonant.
The exact condition of this freezing/protection behavior for a general HMS model turns out to be complicated (see Appendix~\ref{sec:symmetricRandom}).
Even if we start with the already-simplified single-particle orbitals in Sec.~\ref{sec:SPSolvable}, it is still far from being asymptotically solvable.
Therefore, we will make a further simplification: we require that the resonant segments ($[c,d]$ and $[\widetilde{d},\widetilde{c}]$ in Fig.~\ref{fig:solvable}) remain resonant (protected) if and only if configurations of the states in the non-resonant segments ($[a,b]$ and $[\widetilde{b},\widetilde{a}]$) are mirror-symmetric.
We will show that this requirement can be achieved by a particular limit involving the small parameters $\alpha$, $\beta$, and $\gamma$ in the iteration step, while in the single-particle constructions we do not need to tune $\alpha$.

Another issue is what type of many-body chain we should use.
The conceptually most straightforward way is probably to add nearest-neighbor density-density interactions to the single-particle model we constructed in Sec.~\ref{sec:SPSolvable}.
However, the particle conservation of the resulting model leads to several technical difficulties.
First, when we couple segments of chains, the central bond can move at most one particle from one side to the other in first-order perturbation theory. Thus, to allow resonance between two halves with particle numbers differing by two or more, we need either higher-order perturbations or long-range hoppings.
While higher-order perturbations will make error bounding extremely difficult, adding long-range hopping contradicts our goal of studying a local and short-ranged spin chain.
Second, when we discuss the thermodynamic properties of the chain, particle conservation makes the eigenstate labeling and ensemble averaging much more complicated than in a system without additional conservation laws.
Although we expect that the resulting system will be qualitatively the same as what we will construct below (see Appendix~\ref{sec:symmetricRandom} for a non-solvable particle-conserving model), it would be unnecessarily complicated as a model to demonstrate our intuitive idea described in Sec.~\ref{sec:summary}.

Given the reasons above, we will use another notion of the many-body counterpart of a single-particle model, which has been used, for example, in Ref.~\cite{Imbrie2016}, to generalize a proof of Anderson localization to MBL.
In this picture, going from a single-particle to a many-body system is simply a change in the operation of connecting two subsystems with Hilbert space $\mathcal{H}_A$ and $\mathcal{H}_B$ from a direct sum $\mathcal{H}_A\oplus\mathcal{H}_B$ to a tensor product $\mathcal{H}_A\otimes\mathcal{H}_B$.
The many-body counterpart of a ``position eigenstate'' will then become a tensor product basis state of the sites, and that of  ``hopping'' and ``potential'' should be treated as those on the Fock space lattice.
We will follow Ref.~\cite{Imbrie2016} to use the mixed-field Ising model, whose Fock space lattice is a hypercube, such that the transverse-field terms correspond to the hopping terms, and the longitudinal-field and Ising coupling terms correspond to the potential.
On the other hand, the HMS will always refer to the structure in real space, similar to how the disorder is still added in real space in Ref.~\cite{Imbrie2016}. %

We will show that this model has two thermodynamic phases: one is ETH-like and the other is MBL-like, and that they can be separated by a finite-temperature phase transition.
We will also discuss the difference between the usual ETH and MBL phases.

Note that although we will rigorously show some MBL-like behavior in our system, this has nothing to do with demonstrating the existence of a thermodynamic MBL phase in the usual sense. 
Since we will artificially avoid accidental resonances by coupling finite segments of chains with weaker and weaker bonds, we are no longer in the same setting as the usual ``MBL phase'', which typically requires some stability under generic local perturbations. 
Indeed, if we allow tuning asymptotic weak bonds, then thermodynamic MBL certainly ``exists'' as an asymptotically decoupled chain without resonances, but it is not stable under generic perturbations.
One possible consequence of the interplay between the structural resonances caused by HMS and the accidental resonances when we are no longer in the rigorously error-bounded setting will be discussed in Sec.~\ref{sec:rare}.

\subsection{Construction}

We will construct a mixed-field Ising model on an infinite 1D qubit lattice of the following form
\begin{equation}\label{eq:MBHinf}
    H_\infty=\sum_{j=-\infty}^\infty J_j\sigma_j^z\sigma_{j+1}^z + \sum_{j=-\infty}^\infty\left(h_j^z\sigma_j^z+h_j^x\sigma_j^x\right),
\end{equation}
where $\sigma_j^{x,y,z}$ are the Pauli matrices at site $j$.
This is the many-body counterpart of Eq.~(\ref{eq:SPHinf}).
As we explained, we should view $J_j$ and $h^z_j$ terms as the ``potential'' on the Fock space lattice, which is the counterpart of the $V_j$ term on the real space lattice in (\ref{eq:SPHinf}).
Similarly, the $h^x_j$ term is the ``hopping'' on the Fock space lattice, the counterpart of the $t_j$ term in (\ref{eq:SPHinf}).

\subsubsection{Initialization of the building blocks}\label{sec:MBInit}

As in the single-particle case of Sec.~\ref{sec:SPInit}, a set of initial building blocks is introduced for the iterative construction. They will be labeled in the exact same way ($H^\text{res}_1$, $H^\text{nr}_n$, etc) as in the single-particle case.
We require that each block has a non-degenerate many-body spectrum, and that $\sigma^z_L$ of the blocks has all off-diagonal matrix elements $\langle A|\sigma^z_L|B\rangle\neq 0$ (where $A\neq B$ are energy eigenstate labels), and all diagonal matrix elements $\langle A|\sigma^z_L|A\rangle$ are distinct. The reason for these requirements will be illustrated in the next subsection.  Again, our construction works for a general choice of Hamiltonians as the building blocks. However, for simplicity, we will make a specific choice for the Hamiltonians as in the single-particle case.
Let
\begin{equation}
    H=\sum_{j=1}^{L-1} J_j\sigma_j^z\sigma_{j+1}^z + \sum_{j=1}^L\left(h_j^z\sigma_j^z+\delta\cdot h_j^x\sigma_j^x\right),
\end{equation}
with $H$ standing for $H^\text{res}_1$ or $H^\text{nr}_n,n=1,2,\ldots$, and $L$ being the corresponding size ($L^\text{res}_1$ or $L^\text{nr}_n$).
The parameters $J_j$, $h^z_j$, and $h^x_j$ are independent uniform random numbers in $[-1,1]$, and $\delta>0$ is a small number that depends on the block (see Sec.~\ref{sec:SPInit} for the remark on randomness).

As in the single-particle case, to have a clean asymptotic solution of the energy eigenstates, we will require $\delta$ to be small enough so that the energy eigenstates of the initial blocks are close to the $\sigma^z$ eigenstates. 
First, we label the energy eigenstates of the blocks by bitstrings of length $L$, and define the integrals of motion $\tau^z_{1,j}$ of the blocks with eigenvalues being the $j$th digit of the bitstring (with `0' corresponding to eigenvalue $+1$ and `1' to $-1$).
Then the closeness condition can be expressed as
\begin{equation}
    \left|\langle\tau^z_{1,j}\rangle-\langle\sigma^z_{j}\rangle\right|<\epsilon_0
\end{equation}
for each eigenstate of the initial block, analogous to the single-particle case (\ref{eq:SPInitError}).
The corresponding $\sigma^z$ and $\tau^z_1$ eigenstates with bitstring $A$ are denoted by $|A\rangle_0$ and $|A\rangle_1$, respectively.

When discussing the thermodynamic properties, we will need some additional ``regularity'' conditions.
We assume $L^\text{nr}_n\to\infty$ as $n\to\infty$, so that in a proper sequence of the limit $\delta\rightarrow0$, the blocks for large $n$ become a thermodynamic classical Ising chain, which has a well-defined entropy density $s(e)$ as a function of the energy density $e$, and has no long-range correlations. 
We require $\delta$ (separately for each block) to be small enough that these properties still hold as $n\to\infty$.
When discussing approximate LIOMs, we will need an additional requirement that for any temperature $T$, the diagonal matrix element variance
of $\tau^z_{1,j}$ in the canonical ensemble has a uniform lower bound $V(T)>0$, independent of the block and the site index $j$.
This can be achieved by the property of the classical Ising model with bounded field strengths. How these additional properties are used will be explained in the respective sections.

\subsubsection{The iteration step}

Next, we construct the iteration step from $H^\text{res}_n$ to $H^\text{res}_{n+1}$, in parallel with the single-particle case in Sec.~\ref{sec:SPiter}, and again with the same method and symbols illustrated in Fig.~\ref{fig:solvable}.

First, we make a mirror copy of $H^\text{res}_n$ defined by the previous stage in the recursion to become $\widetilde{H^\text{res}_n}$.
Specifically, $\widetilde{H^\text{res}_n}$ is formed by reflecting all the site indices of each spin operator in $H^\text{res}_n$. 
Meanwhile, for each eigenstate $|B\rangle$ of $H^\text{res}_n$ (the label $B$ is a bitstring that is constructed recursively), we label the corresponding eigenstate of $\widetilde{H^\text{res}_n}$ (with the same energy) as $|\widetilde B\rangle$, where $\widetilde B$ is the reversal of the bitstring $B$.

Next, we place the non-resonant building block $H^\text{nr}_n$ (spanned over the sites $a,\ldots,b$) to the left of $H^\text{res}_n$.
We similarly make a mirrored copy to become $\widetilde{H^\text{nr}_n}$.
To make it non-resonant, we add a symmetry-breaking perturbation 
\begin{equation}
    H^\beta_n=\beta\cdot\left[\sum_{j=\widetilde{b}}^{\widetilde{a}-1} J'_j\sigma_j^z\sigma_{j+1}^z + \sum_{j=\widetilde{b}}^{\widetilde{a}}\left(h_j^{\prime z}\sigma_j^z+h_j^{\prime x}\sigma_j^x\right)\right]
\end{equation}
with independent uniform random numbers $J'_j,h^{\prime x,z}_j$ in $[-1,1]$ and a small number $\beta>0$ to be chosen below.
The eigenstates of $\widetilde{H^\text{nr}_n}+H^\beta_n$ are labeled consistently by the reversal of the bitstrings that label $H^\text{nr}_n$.

The four subchains are then coupled by
\begin{equation}
    H^\gamma_n=\gamma\,\sigma^z_b\sigma^z_c,\quad H^{\alpha\gamma}_n=\alpha\gamma\,\sigma^z_d\sigma^z_{\widetilde{d}},\quad \widetilde{H^\gamma_n}=\gamma\,\sigma^z_{\widetilde{c}}\sigma^z_{\widetilde{b}}
\end{equation}
to form the final Hamiltonian
\begin{equation}
    H^\text{res}_{n+1}=H^\text{nr}_n+H^\gamma_n+H^\text{res}_n+H^{\alpha\gamma}_n+\widetilde{H^\text{res}_n}+\widetilde{H^\gamma_n}+\widetilde{H^\text{nr}_n}+H^\beta_n,
\end{equation}
with small numbers $\alpha,\gamma>0$ to be chosen below.

Now, the main difference from the single-particle counterpart in Sec.~\ref{sec:SPiter} is that, in addition to controlling the non-resonances by $\beta$ and the resonances by $\gamma$, we need to tune an additional parameter $\alpha$ to allow a simple condition of freezing and protection.
By definition, $\alpha$ controls the relative magnitude between two coupling Hamiltonians $H_n^{\alpha\gamma}$ and $H_n^\gamma$. 
The smaller $\alpha$ is, the stronger the influence that the non-resonant regions have on the resonant regions.
Intuitively, by making $\alpha$ small enough, any non-symmetric configuration in the non-resonant pair will freeze the resonant pair. 
But we also want to make it large enough so that when the non-resonant configuration is mirror-symmetric, the asymmetry of the interaction (between non-resonant and resonant regions in different halves) induced by the $\beta$ term is small enough to protect the resonance.
We will see that this can be achieved by choosing $\alpha\gg\beta\gg\gamma$.

We will again use first-order degenerate perturbation theory as in the single-particle case, with only $\gamma$ treated as the perturbation parameter, while $\alpha$ and $\beta$ are fixed positive parameters during this perturbation step.
That is, the unperturbed and perturbation parts for $H^\text{res}_{n+1}$ are
\begin{align}
    H^\text{unp}_{n+1} &= H^\text{nr}_n+H^\text{res}_n+\widetilde{H^\text{res}_n}+\widetilde{H^\text{nr}_n}+H^\beta_n,\\
    H^\text{pert}_{n+1} &= H^\gamma_n+H^{\alpha\gamma}_n+\widetilde{H^\gamma_n}.
\end{align}
The unperturbed eigenstates are labeled by $|ABCD\rangle_\text{unp}$, where $A$, $B$, $C$, and $D$ are bitstrings labeling the eigenstates of $H^\text{nr}_n$, $H^\text{res}_n$, $\widetilde{H^\text{res}_n}$, and $\widetilde{H^\text{nr}_n}+H^\beta_n$, respectively.
Notice that the degeneracies between $|ABCD\rangle_\text{unp}$ and $|\widetilde{D}BC\widetilde{A}\rangle_\text{unp}$ are split by $\beta$, and we restrict $\beta$ to be small enough to avoid accidental degeneracies.
Hence, the only degeneracies in $H^\text{unp}_{n+1}$ are twofold:
\begin{equation}
    \{|ABCD\rangle_\text{unp},|A\widetilde{C}\widetilde{B}D\rangle_\text{unp}\},\quad B\neq\widetilde{C},
\end{equation}

Using the perturbation theory, the diagonal term of the $2\times 2$ effective Hamiltonian spanned by the unperturbed states is
\begin{multline}
    \Delta:=\langle ABCD |H^\text{pert}_{n+1}|ABCD\rangle_\text{unp}\\ - \langle A\widetilde{C}\widetilde{B}D |H^\text{pert}_{n+1}|A\widetilde{C}\widetilde{B}D\rangle_\text{unp}\\
    =\gamma\big(\langle A|\sigma_b^z|A\rangle-\langle D|\sigma_{\widetilde{b}}^z|D\rangle\big)\big(\langle B|\sigma_c^z|B\rangle-\langle \widetilde{C}|\sigma_{c}^z|\widetilde{C}\rangle\big)
\end{multline}
and the off-diagonal term is
\begin{equation}
    t:=\langle ABCD |H^\text{pert}_{n+1}|A\widetilde{C}\widetilde{B}D\rangle_\text{unp}=\alpha\gamma\big|\langle B|\sigma_d^z|\widetilde{C}\rangle\big|^2.
\end{equation}
Note that the first factor in $\Delta$ has the limit:
\begin{equation}\label{eq:MBbetalimit}
    \lim_{\beta\to0}\big|\langle A|\sigma_b^z|A\rangle-\langle D|\sigma_{\widetilde{b}}^z|D\rangle\big|=
    \begin{cases}
        0,&\text{if }A=\widetilde{D},\\
        s_{A,D},&\text{if }A\neq\widetilde{D},
    \end{cases}
\end{equation}
where $s_{A,D}\neq 0$ is independent of $\alpha$ and $\gamma$.
Hence, we can restrict the candidate range of $\beta$ to $0<\beta<\beta_0$ (independent of $\alpha$ and $\gamma$) so that when $A\neq\widetilde{D}$, any choice of $\beta$ in this range will lead to,
\begin{equation}
    \big|\langle A|\sigma_b^z|A\rangle-\langle D|\sigma_{\widetilde{b}}^z|D\rangle\big|>\frac{1}{2}s_{A,D} 
\end{equation}
We also require that the $\beta$ in this range does not lead to additional accidental degeneracies that would invalidate the degenerate subspaces $\{|ABCD\rangle_\text{unp},|A\widetilde{C}\widetilde{B}D\rangle_\text{unp}\}, B\neq\widetilde{C}$.
Note that this step is why we require that, in the initial building blocks, the diagonal matrix elements of the boundary term are all distinct.

Before determining $\alpha$, $\beta$, and $\gamma$, we will first discuss why we will get the simple condition of freezing/protection, as long as the scales satisfy $1\gg\alpha\gg\beta\gg\gamma$. 
First, note that when $\alpha$ is small enough, we have (for any choice of $0<\beta<\beta_0$ and $\gamma>0$)
\begin{equation}\label{eq:tllDelta}
    |t|\ll|\Delta|\text{ for all }A\neq\widetilde{D}\text{ and }B\neq\widetilde{C},
\end{equation}
which suggests that asymmetric configurations of the non-resonant region freeze the resonant region (in first-order perturbation, which at this point may or may not be accurate).
Here we have used the property that $\langle B|\sigma_c^z|B\rangle-\langle \widetilde{C}|\sigma_{c}^z|\widetilde{C}\rangle$ is nonzero, which is expected due to the conditions on the initial building blocks as well as the random numbers in the subsequent steps.

Next, after fixing a small enough $\alpha$, we will fix $\beta$. 
Due to the first line of (\ref{eq:MBbetalimit}), as long as $\beta$ is small enough, we have (for any $\gamma$)
\begin{equation}\label{eq:tggDelta}
    |t|\gg|\Delta|\text{ for all }A=\widetilde{D}\text{ and }B\neq\widetilde{C},
\end{equation}
which suggests that symmetric configurations of the non-resonant region protect the resonant region (again, in first-order perturbation).
Here we have used the property that $\langle B|\sigma_d^z|\widetilde{C}\rangle$ is nonzero.
Again, this is due to the conditions on the initial building blocks as well as the random numbers in the subsequent steps.
In particular, this is the technical reason why we need a system without particle conservation.
If particles are conserved in the system, the resulting term $\langle B|\sigma_d^{\pm}|\widetilde{C}\rangle$ would only be nonzero if $B$ and $C$ have the same number of particles (unless we add long-range hopping or go to higher-order perturbations).

Finally, after fixing a small enough $\alpha$ and then a small enough $\beta$ based on $\alpha$, we are ready to fix $\gamma$.
Until now, $t$ and $\Delta$ have been just numbers, and we have not assumed anything about the accuracy of first-order perturbation theory; that is, whether the conditions (\ref{eq:tllDelta}) and (\ref{eq:tggDelta}) on $t$ and $\Delta$ actually lead to the expected freezing/protecting conditions.
However, as the degenerate perturbation theory is only controlled by $\gamma$, and the conditions (\ref{eq:tllDelta}) and (\ref{eq:tggDelta}) only depend on the ratio of $t$ and $\Delta$, we can always choose a small enough $\gamma$ (after fixing $\alpha$ and $\beta$) to make the perturbation theory works as accurately as possible. 
In particular, since we are in a finite-dimensional Hilbert space, we can always make $\gamma$ small enough to avoid any energy level crossings that would induce accidental resonances. 
The perturbed eigenstates $|ABCD\rangle_{n+1}$ of $H^\text{res}_{n+1}$ (to be used in the next iteration, along with their labels) are
\begin{multline}\label{eq:MBpert}
    |ABCD\rangle_{n+1}\\\approx
    \begin{cases}
        |ABCD\rangle_\text{unp},&\text{if }A\neq \widetilde{D}\text{ or }B=C,\\
        \frac{|ABCD\rangle_\text{unp}+|A\widetilde{C}\widetilde{B}D\rangle_\text{unp}}{\sqrt2},&\text{if }A=\widetilde{D}, B>C,\\
        \frac{|ABCD\rangle_\text{unp}-|A\widetilde{C}\widetilde{B}D\rangle_\text{unp}}{\sqrt2},&\text{if }A=\widetilde{D}, B<C.
    \end{cases}
\end{multline}
The choice of $\pm$ in the eigenstates related to comparing $B$ and $C$ as binary numbers is arbitrary and is only a convenient way to label the new eigenstates as binary strings.
Note that in the case where the binary string $ABCD$ contains only a single ``1'' and all other bits are ``0'' so that it can be labeled by a single site $j$, (\ref{eq:MBpert}) reduces to the single-particle counterpart (\ref{eq:SPPert}).

In the above construction, although the values of $\alpha,\beta$ and $\gamma$ are assumed to be so small that there are no many-body energy level crossings happening in the perturbation step, we expect that such a requirement is unnecessary in reality.
Specifically, if thermodynamic MBL is stable, we would expect perturbation theory to work in MBL systems even if there are energy level crossings, as the probability of energy level crossings causing accidental resonances decays exponentially.
Even if thermodynamic MBL is not stable, accidental resonances are still expected to be suppressed well before the strict scales we used (due to the very long thermalization timescale).
Our construction may also have some tolerance against accidental resonances.
However, the interplay between accidental resonances and structural (mirror) resonances is beyond the scope of this work.

\subsubsection{Error bounds}\label{sec:MBerr}

As in the single-particle case in Sec.~\ref{sec:SPerr}, the final choice of the small parameters $\alpha$, $\beta$, and $\gamma$ (for the step from $n$ to $n+1$) will be based on rigorously controlled error bounds.
From the discussion above, we have
\begin{multline}\label{eq:MBPertIdeal}
    |ABCD\rangle_\text{lim}:=\lim_{\alpha\to0}\lim_{\beta\to0}\lim_{\gamma\to0}|ABCD\rangle_{n+1}\\=
    \begin{cases}
        |ABCD\rangle_n,&\text{if }A\neq \widetilde{D}\text{ or }B=\widetilde{C},\\
        \frac{1}{\sqrt2}\big(|ABCD\rangle_n+|A\widetilde{C}\widetilde{B}D\rangle_n\big),&\text{if }A=\widetilde{D}, B>\widetilde{C},\\
        \frac{1}{\sqrt2}\big(|ABCD\rangle_n-|A\widetilde{C}\widetilde{B}D\rangle_n\big),&\text{if }A=\widetilde{D}, B<\widetilde{C}.
    \end{cases}
\end{multline}
Here $|ABCD\rangle_n$ are the eigenstates of the Hamiltonian in the same limit
\begin{equation}
    H^\text{lim}_{n+1}:=\lim_{\alpha\to0}\lim_{\beta\to0}\lim_{\gamma\to0}H^\text{res}_{n+1}=H^\text{nr}_n+H^\text{res}_n+\widetilde{H^\text{res}_n}+\widetilde{H^\text{nr}_n}
\end{equation}
with consistent labeling (as in the single-particle case, this does not cause a notational inconsistency with $|ABCD\rangle_{n+1}$).
For a better description of the ``evolution'' of the eigenstates from $|ABCD\rangle_n$ to $|ABCD\rangle_{n+1}$, we define the unitaries
\begin{equation}\label{eq:Un}
\begin{aligned}
    U_n|ABCD\rangle_n&=|ABCD\rangle_{n+1},\\
    U_n^\text{lim}|ABCD\rangle_n&=|ABCD\rangle_\text{lim}
\end{aligned}
\end{equation}
so that
\begin{equation}
    \lim_{\alpha\to0}\lim_{\beta\to0}\lim_{\gamma\to0}U_n=U_n^\text{lim}.
\end{equation}
Note that the order of limits in the equations above reflects the separation of scales $1\gg\alpha\gg\beta\gg\gamma$.

For the technical requirements in all the subsequent derivations, we need an error bound parameter $0<\epsilon_n<1$ associated with this iteration.
We choose $\alpha,\beta,\gamma$ to make all of the following rigorous conditions true:
\begin{equation}\label{eq:UnError}
    U_n=U_n^\text{lim}+\mathcal{E}_n,\quad \|\mathcal{E}_n\|<\epsilon_n, 
\end{equation}
and for energies:
\begin{equation}
    2|\Delta E^\text{max}_n|<\epsilon_n
\end{equation}
where $|\Delta E^\text{max}_n|$ is the maximal energy shift between corresponding eigenstate index from $H^\text{lim}_{n+1}$ to $H^\text{res}_{n+1}$.
Iterative applications of Eq.~(\ref{eq:MBPertIdeal}), along with the error bounds, lead to the asymptotic solution of our model: each eigenstate can be labeled and expressed exactly up to a controllable error bound, and the energy shifts at each level are also controlled.

To have a controllable accumulated error as we take $n\to\infty$, we will require $\sum_n\epsilon_n<\infty$. %
As in the single-particle case, we will assume informally that we have a sequence of timescales $\{t_n\}$ with $1/\Delta^\text{min}_n\ll t_{n}\ll 1/\Delta^\text{max}_{n+1}$ corresponding to the levels.

\subsubsection{Infinite system}

As in the single-particle case in Sec.~\ref{sec:SPinf}, the infinite system $H_\infty$ is defined based on a sequence of embeddings of $H^\text{res}_n$ into $H^\text{res}_{n'}$ ($n'>n$).

For the physical picture of the extension of chains according to the timescales $t_n$, we need to use local observables instead of local states in the single-particle case.
So the picture becomes that we are looking at Heisenberg evolution of a local observable $O(t)$, and that in a finite chain $H^\text{res}_n$ we can only probe its time evolution up to $t_n$, beyond which we need to extend the chain through the embedding process.

An important consideration when taking the system to infinity in the many-body case is whether the thermodynamic limit exists. One of the conditions is that the boundary effects should be negligible compared to the bulk states. 
In our system, at every $n$, the bulk states in the resonant regions are always influenced by the boundary states of the non-resonant blocks. 
This is contrary to the conventional thermodynamic limit where the boundary effects are ignorable, and the size of the boundaries is much smaller than the bulk.
Although this is the case, we can still characterize the properties of the bulk states using local observables; i.e., the local observables have a well-defined limit in an arbitrarily large system when sufficiently far from the boundary.
The argument is similar to that in Sec.~\ref{sec:SPinf}, where to obtain the limit of a local observable, one should always study it below a timescale $t_n$ at each $n$, and then take $n \rightarrow \infty$.
Related problems of the thermodynamic limit are discussed in different models~\cite{eggarter1974cayley,robert1982spin,jeet2026breakdown}

\subsection{Progressive approximations on infinite lattice}\label{sec:MBapprox}

\begin{figure*}
    \centering
    \includegraphics[scale=1]{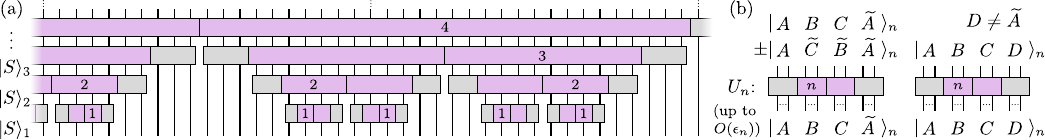}
    \caption{\justifying (a) The unitary circuit that progressively approximates the eigenstates of the many-body solvable model (visualized with $L^\text{res}_1=1,L^\text{nr}_n=n$). The $n$th layer takes the $H^{\infty}_n$ eigenstate $|S\rangle_n$ to the corresponding $H^{\infty}_{n+1}$ eigenstate $|S\rangle_{n+1}$. (b) The approximate action of the gates [see Eq.~(\ref{eq:Unaction})]. A gate with $n$ written on the right denotes its mirror.  }
    \label{fig:circuit}
\end{figure*}

The final infinite Hamiltonian $H_\infty$ can also be approximated by a series of infinite Hamiltonians $H^\infty_n$ consisting of decoupled blocks, in the exact same way as described in Sec.~\ref{sec:SPapprox} and depicted in Fig.~\ref{fig:approx}, with the difference that now we can no longer use operator norm to describe the limit ``$H^\infty_n\to H_\infty$'' due to the divergence of the many-body operator norm.

Similar to the single-particle model, the energy eigenstates of $H^\infty_n$ can be described by the tensor product of the eigenstates of each block (although the energy subspaces are highly degenerate, the most ``local'' bases are the tensor-product ones).
As these eigenstates of each block are labeled by a bitstring, we can concatenate all of them to form a bi-infinite bitstring $S$.
The eigenstates of $H^\infty_n$ are then denoted by $|S\rangle_n$.
The notation of bitstring $ABCD$ we used before to label the eigenstates now becomes a substring of $S$.
Note that an infinite tensor product of Hilbert spaces is ill-defined in general, so the notation $|S\rangle_n$ should be regarded as a formal tensor product of the blocks, and we do not try to define its limit ``$|S\rangle_\infty$'' that would be the eigenstates of $H_\infty$.

A convenient way to describe these eigenstates is to use a unitary circuit consisting of $U_n$ and its mirror $\widetilde{U_n}$ on the $n$th layer, shown in Fig.~\ref{fig:circuit}.
Recall that, by the definition (\ref{eq:Un}) of $U_n$ and the error bounds (\ref{eq:UnError}), we have
\begin{multline}\label{eq:Unaction}
    U_n|ABCD\rangle_n=|ABCD\rangle_{n+1}\\=
    \begin{cases}
        |ABCD\rangle_n,&\text{if }A\neq \widetilde{D}\text{ or }B=\widetilde{C},\\
        \frac{1}{\sqrt2}\big(|ABCD\rangle_n+|A\widetilde{C}\widetilde{B}D\rangle_n\big),&\text{if }A=\widetilde{D}, B>\widetilde{C},\\
        \frac{1}{\sqrt2}\big(|ABCD\rangle_n-|A\widetilde{C}\widetilde{B}D\rangle_n\big),&\text{if }A=\widetilde{D}, B<\widetilde{C}.
    \end{cases}\\
    +\mathcal{E}_n,\quad \|\mathcal{E}_n\|<\epsilon_n
\end{multline}
Now defining the $n$th layer of the circuit [as shown in Fig.~\ref{fig:circuit}(a)] as $U^\infty_n$, we then have
\begin{equation}
    U^\infty_n|S\rangle_n=|S\rangle_{n+1}
\end{equation}
This circuit is viewed as a progressive approximation of the eigenstate of $H_\infty$.
One can view $H^\infty_n$ as an approximation of $H_\infty$ up to a timescale $t_n$.
If we start with the eigenstates $|S\rangle_1$ of the initial building blocks, evolving the circuit from layer $1$ to $n$ produces $|S\rangle_n$, which are approximated eigenstates of $H_\infty$ up to a timescale $t_n$.
Thus, taking the $n\to\infty$ limit corresponds to taking the infinite time limit. 
As we will shortly see, this leads to a unique limit of local observable expectations.

Analogous to how the Pauli operators act on the $\sigma^z$ bitstring eigenstates, we can similarly define the ``progressively dressed'' Pauli operators $\tau^{x,y,z}_{n,j}$, so that they act on the $j$th component of the $|S\rangle_n$ eigenstate in the exact same way as $\sigma^{x,y,z}_j$ on the $\sigma^z$ eigenstate labeled by the bitstring $S$.
Note that we have
\begin{equation}
U^\infty_n\tau^{x,y,z}_nU^{\infty\dagger}_n=\tau^{x,y,z}_{n+1}.
\end{equation}
We will interpret $\{\tau^{z}_{n,j}\}_{j=-\infty}^\infty$ as a complete set of LIOMs of $H^\infty_n$, and approximate LIOMs of $H_\infty$ within the timescale $t_n$. 
In the limit $n\rightarrow \infty$, $\tau^{z}_{n,j}$ is no longer a (quasi-)local operator, analogous to a local state that may participate in an arbitrarily long resonance in the single-particle case.
However, different from the single-particle case, there remain possibilities that $\{\tau^{z}_{n_0,j}\}_{j=-\infty}^\infty$ for a fixed $n_0$ is a good set of approximate LIOMs of $H_\infty$ up to an arbitrarily long timescale, which corresponds to the situation that the majority of energy eigenstates are approximately LIOM eigenstates, while only a vanishing fraction of energy eigenstates are not.
Such a case is referred to as the MBC-L phase, while the opposite scenario will be considered the MBC-E phase. 
We will later see whether we have the MBC-L or MBC-E phase, which depends on the system parameters and temperature.

\subsection{Canonical ensemble}

We will analyze the system in an energy (or temperature)-dependent way; thus, we need to look at some thermodynamic ensemble.
As each $H^\infty_n$ consists of decoupled blocks, it is natural to use the canonical ensemble, which has the property that the ensemble density matrix of a decoupled set of systems is the tensor product of the individual systems. 
That is, the canonical ensemble of $H^\infty_n$ at a temperature $T\neq 0$ can be written as
\begin{equation}
    \rho^\infty_n=\bigotimes_{H:\text{ block}}\frac{e^{-H/T}}{\Tr e^{-H/T}}
\end{equation}
where $H$ runs over the blocks ($H^\text{res}_n$ and $H^\text{nr}_{n'}$ for $n'\geq n$ and their mirror counterparts) of $H^\infty_n$.
We similarly denote the canonical ensemble of those blocks as $\rho^\text{res}_n$, $\rho^\text{nr}_n$, etc, with a fixed $T$ left implicit.

\subsubsection{Interpretation as closed systems}\label{sec:interpretclosedsystems}

Although we can interpret $\rho^\infty_n$ as the equilibrium state when we couple each block of $H^\infty_n$ to a thermal bath, the underlying intention is instead to study a closed system.
Indeed, many-body localization and quantum thermalization are mostly about the properties of an isolated quantum system.
Therefore, we should conceptually treat our system as being isolated. That is, in a microcanonical ensemble at some energy density $e$, rather than being coupled to a bath with temperature $T$.
However, due to the boundary size issue of our system, it is not clear how to rigorously define a microcanonical ensemble, as we would need to make finite cuts of our system and take limits without being dominated by boundary effects.
(The problem we need to finesse is that a finite isolated system necessarily has a boundary, which is problematic for our purpose.)
Instead, we will just make an informal argument about why this interpretation (and using the canonical ensemble) is expected to work.

First, we fix an $n$ and treat $H^\infty_n$ as a set of isolated blocks (not coupled to an external bath).
From textbook statistical mechanics, we know that if we have a system consisting of a large number $N$ of identical blocks, the microcanonical ensemble of the total system at energy $e$ is locally equivalent to the product of canonical ensembles of the blocks at temperature $T$, in the $N\to\infty$ limit, where $T$ is determined such that the expected energy density of a block is $e$.
Our situation is a bit more complicated, as the blocks are not identical and the distribution of blocks within $H^\infty_n$ is highly non-uniform.
Moreover, when $L^\text{nr}_n$ grows fast enough, a finite patch of the system can be dominated by a single $H^\text{nr}_n$ block.
These issues make an unambiguous definition of a thermodynamic limit difficult. For example, the block number limit $N\rightarrow\infty$ may not be the same as the usual system size limit $L\rightarrow\infty$ where the limit is taken by adding sites.
Nevertheless, we will assume some regularity requirements on the initial blocks such that the entropy density function of $H^\text{nr}_n$ converges to a fixed function $s(e)$ fast enough as $n\to\infty$, and that $s(e)$ is strictly concave (which implies a one-to-one correspondence between $e$ and $T$).
In this way, we should expect some form of the central limit theorem to hold, such that with some reasonable choices of a thermodynamic limit, the microcanonical ensemble of $H^\infty_n$ at energy density $e$ will be locally equivalent to $\rho^\infty_n$ at temperature $T=de/ds$.

The above procedure corresponds to taking some kinds of $L\to\infty$ limit first, and next, we also need to take the limit on the depth of the circuit ($n\to\infty$) to complete the construction.
At a given $n$, local observables start to suffer from finite-size effects at a length scale $L^\text{res}_n$.
That is, the effective size of the bulk is around $L^\text{res}_n$.
So taking $L\to\infty$ before $n\to\infty$ implies that the effective bulk can be much smaller than the entire system, which is the case for our system.
We expect (and assume) that the two limits can be taken together in some sense, but providing the exact procedure is beyond the scope of this paper.

Below, all the rigorous proofs will be directly based on the local properties of the canonical ensemble $\rho^\infty_n$ with $n\to\infty$ taken at the end, and only their physical interpretations will be related to the closed system interpretation discussed here.
Our results are based on this reasonable assumption of the equivalence between microcanonical and canonical ensembles for the problem, but we believe that our results are valid beyond this assumption even for the microcanonical ensemble in the thermodynamic limit, although doing that is beyond the scope of the current work.

\subsubsection{Evolution in $n$}\label{sec:rhoEvol}

To analyze the properties of the system at finite $n$ and take $n\to\infty$, we need to consider the relation between $\rho^\infty_n$ and $\rho^\infty_{n+1}$.
To do this, note that $\rho^\infty_{n+1}$ only differs from $\rho^\infty_n$ by merging groups of the tensor factors
\begin{equation}
    \rho^\text{lim}_{n+1}=\rho^\text{nr}_n\otimes\rho^\text{res}_n\otimes\widetilde{\rho^\text{res}_n}\otimes\widetilde{\rho^\text{nr}_n}\mapsto\rho^\text{res}_{n+1}
\end{equation}
which corresponds to four blocks being merged into one in Fig.~\ref{fig:approx}.
This merging consists of two changes: the eigenstate basis rotation due to $U_n$, and the modification of the Boltzmann weights due to energy level shifts from $H^\text{lim}_{n+1}$ to $H^\text{res}_{n+1}$.
To calculate the changes, first we write
\begin{equation}
    \rho^\text{lim}_{n+1}=\sum_{X}q_{X}|X\rangle_n\langle X|
\end{equation}
with Boltzmann weights
\begin{equation}
    q_{X}=\frac{e^{-E^\text{lim}_{n+1,X}/T}}{\sum_{X'}e^{-E^\text{lim}_{n+1,X'}/T}}
\end{equation}
where $E^\text{lim}_{n+1,X}$ is the energy of the state $|X\rangle_n$ in $H^\text{lim}_{n+1}$ (the bitstring $X$ was written as $ABCD$ before; here we use a single letter for brevity).
Now note that $U^\text{lim}_n|X\rangle_{n}$ is an eigenstate of $H^\text{lim}_{n+1}$ as well, with the same energy $E^\text{lim}_{n+1,X}$, as $U^\text{lim}_n$ only rotates the eigenstates in each degenerate subspace of $H^\text{lim}_{n+1}$.
Thus, we have
\begin{equation}\label{eq:Ulimrho}
    U^\text{lim}_n\rho^\text{lim}_{n+1}U^{\text{lim}\dagger}_n=\rho^\text{lim}_{n+1},
\end{equation}
which means that the only change due to eigenstate rotation is from the error term in (\ref{eq:UnError}).
To bound the change, we use the trace distance for the density matrices:
\begin{equation}\label{eq:rhoBasisShift}
\begin{aligned}
    &\left\|U_n\rho^\text{lim}_{n+1}U^{\dagger}_n-\rho^\text{lim}_{n+1}\right\|_1\\
    &\leq\|\mathcal{E}_n\|\, \|\rho^\text{lim}_{n+1}\|_1\, \|U_n^\dagger\|+\|U^\text{lim}_n\|\, \|\rho^\text{lim}_{n+1}\|_1\, \|\mathcal{E}_n^\dagger\|\\
    &<2\epsilon_n
\end{aligned} 
\end{equation}
where $\|\cdot\|_1$ denotes the trace norm.
Next, we calculate the change due to energy level shifts.
The final ensemble is
\begin{equation}
    \rho^\text{res}_{n+1}=\sum_{X}q'_{X}|X\rangle_{n+1}\langle X|
\end{equation}
with new Boltzmann weights
\begin{equation}
    q'_{X}=\frac{e^{-E^\text{res}_{n+1,X}/T}}{\sum_{X'}e^{-E^\text{res}_{n+1,X'}/T}}
\end{equation}
where $E^\text{res}_{n+1,X}$ is the energy of the state $|X\rangle_{n+1}$ in $H^\text{res}_{n+1}$.
As $\rho^\text{res}_{n+1}$ only differs from $U_n\rho^\text{lim}_{n+1}U^{\dagger}_n$ by a shift of Boltzmann weights from $q_{X}$ to $q'_{X}$, what we need to bound is $q'_X-q_X$.
To derive the bound, first note that due to the error bound,
\begin{equation}
    \left|E^\text{res}_{n+1,X}-E^\text{lim}_{n+1,X}\right|<\frac{\epsilon_n}{2},
\end{equation}
we have
\begin{equation}\label{eq:rX}
    e^{-\epsilon_n/2T}<r_{X}:=\frac{e^{-E^\text{res}_{n+1,X}/T}}{e^{-E^\text{lim}_{n+1,X}/T}}<e^{\epsilon_n/2T}.
\end{equation}
The new weights can be expressed in terms of the old weights as
\begin{equation}\label{eq:qpX}
    q'_{X}=\frac{q_{X}r_{X}}{\sum_{X'}q_{X'}r_{X'}}.
\end{equation}
Now, the numerator of (\ref{eq:qpX}) is bounded by (\ref{eq:rX}), and the denominator is a weighted average of such bounds.
Therefore, we arrive at the ratio bound
\begin{equation}
    e^{-\epsilon_n/T}<\frac{q'_X}{q_X}<e^{\epsilon_n/T}.
\end{equation}
To transform this bound into a trace distance bound, we bound the sum of probability differences
\begin{equation}\label{eq:qXdiffsum}
\begin{aligned}
    &\sum_X|q'_X-q_X|\\
    &=\sum_{q_X>q'_X}q_X\left(1-\frac{q'_X}{q_X}\right)+\sum_{q_X<q'_X}q_X\left(\frac{q'_X}{q_X}-1\right)\\
    &<(1-x)(1-e^{-\epsilon_n/T})+x(e^{\epsilon_n/T}-1)\\
    &=2\tanh\frac{\epsilon_n}{2T}
\end{aligned}
\end{equation}
where $x$ is the sum of $q_X$ that $q_X<q'_X$, under the most extreme condition that all the probability shifts are saturated, that is, $(1-x)e^{-\epsilon_n/T}+x e^{\epsilon_n/T}=1$.
So we arrive at the bound on trace distance
\begin{equation}
    \left\|\rho^\text{res}_{n+1}-U_n\rho^\text{lim}_{n+1}U^{\dagger}_n\right\|_1<2\tanh\frac{\epsilon_n}{2T}.
\end{equation}
Combining with (\ref{eq:rhoBasisShift}), we finally arrive at
\begin{equation}\label{eq:rhoFullBound}
\begin{aligned}
    \left\|\rho^\text{res}_{n+1}-\rho^\text{lim}_{n+1}\right\|_1&<2\epsilon_n+2\tanh\frac{\epsilon_n}{2T}\\
    &\sim 2\epsilon_n+\frac{\epsilon_n}{T}.
\end{aligned}
\end{equation}

The above bound implies that for any local observable $O$ (a Hermitian operator supported on a finite number of sites in the infinite chain), its canonical expectation value
\begin{equation}
    \langle O\rangle_n:=\Tr(O\rho^\infty_n)
\end{equation}
has a well-defined limit
\begin{equation}
    \lim_{n\to\infty}\langle O\rangle_n:=\langle O\rangle_\infty.
\end{equation}
This is due to that, for any large enough $n$, $O$ is completely inside an $H^\text{res}_n$ or $\widetilde{H^\text{res}_n}$ block.
So the bound (\ref{eq:rhoFullBound}) implies
\begin{equation}\label{eq:OnEvol}
\begin{aligned}
    \left|\langle O\rangle_{n+1}-\langle O\rangle_n\right|
    &\leq\|O\|\left\|\rho^\text{res}_{n+1}-\rho^\text{lim}_{n+1}\right\|_1\\
    &\lesssim\|O\|\left(2\epsilon_n+\frac{\epsilon_n}{T}\right).
\end{aligned}
\end{equation}
By the requirements on the error bounds, the accumulated shifts converge, leading to a well-defined limit.
The limit $\langle O\rangle_\infty$ provides a well-defined notion for the thermodynamic expectation value of a local observable, even if we cannot directly define the energy eigenstates of $H_\infty$.
Also note that we only consider nonzero temperatures and that the bound converges more slowly for temperatures closer to zero.

\subsubsection{Sampling from the ensemble}

As we will study the eigenstate properties of the system, especially for ETH, which is based on the properties of individual eigenstates and not just ensemble averages, we need to define how to sample an eigenstate from the ensemble.
Conceptually, what we intend to do is to sample an eigenstate from a narrow energy window around an energy density $e$.
However, as we mentioned, there are difficulties in defining the microcanonical ensemble rigorously.
Instead, we will go with the counterpart in the canonical ensemble, to ``choose an eigenstate at temperature $T$''.
To define this, we again use the property that the canonical ensemble of $H^\infty_n$ is a product ensemble of blocks, so the natural definition is to independently sample an energy eigenstate for each block $H$ in $H^\infty_n$, such that an eigenstate $|X\rangle$ of $H$ has a probability of the Boltzmann weight $q_X=e^{-E_X/T}/\sum_{X'}e^{-E_{X'}/T}$ of being chosen.
If we take the formal tensor product of the sampled eigenstates from all blocks, we get an eigenstate $|S\rangle_n$ of $H^\infty_n$.
We will say that this eigenstate $|S\rangle_n$ is being sampled from $\rho^\infty_n$.
As we do not define the eigenstates of $H_\infty$, we will not define the sampling of eigenstates from $H_\infty$.
Rather, we will only study eigenstate properties at a finite $n$, with the limit $n\to\infty$ taken in the end.

As discussed in Sec.~\ref{sec:interpretclosedsystems}, we expect a closed system interpretation of $\rho^\infty_n$ under a suitable regularity condition of the $H^\text{nr}_n$ blocks and a suitable $L\to\infty$ limit.
Under these assumptions, we expect that the sampling from $\rho^\infty_n$ is also equivalent to that from a closed system.
More specifically, as long as we only look at the local properties of the sampled eigenstate, we expect that it is equivalent to randomly choosing an energy eigenstate in a narrow energy window around an energy density $e$, which is determined by $T=de/ds$ from the limit entropy function $s(e)$ of $H^\text{nr}_n$ as $n\to\infty$.

\subsection{Fluctuations of local observable expectations}

To explore the thermalization/localization properties of the many-body system, we study the fluctuations of local observable expectations for eigenstates sampled from $\rho^\infty_n$.
There are two possible scenarios.
If the fluctuations for all local observables go to zero as $n\to\infty$, then we can say that a randomly sampled eigenstate locally looks thermal with probability 1.
Otherwise, a randomly sampled eigenstate will contain local information that distinguishes it from a thermal state.
Under the closed system interpretation of $\rho^\infty_n$ with a suitable way to take the $L\to\infty$ and the $n\to\infty$ limits together, we expect the first scenario to be a form of the weak eigenstate thermalization hypothesis (ETH), and the second scenario to exhibit MBL-like behavior.
In this subsection, we derive the key recursion relations of the evolution of the diagonal matrix element variance 
with respect to the level index $n$, which is then used in the next subsection to characterize the two MBC phases.
Then in Sec.~\ref{sec:mobilityedge}, we will show that they can exist as finite-temperature phases in some situations.

We fix a local observable $O$ and consider sampling a state $|S\rangle_n$ from $\rho^\infty_n$.
The quantum expectation value, $_n\langle S|O|S\rangle_n$, is then considered a classical random variable.
The classical expectation value of this random variable is the same as the ensemble average. That is,
\begin{equation}
    \mathbb{E}_{|S\rangle_n\sim\rho^\infty_n}[_n\langle S|O|S\rangle_n]=\langle O\rangle_n,
\end{equation}
where $\sim$ denotes that the eigenstate is drawn from the corresponding ensemble.
We study the variance of this classical random variable, denoted as the diagonal matrix element variance:
\begin{equation}
\begin{aligned}
    \operatorname{Var}_n[O]&:=\operatorname{Var}_{|S\rangle_n\sim\rho^\infty_n}[_n\langle S|O|S\rangle_n]\\
    &=\mathbb{E}_{|S\rangle_n\sim\rho^\infty_n}[(_n\langle S|O|S\rangle_n-\langle O\rangle_n)^2].
\end{aligned}
\end{equation}
We will derive a recursion relation for the diagonal matrix element variance between $n$ and $n+1$, and then characterize its behavior in the $n\to\infty$ limit.

\subsubsection{Evolution in $n$ for a fixed local observable}

In this subsection, we derive a recursion relation from $\operatorname{Var}_n[O]$ to $\operatorname{Var}_{n+1}[O]$ with a fixed local $O$.
As in Sec.~\ref{sec:rhoEvol}, we consider the merging of four blocks into one from $n$ to $n+1$, where we have two changes: the basis rotation caused by $U_n$ and the shifts of Boltzmann weights.

We choose a large enough $n$ so that the support of $O$ is within an $H^\text{res}_n$ or $\widetilde{H^\text{res}_n}$ block.
Without loss of generality, we further assume that $O$ is located in an $H^\text{res}_n$ block.
The mirror of $O$ in the $\widetilde{H^\text{res}_n}$ block is referred to as $\widetilde{O}$.
Now the variance can be written as
\begin{equation}
    \operatorname{Var}_n[O]=\operatorname{Var}_{|X\rangle_n\sim\rho^\text{lim}_{n+1}}[_n\langle X|O|X\rangle_n]
\end{equation}
where $X=ABCD$ is the bitstring labeling the eigenstates of $H^\text{lim}_{n+1}$. The label $ABCD$ indicates that the eigenstate is block-wise from its four decoupled blocks.

The first step is to do the change of basis from $|X\rangle_n$ to $U^\text{lim}_n|X\rangle_n=|X\rangle_\text{lim}=|ABCD\rangle_\text{lim}$ (in the notations of Sec.~\ref{sec:MBerr}).
Although this does not change the density matrix itself [see (\ref{eq:Ulimrho})], it does affect the variance, as we are now sampling using a different basis in each degenerate subspace $H^\text{lim}_n$.
By (\ref{eq:MBPertIdeal}), we have
\begin{multline}\label{eq:varlimObranches}
    _n\langle X|U^\text{lim}_nO\,U^{\text{lim}\dagger}_n|X\rangle_n\\=
    \begin{cases}
        _n\langle X|O|X\rangle_n,&\text{if }A\neq \widetilde{D},\\
        \frac{1}{2}(_n\langle X|O|X\rangle_n +\,_n\langle X|\widetilde{O}|X\rangle_n),&\text{if }A=\widetilde{D}.
    \end{cases}
\end{multline}
The bitstrings $A$, $B$, $C$, and $D$ are independently sampled. And since $O$ ($\widetilde O$) is only supported on the $H^\text{res}_{n}$ ($\widetilde{H^\text{res}_n}$) block, $_n\langle X|O|X\rangle_n$ ($_n\langle X|\widetilde{O}|X\rangle_n$) only depends on $B$ ($C$).
Moreover, the distribution of $B$ is exactly the same as $\widetilde{C}$.
Thus, we have
\begin{multline}\label{eq:VarlimO}
    \operatorname{Var}_{|X\rangle_n\sim\rho^\text{lim}_{n+1}}[_n\langle X|U^\text{lim}_nO\,U^{\text{lim}\dagger}_n|X\rangle_n]\\
    =(1-p_n)\operatorname{Var}_n[O]+p_n\frac{\operatorname{Var}_n[O]}{2},
\end{multline}
where $p_n$ is the probability that $A=\widetilde{D}$. Since $A$ and $\widetilde{D}$ have the same distribution, equivalently, $p_n$ can be understood as the probability that two individual samplings from $\rho^\text{nr}_n$ result in the same state. Thus, we have
\begin{equation}
    p_n=\Tr \left(\frac{e^{-H^\text{nr}_{n}/T}}{\Tr e^{-H^\text{nr}_{n}/T}}\right)^2=e^{-S_{2,n}^\text{nr}}
\end{equation}
where $S_{2,n}^\text{nr}$ is the thermodynamic R\'enyi-$2$ entropy of $H^\text{nr}_{n}$ at temperature $T$.
Eq.~(\ref{eq:VarlimO}) has a clear physical interpretation: the variance of the local observable remains the same if a resonance does not occur at the $n$th level, and it splits in half if a resonance occurs due to the local state being averaged with its mirror counterpart.

The rest of the derivation towards $\operatorname{Var}_{n+1}[O]$ is to apply the error terms.
First, by the error term in $U_n$, and the general inequality for random variables $Y$ and $Z$
\begin{equation}
    \big|\operatorname{Var}[Y+Z]-\operatorname{Var}[Y]\big|\leq\operatorname{Var}[Z]+2\sqrt{\operatorname{Var}[Y]\operatorname{Var}[Z]}
\end{equation}
we have
\begin{multline}\label{eq:VarRotation}
\Big|\operatorname{Var}_{|X\rangle_n\sim\rho^\text{lim}_{n+1}}[_n\langle X|U_nO\,U^{\dagger}_n|X\rangle_n]\\
-\operatorname{Var}_{|X\rangle_n\sim\rho^\text{lim}_{n+1}}[_n\langle X|U^\text{lim}_nO\,U^{\text{lim}\dagger}_n|X\rangle_n]\Big|\\
<4\epsilon_n(1+\epsilon_n)\|O\|^2
\end{multline}
Next, for the shift of the Boltzmann weight, by (\ref{eq:qXdiffsum}) and the general inequality for a random variable $Y$ under two different probability distributions $\{q_i\}$ and $\{q'_i\}$
\begin{equation}
    \big|\operatorname{Var}_q[Y]-\operatorname{Var}_{q'}[Y]\big|\leq\frac{(\operatorname{range}Y)^2}{2}\sum_i|q'_i-q_i|,
\end{equation}
we have
\begin{multline}\label{eq:VarShift}
\Big|\operatorname{Var}_{n+1}[O]
-\operatorname{Var}_{|X\rangle_n\sim\rho^\text{lim}_{n+1}}[_n\langle X|U_nO\,U^{\dagger}_n|X\rangle_n]\Big|\\
=\Big|\operatorname{Var}_{|X\rangle_{n+1}\sim\rho^\text{res}_{n+1}}[_{n+1}\langle X|O|X\rangle_{n+1}]\\
-\operatorname{Var}_{|X\rangle_{n+1}\sim U_n^\dagger\rho^\text{lim}_{n+1}U_n}[_{n+1}\langle X|O|X\rangle_{n+1}]\Big|\\
<4\|O\|^2 \tanh\frac{\epsilon_n}{2T}.
\end{multline}
Combining (\ref{eq:VarlimO}), (\ref{eq:VarRotation}), and (\ref{eq:VarShift}), we obtain the final recursion relation:
\begin{equation}\label{eq:VarRecursion}
    \operatorname{Var}_{n+1}[O]=\left(1-\frac{p_n}{2}\right)\operatorname{Var}_{n}[O]+\delta_n
\end{equation}
with error term
\begin{equation}
\begin{aligned}
    |\delta_n|&<4\|O\|^2\left(\epsilon_n+\tanh\frac{\epsilon_n}{2T}+\epsilon_n^2\right)\\
    &\sim 2\|O\|^2\left(2\epsilon_n+\frac{\epsilon_n}{T}\right).
\end{aligned}
\end{equation}
This property will be used in the next subsection to characterize the two phases of MBC.

\subsubsection{Evolution in $n$ for approximate LIOMs}

As we are going to discuss the ETH/MBL behaviors in $H_\infty$, one important set of local observables is the approximate LIOM operators $\tau^z_{n,j}$, defined as returning the $j$th bit of the $H^\infty_n$ eigenstate $|S\rangle_n$ (with `0' corresponding to eigenvalue $+1$ and `1' to $-1$).
In addition to evolving the variance of $\tau^z_{n_0,j}$ for a fixed $n_0$ in $n$, which is a special case of (\ref{eq:VarRecursion}), we are also interested in the recursion relation from $\operatorname{Var}_{n}[\tau^z_{n,j}]$ to $\operatorname{Var}_{n+1}[\tau^z_{n+1,j}]$ (and also the drift of the expectation values).

Note that this case is actually simpler than deriving (\ref{eq:VarRecursion}).
First, if the $j$th LIOM is within a $H^\text{nr}_{n'}$ or $\widetilde{H^\text{nr}_{n'}}$ block in $H^\infty_n$ with $n'>n$, then its evolution is trivial
\begin{equation}
    \langle\tau^z_{n+1,j}\rangle_{n+1}=\langle\tau^z_{n,j}\rangle_{n},\quad\operatorname{Var}_{n+1}[\tau^z_{n+1,j}]=\operatorname{Var}_{n}[\tau^z_{n,j}].
\end{equation}
Now, if the $j$th LIOM is in one of the four types of blocks in $H^\infty_{n}$ that merge into an $H^\text{res}_{n+1}$ block in $H^\infty_{n+1}$, then we need to use a similar derivation as before.
However, the eigenstate rotation step becomes trivial, as $U_n\tau^z_{n,j}U^\dagger_n$ is exactly $\tau^z_{n+1,j}$.
Therefore, we only need to consider the shift in Boltzmann weights, resulting in
\begin{equation}\label{eq:vartauEvol}
\begin{aligned}
    \langle\tau^z_{n+1,j}\rangle_{n+1}&=\langle\tau^z_{n,j}\rangle_{n} + \delta'_n\\
    \operatorname{Var}_{n+1}[\tau^z_{n+1,j}]&=\operatorname{Var}_{n}[\tau^z_{n,j}] + \delta''_n
\end{aligned}
\end{equation}
with error bounds
\begin{equation}
    |\delta'_n|<2\tanh\frac{\epsilon_n}{2T},\quad|\delta''_n|<4\tanh\frac{\epsilon_n}{2T}.
\end{equation}
There is no decay of variance [compared to (\ref{eq:VarRecursion})], as the $\tau^z_{n,j}$ operators themselves grow longer in $n$ to absorb the effect of resonances, unlike a fixed local operator $O$ that stays the same during a resonance.

\subsection{Two types of MBC phases}

Having derived the recursion relation (\ref{eq:VarRecursion}), we can now characterize the two types of MBC phases by whether $\operatorname{Var}_{n}[O]\to 0$ as $n\to\infty$, that is, whether (almost) all eigenstates sampled at temperature $T$ locally look like the canonical ensemble at $T$.
Due to the convergence of accumulated errors, we can see that it is equivalent to whether the infinite product $\prod_n(1-p_n/2)$ converges to a finite value.
More specifically, pick a large enough $n_0$ beyond which (\ref{eq:VarRecursion}) holds, and define the partial product  $P_n=\prod_{n'=n_0}^{n-1}(1-p_{n'}/2)$, then we have, for $n\geq n_0$,
\begin{equation}
    \operatorname{Var}_{n}[O]=P_n\operatorname{Var}_{n_0}[O]+\sum_{n'=n_0+1}^{n}\frac{P_n}{P_{n'}}\delta_{n'-1}.
\end{equation}
Hence, we need to distinguish between two cases depending on whether $P_n\to0$, or equivalently, whether $\sum_n p_n$ diverges.
If $P_n\to0$, then since $P_n/P_{n'}\leq1$ and $\sum_n\delta_n$ converges, the limit of $\operatorname{Var}_{n}[O]$ is bounded by an arbitrarily small value by choosing a large $n_0$.
Therefore, we have $\operatorname{Var}_{n}[O]\to 0$ as $n\to\infty$.
We call this ETH-like phase the \emph{many-body critically extended} (MBC-E) phase.
That is,
\begin{equation}
    \sum_np_n=\infty\implies\text{MBC-E (ETH-like)}
\end{equation}
We will discuss the difference from the usual ETH phase in Sec.~\ref{sec:mobilityedge}.

Conversely, if $P_n\to P_\infty>0$, and suppose additionally that
\begin{equation}\label{eq:Varn0large}
    \operatorname{Var}_{n_0}[O]>\frac{1}{P_\infty}\sum_{n=n_0}^{\infty}|\delta_{n}|.
\end{equation}
Then $\operatorname{Var}_{n}[O]$ does not go to zero as $n\to\infty$.
That is, the observable $O$ can distinguish a randomly sampled eigenstate from the canonical ensemble.
To construct an extensive set of local observables that satisfy (\ref{eq:Varn0large}), we will use the approximate LIOMs $\tau^z_{n,j}$.
Note that there is a lower bound $\operatorname{Var}_1[\tau^z_{1,j}]>V$ independent of $j$, so by choosing $n_1$ such that $V-\sum_{n=n_1}^\infty\delta_n'>V'>0$, and by (\ref{eq:vartauEvol}), we have
\begin{equation}
    \operatorname{Var}_{n}[\tau^z_{n,j}]>V'
\end{equation}
for all $n>n_1$ with $j$ in an $H^\text{nr}_{n'}$ block with $n'>n_1$.
Then we pick $n_0>n_1$ so that $\sum_{n=n_0-1}^\infty|\delta_n|/P_\infty<V'$.
Now, (\ref{eq:Varn0large}) is satisfied for all such $\tau^z_{n_0,j}$, giving an extensive set of local observables that violate the ETH-like property\footnote{
Note that if the temperature is not too small, one can actually choose $n_1=1$ and have a complete set of such observables (that is, for all $j$).
But for extremely low temperatures, there remains the possibility that some initial building blocks with low $n$ are dominated by the ground state, giving a variance so small that it is indistinguishable from the error bound.
}.
We call this MBL-like phase the \emph{many-body critically localized} (MBC-L) phase.
In other words,
\begin{equation}
    \sum_np_n<\infty\implies\text{MBC-L (MBL-like)}.
\end{equation}

While there are many other diagnostics that are usually applied to MBL that may also apply to MBC-L, we will not provide a rigorous derivation for each one.
For example, in the MBC-L phase, the connected correlation function between two observables $O_i$ and $O_j$ supported around sites $i$ and $j$ decays to zero for fixed $i$ with $j\to\pm\infty$.
This can be seen by noticing that for faraway $i$ and $j$, the blocks they belong to will not be coupled until some $n_0$ (which goes to infinity as $j\to\pm\infty$), giving a zero correlation.
Now, when $i$ and $j$ first get coupled at level $n_0$, with a probability $1-p_n$, there is no resonance, and therefore the only contribution to the correlation is the error term.
The bound of the correlation is then derived by the convergence of $p_n$, Eqs.~(\ref{eq:rhoFullBound}), (\ref{eq:VarRotation}), and (\ref{eq:VarShift}) for the shift at level $n_0$, and Eqs.~(\ref{eq:OnEvol}) and (\ref{eq:VarRecursion}) for the bound of the shifts caused by subsequent levels.
As another example, the entanglement entropy exhibits area-law scaling with the subsystem size (for almost every energy eigenstate).
This can similarly be shown by noticing that a cut in the chain only gets an $O(1)$ entanglement growth when it is in a resonant block and a resonance occurs. Thus, by a similar error-bounding argument as above, one can show that the growth is limited by the convergence of $p_n$.
We will discuss the difference between the MBC-L phase in our model and the usual MBL phase in Sec.~\ref{sec:mobilityedge}.

\subsubsection{Interpretation and single-particle analogy}

One can compare the above results with the single-particle case in Sec.~\ref{sec:splittingweights}, where the system becomes singular continuous.
In the single-particle case, the ``local observable'' is just the local state projector, whose expectation value becomes the weight discussed in Sec.~\ref{sec:splittingweights}.
Since the single-particle eigenstates delocalize as $n\to\infty$, the expectation value decays to zero in such a limit.
Hence, the weight can be considered as the single-particle analogy of the diagonal matrix element variance of local observables.

Now, the first term in Eq.~(\ref{eq:VarlimO}), where a resonance is created, is analogous to the splitting of weights in Sec.~\ref{sec:splittingweights}.
Thus, the mechanism that leads to the absence of a pure-point spectrum in the single-particle model is the reason for the ETH-like behavior in the many-body case.
On the other hand, we mention that the mechanism that leads to the absence of an absolutely continuous spectrum in the single-particle case is the reason for the separation between the resonance timescales $t_n$ at different $n$.
Such a separation is expected to result in differences in the thermalization properties between MBC-E and the usual ETH phase.
For example, energy transport can be arbitrarily slow due to the fast divergence of the timescales.
Therefore, the direct analogy of the singular continuity in our many-body system is the ETH-like but non-thermal-like behavior of MBC-E.

However, the collective freezing effect has no single-particle analogy, whose consequence is that the splitting becomes conditional, depending on the probability $p_n$ in Eq.~(\ref{eq:VarlimO}). In the single-particle case, the weights always get split in half (up to an error term) from $n$ to $n+1$ ($n\geq n_0$), and eventually, there will be an infinite number of splittings when $n \rightarrow \infty$. As an analogy, in the many-body case, this corresponds to $p_n=1$. While the MBC-L phase appears only when there is a finite number of splittings such that the system is MBL-like, there is no single-particle analogy for it.
Therefore, we may say that MBC-L (as well as its transition from MBC-E) is purely a many-body phenomenon.

\subsection{Mobility edges, scars, and inverted scars}\label{sec:mobilityedge}

In the last subsection, we showed that the many-body systems contain two phases: MBC-E (ETH-like) and MBC-L (MBL-like).
Now we will show that instead of the entire spectrum being either ETH-like or MBL-like, our model can have a much richer spectrum.
First, we will show that MBC-E and MBC-L phases can exist in the same system as a finite-$T$ transition.
In other words, a many-body mobility edge exists.
Then, we will show that even within the same phase, there exist rare eigenstates that behave like those in the opposite phase. 
We will interpret these rare eigenstates in the MBC-E and MBC-L phases as many-body scars and many-body inverted scars, respectively~\cite{shiraishi2017systematic,srivatsa2020manybody,Iversen2022escaping,chen2024inverting,srivatsa2023mobility}.

\subsubsection{Many-body mobility edges}

Recall that whether our system is in MBC-E or MBC-L is determined by the convergence of $\sum_n p_n=\sum_n \exp(-S_{2,n}^\text{nr})$, where $S_{2,n}^\text{nr}$ is the thermodynamic R\'enyi-$2$ entropy of $H^\text{nr}_{n}$ at temperature $T$.
Hence, one can have a finite-$T$ phase transition if its convergence depends on $T$.
We will show that this happens if $L^\text{nr}_n$ is logarithmically divergent.

Take $L^\text{nr}_n=\lceil p\ln n\rceil$ for a fixed $p>0$.
Since the classical Ising model is always in the paramagnetic phase at finite $T$ with a finite correlation length (assuming the $\delta$ terms in the initial building blocks decay to zero with $n$ fast enough), by the extensiveness of R\'enyi-$2$ entropy we have, at large $n$
\begin{equation}
    S_{2,n}^\text{nr} (T)=L^\text{nr}_ns_{2,n}^\text{nr} (T)+O(1)
\end{equation}
for some R\'enyi-$2$ entropy density function $s_{2,n}^\text{nr} (T)$, which goes to $0$ as $T\to 0$ and goes to $\ln 2$ for $T\to\infty$.
The series then becomes
\begin{equation}
    \sum_ne^{-S_{2,n}^\text{nr} (T)}\sim\sum_n n^{-ps_{2,n}^\text{nr} (T)}
\end{equation}
which converges if and only if $ps_{2,n}^\text{nr} (T)>1$.
That is, by tuning $p$, we have a tunable critical temperature $T_c$ with $ps_{2,n}^\text{nr} (T_c)=1$ above (below) which the system is in MBC-L (MBC-E).
Assuming the closed-system interpretation in Sec.~\ref{sec:interpretclosedsystems} holds, this finite-$T$ phase transition corresponds to a thermodynamic many-body mobility edge (MBME) in the energy spectrum.
The counterintuitive phenomenon that high temperatures lead to localization can be viewed as an example of quantum inverse freezing~\cite{thomas2018quantum}.

On the other hand, if $L^\text{nr}_n$ grows asymptotically faster (slower) than $\log n$, the system will be in MBC-L (MBC-E) for all nonzero temperatures (including infinity).
That is, there is no many-body mobility edge in such cases.

\subsubsection{Many-body scars}

In the MBC-E phase, we have shown that for any local observable $O$, the diagonal matrix element variance $\operatorname{Var}_n[O]$ goes to zero as $n\to\infty$.
This guarantees that the fraction of sampled eigenstates with non-thermal expectation values goes to zero when $n\to\infty$.
However, this does not mean that \emph{all} eigenstates will have thermal expectation values.
Indeed, whether there is a resonance at level $n$ is controlled by the state of the non-resonant region, not inherently by the system parameters.
Therefore, one can construct states whose configurations in the non-resonant regions beyond some level $n_0$ are chosen to always be asymmetric.
For such a rare state, Eq.~(\ref{eq:varlimObranches}) always chooses the first value above $n_0$, so that $_n\langle X|O|X\rangle_n$ gets fixed at $n_0$ (up to error terms) and does not approach the thermal value.
Moreover, the conditions that produce this rare state do not asymptotically change the energy density.
To see this, one can intuitively think of a Boltzmann sampling of the blocks at level $n_0$. To produce rare states, we just need that a pair of non-resonant blocks never gets symmetric states in the samples, which does not modify the temperature (a more rigorous construction is presented below).
Therefore, under the closed system interpretation, every energy window is expected to contain some rare state, so that the MBC-E phase only satisfies the weak ETH, not the strong one.
These rare states can thus be interpreted as a type of quantum many-body scars.

Now we present a construction of a scarred state $|S'\rangle$ given a regular energy eigenstate $|S\rangle$ (where $S$ and $S'$ are the bitstring labels) based on swapping non-resonant blocks.
Under a suitable definition of the energy density of the infinite chain, we expect that this construction allows us to make the energy density of $|S'\rangle$ arbitrarily close to that of $|S\rangle$, assuming that the latter has a well-defined energy density.
First, fix a region $R$ and a large enough level $n_0$ so that $R$ is in a resonant block, and then cut the circuit that produces $|S\rangle$ at layer $n_0$ to get $|S\rangle_{n_0}$.
As the Hamiltonians at this stage are decoupled finite blocks, an $H^\text{nr}_{n_0}$ block with a given configuration almost certainly appears an infinite number of times within $S$, which means that it is possible to exchange the configuration of two $H^\text{nr}_{n_0}$ blocks (by exchanging the corresponding substrings in $S$) so that the configurations in the $H^\text{nr}_{n_0}$, $\widetilde{H^\text{nr}_{n_0}}$ pair whose corresponding resonant region contains $R$ are not mirror symmetric.
Similarly, this can be done for the $H^\text{nr}_{n}$, $\widetilde{H^\text{nr}_n}$ pair for each $n>n_0$.
The resulting binary string after all of these block swaps becomes $S'$, whose corresponding eigenstate $|S'\rangle$ violates ETH due to the lack of resonance beyond $n_0$ for a region containing $R$ (i.e.\ Eq.~(\ref{eq:varlimObranches}) always goes through the second branch beyond $n_0$).

Although an infinite shuffling of a sequence in general may lead to convergence issues or change the asymptotic energy density, there are no such issues here.
To see this, note that each swap occurs at a pair of non-resonant blocks at a different $n$, so they correspond to non-overlapping substrings in $S$, making $S'$ well-defined.
Secondly, as we assume that $L^\text{nr}_n\to\infty$ as $n\to\infty$, in order to make the configuration of $H^\text{nr}_{n}$ and $\widetilde{H^\text{nr}_n}$ different, we can choose the swaps so that the energy density difference $\Delta e_n$ between the $H^\text{nr}_{n}$ block before and after the swap satisfies $\Delta e_n\to 0$ as $n\to\infty$.
Therefore, under a suitable closed system interpretation, these swaps of non-resonant blocks do not modify the energy density at level $n_0$ and that $|S\rangle_{n_0}$ and $|S'\rangle_{n_0}$ have the same energy densities
\footnote{This follows from the fact that for sequences $a_k$ and $b_k$ related by swapping terms at disjoint indices $(k_m,k'_m)$ (i.e.\ $b_{k'_m}=a_{k_m}$, $a_{k'_m}=b_{k_m}$ and otherwise $a_k=b_k$) such that $|a_{k'_m}-a_{k_m}|\to 0$ as $m\to\infty$, the averages $\sum_{k=1}^la_k/l$ and $\sum_{k=1}^lb_k/l$ converge to the same value as $l\to\infty$ (if either converges).}.

Now, by the error bounds, the energy density shifts due to the circuit from level $n_0$ to infinity converge, so that the energy density between $|S\rangle_{n_0}$ and $|S\rangle$, and those between $|S'\rangle_{n_0}$ and $|S'\rangle$ can be made arbitrarily small under a suitable closed system interpretation.
This identifies $|S'\rangle$ as the required scar state.

The existence of scar states also means that there is no uniform timescale for thermalization, since a normal state may locally look like a scar state up to a finite number of levels.
Indeed, given a local observable $O$ and any level index $n$, there is a finite fraction of states that $O$ does not participate in any resonances before level $n$, so the variance of $O$ decays only after that.
That is, for any $n$, there is a finite fraction of states whose thermalization timescale is longer than $t_n$ (although the fraction goes to zero as $n\to\infty$).
Note that this phenomenon occurs even if the system is in MBC-E at all temperatures. 

Another related property that makes MBC-E different from ETH is that, although the entanglement entropy scales as volume law (in the sense of scaling subsystem size in an infinite chain) due to the local density matrix being thermal, if we take a finite system of length $L$ and cut it in half, from the structure of the circuit in Fig.~\ref{fig:approx} one can see that the entanglement entropy cannot grow faster than $O(\log L)$.
In other words, the volume law is only manifested when we first take the system size to infinity.
This can be interpreted as meaning that, due to $t_n$ growing too fast, a finite system is not enough to probe the full thermalization time, making the effective bulk much smaller than the system size.

\subsubsection{Many-body inverted scars}

The above argument regarding many-body scars in the MBC-E phase can be carried over to the MBC-L phase as well.
That is, given a regular state $|S\rangle$ in MBC-L, we can construct a rare state $|S'\rangle$ by constraining an infinite number of pairs of non-resonant blocks to have symmetric configurations to make it thermal-like.
This condition similarly does not change the energy.
The more rigorous block-swapping construction of the rare states also mostly applies in this case, by swapping the $H^\text{nr}_n$ block for an infinite number of $n$ above $n_0$ with another block that is symmetric with $\widetilde{H^\text{nr}_n}$, with the difference that when showing that the energy density does not change by the shuffling, we need to use the fact that the original energy density difference between the pair is expected to go to zero for a generic choice of $|S\rangle$.
Under the closed system interpretation, this implies that there exists a set of energy levels of a vanishing fraction, which is dense in the spectrum of MBC-L and ETH-like.
In other words, the quantum expectation values for such rare states converge to the ensemble average of all nearby eigenstates (not just the rare ones).
We interpret them as many-body inverted scars.
Note that although the construction of the inverted scar state begins with a region $R$, the fact that it is thermal-like does not depend on whether the observable is related to $R$, as any $R'$ eventually becomes in the same resonant region as $R$ for a large enough $n$.
In other words, being an inverted scar state (as well as a scar state) is a global property, rather than related to a particular region.

The existence of inverted scar states also implies that the MBC-L phase does not have a natural length scale as in usual MBL phases, as a generic state can locally resemble an inverted scar state.
In a usual MBL phase with a complete set of LIOMs, any given local region $R$ is at most involved in resonances with a finite number of sites, as determined by the LIOM supports.
However, in the MBC-L phase, for any level index $n$, there is a finite fraction of states in which $R$ is involved in the resonances within a whole $H^\text{res}_n$ block at level $n$, meaning that there is no upper bound on the length scale over which $R$ can be involved in a resonance as $n\rightarrow \infty$.
This occurs even if the system is in MBC-L at all nonzero temperatures.

The difference between MBC-L and the usual MBL phases can be detected by other diagnostics as well.
In MBC-L, although the correlation functions between observables $O_i$ and $O_j$ decay to zero when $|i-j|\to\infty$ on average, one can select rare states and observables such that $i$ and $j$ are a pair of mirror sites for an arbitrarily large $n_0$ with the state chosen to be resonant at that level, which then results in an $O(1)$ correlation.
For the entanglement entropy, there is no uniform upper bound on the area law constant for a fixed cut in MBC-L.

Another interesting property of the MBC-L phase is that, although there are approximate LIOM with non-thermal expectation values, there does not exist any true LIOM in the system in the conventional sense of commuting with the Hamiltonian, being quasilocal, and having discrete eigenvalues.
To see this, suppose $O$ is a LIOM that has two closest eigenvalues $\lambda$ and $\lambda'$, and a primary part $O_1$ supported in a region $R$ (with $\|O-O_1\|\ll|\lambda-\lambda'|$).
One can then construct an energy eigenstate with a mirror resonance that superposes the local states in $R$ with eigenvalues $\lambda$ and $\lambda'$; thus, the resulting expectation value $\langle O\rangle$ under such an eigenstate is between $\lambda$ and $\lambda'$, contradicting the assumption that $O$ is a LIOM.

On the other hand, the $\tau^z_{j,n}$ operators in the $n\to\infty$ limit behave roughly like a complete set of ``LIOMs'' in an infinite-temperature MBC-L phase, in the sense that they are true integrals of motion but with an unconventional notion of locality.
To compare this with the usual definition of LIOMs by a Pauli expansion, we first write $U_n$ as
\begin{equation}
    U_n=(1-M_n)+R_nM_n+\mathcal{E}_n 
\end{equation}
where $M_n$ is the orthogonal projector onto the $A=\widetilde{D}$ subspace in the notation of Eq.~(\ref{eq:Unaction}) and $R_n$ is a unitary operator that takes $|BC\rangle_n$ to a superposition according to the rule in Eq.~(\ref{eq:Unaction}).
Now for each $j$, for a large enough $n$ such that $\tau^z_{j,n}$ is supported in a resonant region, by conjugating with $U_n$ we have\footnote{Note that $M_n=p_n+\sum_Q\prod_{j\in Q}\sigma^z_j\sigma^z_{\tilde{j}}$.}
\begin{multline}\label{eq:UnConj}
    \tau^z_{j,n+1}=(1-p_n)\tau^z_{j,n}+p_n R^\dagger \tau^z_{j,n}R\\+(R^\dagger \tau^z_{j,n}R-\tau^z_{j,n})\sum_Q\prod_{j\in Q}\sigma^z_j\sigma^z_{\tilde{j}}+O(\epsilon_n)
\end{multline}
where $p_n=2^{-L^\text{nr}_n}$ is the infinite-temperature resonance probability and $Q$ runs over all nonempty subsets of sites in the $H^\text{res}_n$ block.
Iterating to $n\to\infty$, we obtain a formal Pauli expansion of $\tau^z_{j,\infty}$, which is a formal integral of motion of $H_\infty$:
\begin{equation}\label{eq:tauInf}
    \tau^z_{j,\infty}=\sum_{a\leq j\leq b} O_{a,b}
\end{equation}
where each $O_{a,b}$ is a weighted sum of Pauli ($\sigma$) strings, each with the convex hull of support being $[a,b]$ (i.e.\ acts nontrivially on $a$ and $b$ and acts trivially outside $[a,b]$).
Note that by the first line of (\ref{eq:UnConj}) and the convergence of $\prod_n(1-p_n)>0$ in the MBC-L phase, the coefficients of such local Pauli strings converge to generally nonzero values in the $n\to\infty$ limit, making the expansion well-defined.
However, instead of having $\|O_{a,b}\|$ decaying with $|a-b|$ (which is the case for LIOMs in  MBL~\cite{Serbyn2013,Ros2015,Chandran2015,Imbrie2017}), in the MBC-L phase, there is always a finite fraction of states with $O(1)$ eigenvalues, no matter how large $|a-b|$ is, preventing $\|O_{a,b}\|$ from decaying and excluding $\tau^z_{j,\infty}$ from being LIOMs.
These rare states are exactly the ones that locally resemble an inverted scar state in $[a,b]$ (i.e.\ with $M_n$ eigenvalue 1 for the largest $n$ in $[a,b]$), whose fraction goes to zero as $|a-b|\to\infty$.
Intuitively, instead of the nonlocal terms in the expansion (\ref{eq:tauInf}) being smaller and smaller, they are instead more and more ``inactive''. That is, they are relevant only when rarer and rarer long-range inverted scarring happens.
We interpret this as a non-conventional notion of locality (a related concept of ``small-plus-large'' operators is used in Ref.~\cite{Yang2026}).
It will be interesting to see how the common analytical results of MBL based on the assumption of having a complete set of LIOMs change if we modify the notion of locality in this way, but a concrete formalism is out of the scope of this paper.

\section{Consequences for several studied models}\label{sec:nonideal}

In this section, we go beyond the solvable limit and discuss the possible consequences of our findings for three previously studied quasiperiodic models.
Two examples will be the many-body consequences in the single-particle critical phases: the Fibonacci quasicrystal and the extended Aubry-Andr\'e-Harper (EAAH) model.
The third example will be the consequence in the localized (non-critical) phase of the interacting Aubry-Andr\'e (AA) model, due to the existence of rare mirror regions.
Since these models are not in the solvable limit, the results in this section will not be rigorous, but will rather be a combination of the heuristic extension from the constructed solvable limit and some speculation on the interplay between structural and accidental resonances.

The mirror centers for a general model are often emergent from the underlying quasiperiodic structure, and these emergent mirror centers may not have a clean hierarchical structure like those in the constructed solvable model, and they may also mix with other structures (such as repetition without mirroring).
Therefore, our rigorous constructions usually cannot be directly applied to the general models.

Instead, we study a particular site $j_0$ or region $R_0$ in the chain and ask whether it is participating in (hierarchical) mirror resonances, ignoring that they may also participate in other kinds of resonances.
In other words, we regard the HMS as an effective description of some realistic quasiperiodic models.
The mirror centers of these resonances are marked as $x_n$ (which can either be locations of sites or bonds), so that $j_0$ first participates in the resonance around $x_1$ at a short timescale, and then the one around $x_2$ at a longer timescale, and so on.
For each level index $n$, we recursively define the region $R_n$ to be the union of $R_{n-1}$ and its mirror image $\widetilde{R_{n-1}}$ around $x_n$.
Then one can generalize the resonance probability $p_n$ discussed in Sec.~\ref{sec:MBSolvable} as the probability that $R_n$ is contained in a many-body mirror resonance around $x_n$, thus generalizing the MBC-E and MBC-L phases (with the caveat that the resonances at different $n$ may not be independent and that the region may involve an infinite number of non-mirror resonances as well).

\begin{figure*}
    \centering
    \includegraphics[scale=1]{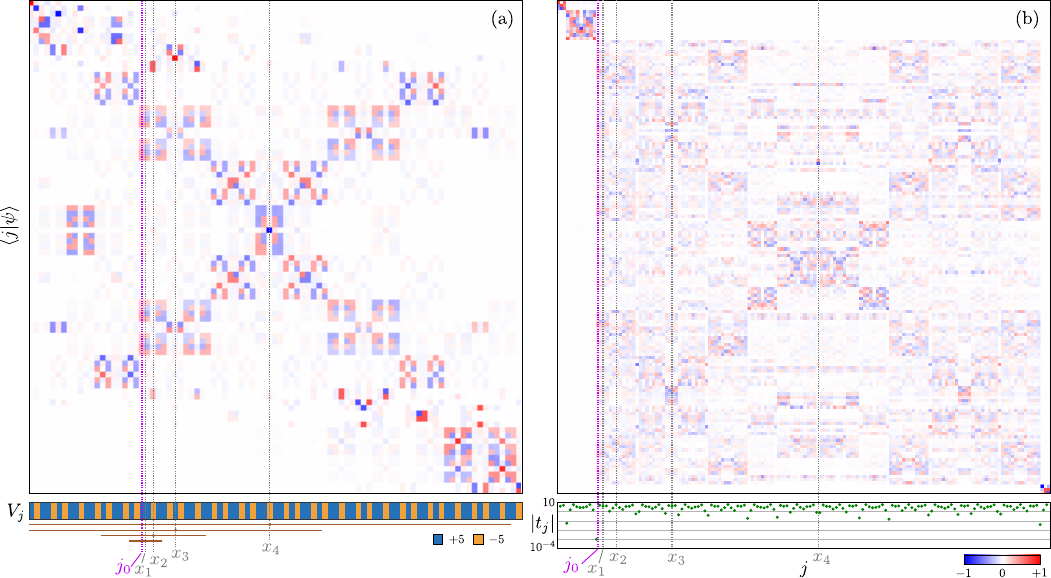}
    \caption{\justifying Demonstration of the single-particle HMS in (a) the Fibonacci and (b) the EAAH models. The heatmaps show the eigenstates of the models, sorted to mimic Fig.~\ref{fig:eigenstates}. The site $j_0$ participates in the mirror resonances centered at $x_1$ to $x_4$ hierarchically. Brown bars in (a) mark the palindromes centered at $x_n$ in the Fibonacci word. The lower panel in (b) shows the spatial profile of the absolute values of the hoppings $|t_j|$. The parameters are (a) $m=10$ ($L=89$), $W=5$; (b) $L=150$, $\alpha=\frac{\sqrt{5}-1}{2}$, $\mu=5$, $V=1$, $\phi=3.4975648$. }
    \label{fig:fib_eaah}
\end{figure*}

\subsection{Fibonacci quasicrystal}\label{sec:fibonacci}

The Fibonacci quasicrystal~\cite{Jagannathan2021} can be constructed from the \emph{Fibonacci word} $C_m$, which is a string consisting of two letters $\mathrm{A}$ and $\mathrm{B}$, constructed as follows
\begin{equation}
\begin{aligned}
    C_0&=\mathrm{B}\\
    C_1&=\mathrm{A}\\
    C_m&=C_{m-1}C_{m-2}\text{ for }m\geq2
\end{aligned}
\end{equation}
where $C_{m-1}C_{m-2}$ denotes concatenation. So the next few terms are:
\begin{equation}
\begin{aligned}
    C_2&=\mathrm{AB}\\
    C_3&=\mathrm{ABA}\\
    C_4&=\mathrm{ABAAB}\\
    C_5&=\mathrm{ABAABABA}
\end{aligned}
\end{equation}
Note that the length of $C_m$ is the $m$th Fibonacci number $F_m$.
The single-particle 1D Fibonacci quasicrystal with length $F_m$ is then defined by the following hopping and potential, in the notation of Eq.~(\ref{eq:SPHinf}) with open boundary conditions:
\begin{equation}
    t_j=1,\quad V_j=\begin{cases}
        +W&\text{if }(C_m)_j=\mathrm{A}\\
        -W&\text{if }(C_m)_j=\mathrm{B}
    \end{cases}
\end{equation}
where $(C_m)_j$ denotes the $j$th letter in the string $C_m$.
The infinite Fibonacci quasicrystal is constructed from the finite ones using a procedure similar to that in Sec.~\ref{sec:SPinf}, whose spectrum has been shown to be singular continuous for all $W\neq 0$~\cite{Kohmoto1983,Suto1987,sutHo1989singular,Damanik2016,Jagannathan2021}.

\subsubsection{Origin of the HMS}

The HMS in the single-particle Fibonacci model is demonstrated in Fig.~\ref{fig:fib_eaah}(a).
Such mirror resonances originate from the fact that the Fibonacci words contain long \emph{palindromes}, which are substrings that are identical to their reversal.
Indeed, one can show by an inductive argument that every $C_m,m\geq3$ with the last two letters removed is a palindrome, which we will call $P_m$ (see, e.g., Ref.~\cite{Fici2015}).
As $C_m$ for large $m$ itself contains many copies of $C_{m'}$ for $m'<m$, one can see that half of the palindrome itself contains many smaller palindromes, and so on.
Conversely, on an infinite Fibonacci chain, every $C_m$ itself is embedded everywhere within some $C_{m'}$ with arbitrarily large $m'>m$, thus forming an HMS to infinity.

There is some difference between the HMS of the Fibonacci model and our constructed solvable model.
Firstly, the would-be ``non-resonant regions'' themselves contain many mirror resonances, as demonstrated by the multiple peaks in the region to the left of $j_0$ in Fig.~\ref{fig:fib_eaah}(a), as opposed to Fig.~\ref{fig:eigenstates}, in which each non-resonant particle only has one peak in the non-resonant region. 
Secondly, there are non-mirror structural resonances in this model, as demonstrated by the states near the center of Fig.~\ref{fig:fib_eaah}(a), which have a unit of four peaks around the site $x_4$ and a similar unit of four additional peaks near the left edge of the chain, due to a more complicated repetitive structure of the Fibonacci words.

We remark that an alternative labeling of the sites, called the \emph{conumber}, has been known to give a cleaner description of the eigenstates in the Fibonacci quasicrystal~\cite{Ashraff1988,Mace2016,Jagannathan2021}.
However, it is not obvious whether it can be carried over to our many-body model, as the interaction is local in real space, not in conumber space.

\subsubsection{Many-body consequence}

To study the many-body consequences, we need to estimate the resonance probability $p_n$.
Without a clear non-resonant region, it cannot be estimated from some entropy of ``$L^\text{nr}_n$'' as in the constructed solvable model.
Instead, we note that since this model has no weak bonds, a better approximation is to treat the many-body theory on the palindrome $P_m$ as a mirror-symmetric random MBL studied in Ref.~\cite{Li2025}, where a generalization to include the freezing/protection effects is presented in Appendix~\ref{sec:symmetricRandom}.
We will give an intuition on how the scaling should be as follows.

Roughly speaking, a many-body resonance (or synchronized oscillation) has a level splitting (or oscillation frequency) $\omega_\text{sync}$ which is exponentially small in the physical length of the resonance, that is, the size $|R_n|$ of the region $R_n$ (in some situations it can be smaller than exponential).
For this resonance to be protected, the energy difference between the two oscillating many-body configurations must also be exponentially small.
Unlike the constructed solvable model that uses small error terms in the non-resonant region, in the Fibonacci model, since each site has a potential being either $+W$ or $-W$, any symmetry-breaking term must be of a strength $2W$.
We will look at how this affects the protection condition around a palindrome $P_m$.
Inside a palindrome $P_m$, there is exactly no symmetry-breaking term; outside of it, in general, $P_m$ can be symmetrically extended to a longer palindrome $P'_m$, but the pair of sites immediately beyond $P'_m$ must be different, and thus $2W$ symmetry-breaking.
In the analysis below, we call this pair of symmetry-breaking sites $a$ and $\widetilde{a}$, the range of mirror resonance from $b$ to $\widetilde{b}$, and the mirror center to be $x$ (so that we have $a<b<x<\widetilde{b}<\widetilde{a}$ and $R_n=\{\widetilde{b},\ldots,b\}$).

Two conditions are required for the many-body resonance to be protected.
First, as we can think of $a$ and $\widetilde{a}$ as introducing a symmetry-breaking tail, in order for the tail to be small enough at $b$ and $\widetilde{b}$, we roughly require $b-a\gtrsim\widetilde{b}-b$.
Secondly, if there is a site $a'$ between $a$ and $b$ that has a different approximate LIOM configuration from $\widetilde{a'}$, it similarly introduces a symmetry-breaking tail that can affect $b$ and $\widetilde{b}$.
Hence, we roughly require an $O(\widetilde{b}-b)$ number of sites within $[a,b]$ to have symmetric configurations with their mirror images.
The first condition can always be fulfilled, as $R_{n-1}$ must be contained near the center of $F_m$ with some large enough $m$ on the infinite chain, and therefore near the center of a long palindrome $P'_m$ whose center can be chosen as $x_n$.
For the second condition, as it requires an $O(|R_n|)$ number of approximate LIOMs to be symmetric, it gives a $p_n$ that is exponentially small in $|R_n|$, which itself grows exponentially with $n$.

The above argument strongly suggests that the interacting Fibonacci chain is deep in the MBC-L phase if the potential strength is large enough.
Moreover, it implies that a system size exponential in $n$ is required to see the physical effect of $n$ levels in the hierarchy and that the many-body inverted scars are exponentially rare in the many-body spectrum.
This result is consistent with the small-size numerical observations in Refs.~\cite{mace2019manybody,varma2019diffusive} that a strongly disordered interacting Fibonacci model behaves mostly like MBL but with rare delocalized dynamics.

Of course, there remains the possibility that accidental resonances always dominate at large $n$, driving the system into the ETH phase in the thermodynamic limit.
Ruling out this possibility is at least as difficult as solving the problem of the thermodynamic stability of random MBL, and is therefore out of the scope of this paper.

\subsection{Extended Aubry-Andr\'e-Harper model}

The single-particle extended Aubry-Andr\'e-Harper (EAAH) model is defined by the following hopping and potential, in the notation of Eq.~(\ref{eq:SPHinf}):
\begin{equation}
    \begin{aligned}
    t_j&=1+\mu\cos\left[2\pi\alpha\left(j+\frac{1}{2}\right)+\phi\right]\\
    V_j&=V\cos(2\pi\alpha j+\phi),
    \end{aligned}
\end{equation}
where $\alpha$ is an irrational number (commonly chosen to be the golden ratio) and $\mu,V>0$.
The model is shown to be in the extended phase for $\mu<1, V<2$, the localized phase when $V>2, 2\mu<V$, and the critical phase when $\mu>1, V<2\mu$~\cite{Hatsugai1990,Han1994,Takada2004,Liu2015,Avila2017,wang2021manybody}.

\subsubsection{Origin of the HMS}

In this subsection, we will give an intuitive argument on how the HMS emerges from the EAAH model. 
As we show in the constructed solvable model, the HMS gives rise to singular continuity (or critical phase) in the single-particle case. 
Assume that the HMS dominates the critical phase of the EAAH model, and consider an extreme situation where $\mu \gg V$ in the critical phase; we would conclude that the HMS of the EAAH model should be dominated by the hopping, and we would ignore the effects of the potential $V_j$ for simplicity. The intuition is as follows.

First, due to the irrationality of $\alpha$, there will always be a set of approximate mirror-symmetric points of $t_j$, which is shown as $x_i$ in Fig.~\ref{fig:fib_eaah}(b). 
The resonant block is achieved by these points. 
To achieve the non-resonant blocks, note that when $\mu>0$, there will be arbitrarily weak bonds $|t_j|$ in an infinite chain, again due to the irrationality of $\alpha$.
Combining the approximate symmetry point and the weak bonds together, we consider the tunneling of a particle in a region $R$ from one side to the other across an approximate symmetric point.
The particles will go through several weak bonds (pairs). 
Since the symmetric point is not exact, once the detuning of it is larger than the multiplication of the weak bonds (assuming that the multiplication of the bonds in between the pair of the mirror sites is much larger than the weak bond), the particle cannot successfully tunnel to the other end. 
Thus, $R$ can be considered a non-resonant region. The existence of the non-resonant region is shown in Fig.~\ref{fig:fib_eaah}(b), where the eigenstate shows mirror resonance associated with the approximate mirror center $x_4$ and is almost ``cut off'' at the weak bonds close to both ends of the chain.
For smaller levels (e.g., levels related to $x_1,x_2$ and $x_3$), as the ``weak bonds'' are less weak, the resonant and non-resonant regions become less distinctive.
In the end, the HMS should emerge from iteratively finding the weak bonds and the approximate symmetry points.

As in the Fibonacci model, the ``non-resonant regions'' themselves contain mirror resonances.
On the other hand, it is more similar to our solvable model in that every level in the HMS contains at least a pair of weak bonds away from the mirror center.

\subsubsection{Many-body consequence}
A major difference of the EAAH model in the critical phase compared to our constructed solvable model is that 
we cannot assume the ``non-resonant regions" of the EAAH model to be many-body localized in the interacting case. 
Instead, the EAAH model in the critical phase should be roughly viewed as a series of ergodic blocks linked by weak bonds, since the potential is not strong enough for localization compared to the hopping inside one block.
In such a case, the many-body resonance gaps $\omega_\text{sync}$ in the resonant regions may not be that small, and the influence of the non-resonant regions may be significantly weaker compared to systems with a background MBL phase, loosening the conditions of the protections.

However, estimating the growth rates of $\omega_\text{sync}$ (and thus the protection conditions) is extremely difficult.
At small sizes, due to the lack of multiple weak bonds, the system essentially behaves like an extended phase occasionally cut off by a few isolated weak bonds.
Even the $L=150$ chain in Fig.~\ref{fig:fib_eaah}(b) is only barely enough to demonstrate the hierarchical cut-off strengths of the weak bonds.
Therefore, we do not attempt to estimate $\omega_\text{sync}$ in the interacting many-body case.

Although we cannot make a strong conclusion here, the above argument suggests that the interacting EAAH model, or possibly some variants of it, may potentially be in the MBC-E phase, or even have an MBME.
This model being in MBC-E is consistent with the small-size numerical findings in Ref.~\cite{wang2021manybody} (in which the term ``MBC'' first appears in the literature), as it aligns with the small-size picture of an ETH phase being occasionally cut off by isolated weak bonds.
However, we believe that the non-thermal extended phenomenology observed in Ref.~\cite{wang2021manybody} may be more directly explained by the prethermal model in Ref.~\cite{Tu2024}, in which it is called a non-ergodic extended (NEE) regime.
Specifically, since the interaction strength in Ref.~\cite{wang2021manybody} is a constant across each bond, the information spreading timescale for a small-size system is expected to be much less suppressed than the particle spreading timescale, the latter being strongly suppressed by the occasional weak hopping terms.
This separation of timescales resembles the one discussed in Ref.~\cite{Tu2024} (although from a different mechanism), which is proposed to be the underlying cause of the NEE phenomenology in small-size numerics.

We remark that having arbitrarily weak bonds is not the only way to avoid the difficult conditions in the protection of many-body resonances.
Models with parent flat bands, such as Ref.~\cite{Kumar2026}, can achieve an effective model with arbitrarily weak bonds by having the hopping amplitude almost canceled by symmetry (across different bands).
Therefore, we propose that multi-flat-band models with quasiperiodic modulation may also lead to MBC-E phases and MBMEs.

\subsection{Aubry-Andr\'e model: rare region effects}\label{sec:rare}

The single-particle Aubry-Andr\'e (AA) model is defined by the following hopping and potential, in the notation of Eq.~(\ref{eq:SPHinf}):
\begin{equation}
    t_j=\frac{1}{2},\quad
    V_j=W\cos(2\pi\alpha j+\phi),
\end{equation}
where $\alpha$ is an irrational number (commonly chosen to be the golden ratio) and $W>0$ is called the disorder strength.
The model is shown to be in the localized phase for $W>1$ and the extended phase when $W<1$~\cite{Aubry1980, Harper1955}.
Although it does not have a critical phase in a wide range of $W$\footnote{
Although there is a critical point at $W=1$, it does not hold much interest in our context. As the phase boundary is shifted by the interaction, it is unclear whether we still have a stable HMS at the ETH-MBL boundary.
}, in the localized phase, there is a set of fine-tuned points of $\alpha$, or a generic $\alpha$ but with a measure-zero set of fine-tuned $\phi$, at which the spectrum becomes purely singular continuous~\cite{gordon1976point,Avron1982SingularCS,Jitomirskaya1994,Jitomirskaya2021}.
For the latter case, we call such a fine-tuned $\phi$ a \emph{rare initial phase}, which plays a central role in this subsection.

Although a rare initial phase itself is fine-tuned, its existence implies the occurrence of delocalized \emph{rare regions} in a generic, non-fine-tuned $\phi$ in a long enough chain.
Roughly speaking, an infinite AA chain with a fixed initial phase $\phi$ can locally look like one with any other $\phi'$, since translating $j$ is equivalent to rotating $\phi$ by an irrational angle, which can bring it arbitrarily close to any value in $[0,2\pi)$.
That is, in an infinite AA chain with any initial phase, there exist arbitrarily long regions within which all eigenstates resemble critical states.
This is similar to the existence of arbitrarily long low-disorder rare regions in the Anderson model~\cite{Anderson1958}.

In the Anderson case, even if such regions are extremely rare, in the presence of interactions, they lead to profound effects on the thermodynamic stability of MBL, which is called the avalanche instability~\cite{thiery2018many,morningstar2022avalanches, sels2022bath}.
Roughly speaking, the low-disorder region first thermalizes with interactions, and then it acts as a thermal bath to continue thermalizing nearby spins.
In certain situations, this process may continue indefinitely, leading to the thermalization of the entire chain and the thermodynamic instability of MBL.

The avalanche instability of the interacting AA model has been numerically studied in Ref.~\cite{tu2023avalanche}.
The result suggests that, although a small-size AA model appears MBL for $W\gtrsim 2.0$, if a local thermal seed exists in an infinite AA chain with $W\lesssim 7.5$, an avalanche will occur and destroy the MBL.
Although Ref.~\cite{tu2023avalanche} also concludes that there are no asymptotically long thermal regions in the AA model, the result comes from a renormalization group method that does not take into account structural resonances and thus may be inaccurate for the AA model according to our current understanding.
Hence, whether a local thermal seed exists for $2.0\lesssim W\lesssim 7.5$ to initiate the avalanche remains unknown.

Below, we argue that, due to the existence of rare initial phases, the interacting AA model does contain arbitrarily large delocalized rare regions at the eigenstate level, in which local expectation values are thermal.
However, as different energy eigenstates have such types of rare regions in different places, it is not clear whether they can really lead to avalanche instability.
We also discuss the possibility of rare regions that are common to all eigenstates due to the interplay between structural resonances and accidental resonances.
Although the existence of the latter types of rare regions is more speculative, their existence would very likely imply avalanche instability.
Either way, our results question the common belief that quasiperiodic MBL does not have avalanche instability due to the lack of low-disorder rare regions.

We remark that having an avalanche from a rare region in an AA chain is not the only possibility that may destroy the thermodynamic MBL of the AA model.
It may very well be the case that the ETH-MBL crossover drifts smoothly as we increase system size due to resonances from ``common'' regions that happen everywhere as the system size grows (as opposed to rare regions that are unlikely to appear until a very large size).

\subsubsection{Rare region due to a single mirror center}

Before we make the main argument about the rare region due to HMS, we first discuss whether an avalanche may be initiated by a much simpler mechanism---due to a single approximate mirror center. If so, it would be meaningless to discuss the much more complicated rare region due to HMS.

Ref.~\cite{Li2025} shows that a single exact mirror center is enough to lead to structural many-body resonances, and an extension of the theory to the symmetric-broken case (see Appendix~\ref{sec:symmetricRandom}) shows that an approximate one can also lead to many-body resonances with a finite probability.
Since an infinitely long AA chain contains an infinite number of arbitrarily good approximate mirror centers, this implies the existence of arbitrarily long many-body resonances around single mirror centers (with the caveat regarding accidental resonances).

However, mirror-type resonances are unlikely to induce avalanches, as information is only mixed between two parts of the system related by mirroring, rather than spreading out.
There remains the possibility that the interplay between such mirror resonances and accidental resonances may thermalize the region near the mirror center, for which we are not able to give a definite answer.
But heuristically, if we think of accidental resonances as a graph in the spin configuration space, it seems unlikely that having a set of additional two-point connections between mirrored configurations can help much with the proliferation of multiple-point connections due to accidental resonances.

A related scenario is the \emph{anti-mirror symmetry} MBL studied in Ref.~\cite{Kloss2023}.
An anti-mirror center is similar to a mirror-symmetry one, except that the potential is flipped.
That is, $V_{i}\approx-V_{x+i}$ as opposed to a mirror symmetry $V_{i}\approx V_{x+i}$.
Ref.~\cite{Kloss2023} shows that in the presence of \emph{exact} anti-mirror symmetry at half filling, spin-spin correlators will show correlations between all pairs of sites, not just the mirror pairs, which they interpret as some form of delocalization.
Since approximate anti-mirror centers exist in the AA model as well, this raises the question of whether such ``delocalization'' may lead to thermalization.
However, we find that their result relies fundamentally on closed systems at exact half-filling.
That is, the correlation ultimately arises from the fact that if we randomly draw a spin configuration at half filling, there is an $O(1/L)$ correlation between any two sites.
Since there is no such exact-half-filling condition locally in an infinite system, their result does not apply here.

\subsubsection{Rare regions due to HMS}\label{sec:rareHMS}

Even if a thermal region is unlikely to emerge from a single mirror (or anti-mirror) center, having a hierarchy of mirror centers together changes the story.

The original proof of the existence of a rare initial phase in the single-particle AA model in Ref.~\cite{Jitomirskaya1994} is exactly based on the construction of what we call HMS (and is, to the best of our knowledge, the first time that the idea of HMS appears in the context of critical states).
In our language, the intuition behind the construction is that given any site $j_0$, one can adjust the initial phase $\phi$ step by step so that a sequence of hierarchical mirror centers $x_1,x_2,\ldots$ appears.
Specifically, at the $n$th step, we have an initial phase $\phi_n$ that produces the mirror centers $x_1,\ldots,x_n$, with an ``error tolerance'' $\Delta\phi_n$ such that within $[\phi_n-\Delta\phi_n,\phi_n+\Delta\phi_n]$ those mirror centers successfully create resonances between $j_0$ and its mirror pairs.
Then, since there are always arbitrarily good approximate mirror centers over the entire chain, one can choose one of them to be $x_{n+1}$, and slightly tune the initial phase with a new tolerance such that $[\phi_{n+1}-\Delta\phi_{n+1},\phi_{n+1}+\Delta\phi_{n+1}]\subseteq[\phi_n-\Delta\phi_n,\phi_n+\Delta\phi_n]$
to make $x_{n+1}$ a mirror center supporting the resonances between $j_0$ and $2x_{n+1} - j_0$, so that the new resonant region covers that from $x_n$, thus continuing the hierarchy. 

Now, such a construction of rare initial phases relies only on the fact that (1) there are arbitrarily good single mirror centers in the chain, and (2) the length of mirror resonances can be made arbitrarily long as the mirror center becomes better and better.
Hence, there is a direct generalization to the many-body case (a detailed construction is presented in Appendix~\ref{sec:rareConstruction}).
However, there is a caveat regarding accidental resonances.
While we have been using arbitrarily weak bonds to avoid accidental resonance in our constructed solvable model, in the AA model, there are no weak bonds. Especially in the moderate disorder regime $2.0\lesssim W\lesssim 7.5$ of the AA model susceptible to an avalanche, accidental resonances are expected to play an important role.
One way to mitigate the effect of accidental resonances is to have a large spatial separation between different levels in the hierarchy, which is, in some sense, similar to having weak bonds due to the relevant approximate LIOMs only coupled far away via their exponential tails.
In our context of constructing a rare initial phase, this corresponds to making a choice of $x_{n+1}$ that avoids the resonant region $R_n$ being too close to it, as well as being too close to the boundary of the resonant region related to $x_{n+1}$.

Since the interacting AA model does not have weak bonds, the rare initial phases are expected to produce an MBC-L behavior, by a similar argument as in the discussion of the Fibonacci model.
This implies the existence of arbitrarily long MBC-L-like rare regions for an arbitrary initial phase.
Although an MBC-L-like rare region behaves like MBL on average, since there are an infinite number of them, there are almost certainly an infinite number of inverted scar states.
This implies that a generic energy eigenstate of the interacting AA model, no matter how large $W$ is, contains arbitrarily long rare regions within which the expectation values of local observables are thermal-like.
However, as the eigenstates inside a rare region have different energies, and since the inverted scar states are rare, it is not clear whether such types of rare regions are relevant to avalanche instability.
Note that the numerical finding in Ref.~\cite{Faulend2026} can be viewed as a very-small-sized version of such a rare region.

Another type of HMS rare region comes from the rare initial phases, which are deliberately constructed to be unstable and are much more likely to lead to an avalanche (if they exist).
Recall that in the construction of a rare initial phase, we actually have some freedom in choosing the mirror centers $x_{n+1}$.
Therefore, we may deliberately choose them to be very close to one end of $R_n$, or let the boundary of the resonant region related to $x_{n+1}$ be close to the other end of $R_n$, or both (see Appendix~\ref{sec:rareConstruction} for more details), so that the HMS becomes unstable. That is, accidental resonances dominate structural resonances after some levels.
Although we cannot make a concrete conclusion, it is likely that once configurations not related by mirror symmetry start to resonate due to inadequate boundaries between different levels in the hierarchy, energy levels in the rare regions will start to randomly repel each other, leading to thermalization (of all eigenstates) within the rare region, and therefore the possibility of the onset of an avalanche for $2.0\lesssim W\lesssim 7.5$.
Note that this is different from the single-mirror-center scenario, where structural resonances are always between two configurations; here, structural resonances can involve up to $2^n$ configurations at level $n$, making them more powerful in connecting some otherwise disconnected configuration graphs.

\section{Conclusion}\label{sec:conclusion}

We have constructed asymptotically solvable models for both single-particle and many-body critical phases based on the hierarchical mirror structure, a simplification of the real-space structure commonly found in single-particle critical phases, and presented a rigorously controlled perturbation theory.
For the single-particle case, we have proved the singular continuity of the model based on the Cantor set structure of the energy splitting due to the perturbations.
For the many-body case, we have shown that the direct generalization of the singular continuity mechanism leads to the MBC-E phase, while the interaction-induced collective freezing effect leads to a new MBC-L phase that has no non-interacting counterpart.
Although the two phases resemble the familiar ETH and MBL phases, there are rare states that exhibit opposite behaviors compared to the phases in which they are embedded, interpreted as many-body scars and inverted scars, respectively.
Moreover, the two phases can be separated by a finite-$T$ transition, interpreted as an MBME.
These results show that the many-body counterpart of the singular continuous spectrum, commonly found in certain types of quasiperiodic models, has different possibilities with rich behaviors due to interactions.

Based on our findings, we revisited the previous numerical results on the effects of interaction in single-particle critical phases.
In particular, we propose that the MBL-like behavior with rare dynamics in the Fibonacci quasicrystal~\cite{mace2019manybody,varma2019diffusive} can be explained by it being in the MBC-L phase, and the extended-but-nonthermal behavior in the EAAH model~\cite{wang2021manybody} can be explained as a small-size manifestation of the MBC-E phase.
We also propose that some EAAH-type quasiperiodic models may have MBMEs similar to our solvable model.
We also revisited the problem of avalanche instability of AA-type quasiperiodic models and proposed that an avalanche may be caused by HMS-type rare regions that locally resemble MBC-L chains, possibly due to an interplay between structural and accidental resonances.
Such rare regions can be viewed as a generalization of the numerical finding in Ref.~\cite{Faulend2026}.

Although we only study the HMS in 1D chains, we expect direct generalizations to a more general type of approximate symmetry structure in higher-dimensional quasiperiodic systems (or quasicrystals).
Note that in the higher dimensional projection picture of quasiperiodicity~\cite{Duneau1985,Jagannathan2021}, the mirror center in the quasiperiodic chains studied in Sec.~\ref{sec:nonideal} can be described as the 1D cut going through a $C_{2v}$ symmetric point in the parent 2D lattice, which has a direct generalization to higher dimensional quasicrystals from a more general class of point-group symmetry of the parent lattice.
One advantage of a higher dimensional hierarchical $m$-fold rotational structure over 1D HMS is that structural resonances in a higher level no longer need to go through a virtual path that visits every bottleneck (e.g.\ mirror centers) caused by lower levels; hence, we expect $\omega_\text{sync}$ to be larger and less prone to symmetry-breaking defects.
Beyond such crystal-like symmetries, there are also structural resonances in quasiperiodic systems caused by approximate repetition without mirroring, which is also known to result in a single-particle singular continuous spectrum~\cite{gordon1976point,Avron1982SingularCS,Jitomirskaya2021}, and we also expect a generalization in interacting many-body systems.
For all of the above generalizations, we expect that there may also be two types of corresponding many-body phases: one whose many-body structural resonances are a synchronized version of the single-particle counterpart (resembling MBC-E), and another whose collective freezing effect is too strong such that the structural resonances do not reach infinity except for rare states (resembling MBC-L).

For a quasiperiodic system in general, both structural and accidental resonances exist.
Hence, when interactions are added to a single-particle critical system, there is an additional possibility that the system becomes ETH rather than MBC-L or MBC-E, as considered in this paper, similar to how an Anderson-localized system at low disorder becomes ETH upon adding interaction.
Similarly, when interactions are added to a non-critical system with rare regions, there are possibilities beyond MBL being stable or being destroyed by an avalanche.
Indeed, the numerical results in Ref.~\cite{Tu2024} suggest that there is also the possibility that a quasiperiodic system thermalizes everywhere from accidental resonances alone.
Combining the above discussion, we propose the following possibilities for the fate in the thermodynamic limit for a general quasiperiodic system with slow dynamics from many-body resonances:
(1) If structural resonances dominate and proliferate everywhere, the system becomes either MBC-L or MBC-E.
(2) If accidental resonances dominate and proliferate everywhere, the system is driven into ETH similar to the prethermal regime of random MBL.
(3) If structural resonances proliferate only in rare regions, but foster the accidental resonances to dominate and spread, the system is driven into ETH similar to the avalanche scenario in random MBL.
(4) If neither structural nor accidental resonances proliferate, the system is thermodynamically MBL.

Beyond common quasiperiodic systems, one may also be interested in realizing an MBC system directly based on our constructed solvable model.
As our construction in Sec.~\ref{sec:MBSolvable} is intended to be an existence proof rather than optimizing for realizability, some care must be taken in choosing the types of coupling and the parameters to avoid $\omega_\text{sync}$ from being too small, either for numerical or experimental realizations.
We expect that the most promising type of system is probably the one with particle conservation and long-range coupling to move multiple particles at a time, at a low filling fraction, and that the MBC-L to MBC-E transition is probably most easily observed by tuning the chemical potential rather than the temperature.
Also, a higher-dimensional generalization, if realizable, is expected to be more stable against defects/errors from the discussion above.
For the system size issue in numerics due to the Hilbert space size being doubly exponential in $n$ (which happens in normal quasiperiodic systems as well), we propose that some tensor-network-based method with the block structure designed to match the underlying mirror structure may be able to simulate this type of model, with enough levels in the hierarchy to demonstrate the properties.

\acknowledgements

The authors thank David Long, \mbox{DinhDuy} Vu, Laura Shou, Dominic Else, and Tian-Hua Yang for useful discussions.
This work is supported by the Laboratory for Physical Sciences.

\appendix

\section{Freezing and protection in mirror-symmetric random MBL}\label{sec:symmetricRandom}

In this appendix, we extend the theory of mirror-type many-body resonances in Ref.~\cite{Li2025} by including the symmetry-breaking terms, which give a more accurate description of the freezing and protection conditions than the simplified one shown in Fig.~\ref{fig:intro}.

We mainly follow Ref.~\cite{Li2025} to use a 1D random-field XXZ chain that is mirror symmetric about a bond in the middle, with globally random symmetry-breaking perturbations.
But we will also discuss in the end what will change if we consider models with weak bonds, with non-uniform symmetry-breaking terms, without particle conservation, or with long-range coupling. %
We also discuss the applicability of this theory in the case of having multiple levels of HMS.

\subsection{The model}

We start from a random-field XXZ chain with a boundary at the right
\begin{equation}
    H_0 = \frac{1}{4}\sum_{j< c} \left(\sigma_j^x \sigma_{j+1}^x + \sigma_j^y \sigma_{j+1}^y + \Delta \sigma_j^z \sigma_{j+1}^z \right) + \frac{1}{2}\sum_{j\leq c} h_j \sigma_j^z, \\
\end{equation}
where the site index $j$ is an integer, $c$ is the right-most site, $\sigma^{x,y,z}_j$ are the Pauli operators at site $j$, and $h_j$ are independent uniform random numbers in $[-W,W]$.
Note that in our context, whether there is a left boundary of the chain does not really matter, and we will keep our notation compatible with either.
We assume the interaction strength $\Delta>0$ and the disorder strength $W>0$ are enough for $H_0$ to be MBL (either an infinite-size MBL, if it exists, or a finite-size MBL).
Then we make a mirror copy of this chain around the mirror center $c+\frac{1}{2}$ to become $\widetilde{H_0}$, where the tilde indicates replacing every action on site $j$ with that on the mirror site $\widetilde{j}=2c-j+1$.
Then the two chains are coupled by
\begin{equation}
    H_{\rm cp} = \frac{1}{4} \left(\sigma_{c}^x \sigma_{\widetilde{c}}^x + \sigma_{c}^y \sigma_{\widetilde{c}}^y + \Delta \sigma_{c}^z \sigma_{\widetilde{c}}^z\right), \\
\end{equation}
to become a mirror-symmetric random XXZ chain.
Finally, we add a small symmetric-breaking term throughout the chain
\begin{equation}
    H_\epsilon=\frac{\epsilon }{2}\sum_j h'_j \sigma_j^z,
\end{equation}
where $h'_j$ are again independent uniform random numbers in $[-W,W]$.
The full Hamiltonian is then
\begin{equation}
    H=H_0+\widetilde{H_0}+H_\text{cp}+H_\epsilon.
\end{equation}

To describe the mirror resonances of this system, we follow the approach of Ref.~\cite{Li2025} to define two effective chains.
The first is the $\tau$ chain consisting of the LIOMs of $H_0$ and $\widetilde{H_0}$~\cite{Serbyn2013,Ros2015,Chandran2015,Imbrie2017}:
\begin{equation}
    H_0=\sum_{i\leq c} h_i \tau^z_i
    +\sum_{i,j\leq c}J_{ij}\tau^z_i\tau^z_j
    +\sum_{i,j,k\leq c}J_{ijk}\tau^z_i\tau^z_j\tau^z_k+\cdots
\end{equation}
and $\tau^z_{\widetilde{j}}:=\widetilde{\tau^z_j}$.
The LIOMs $\tau^z_j$ have exponentially decaying tails in real space within the left/right side of the system according to the localization length $\xi$.
The coefficients $J_{ijk\cdots}$ also decay exponentially with the largest distance within $i,j,k\ldots$.

The second effective chain, the $\eta$ chain, is introduced to combine each symmetric pair $\tau^z_{j}$ and $\tau^z_{\widetilde{j}}$ of the $\tau$ chain into one site, in order to describe the Hilbert space of the possible mirror resonances of a $\tau$ configuration.
More specifically, consider a $\tau$ configuration $|S\rangle$ labeled by the $\tau^z_j$ eigenvalues $(-1)^{s_j},s_j=0,1$, all of the possible mirror resonant counterparts of the state (no matter whether the oscillations for different $j$ are synchronized or not) belong to the subspace that swaps some of the pairs $s_j,s_{\widetilde{j}}$.
We say that a site $j$ is \emph{active} if $s_j\neq s_{\widetilde{j}}$.
Then the set of configurations of possible mirror resonances from $|S\rangle$ can be described by an effective chain in which each site corresponds to a pair of active sites in the $\tau$ chain.
We define the Pauli operators of the effective chain by $\eta^{x,y,z}_j$, with site labels $j$ matching the active site labels in $H_L$.
An eigenvalue $\pm1$ of $\eta^z_j$ corresponds to $s_j=-s_{\widetilde{j}}=\pm1$.

The task of describing the many-body mirror resonances of $H$ is then reduced to finding an effective Hamiltonian $H_\text{eff}$ in the $\eta$ chain (from the effect of coupling and symmetry breaking), and to finding the integrals of motion in the $\eta$ chain itself.
Given an initial configuration $|S\rangle$, since the set of integrals of motion (IOMs) in the $\eta$ chain describes the pattern of mirror oscillations as time evolves, we can deduce the form of many-body resonances in the original $\sigma$ chain.

\subsection{Exact mirror symmetry}

Here we briefly review the results of Ref.~\cite{Li2025} in the exactly symmetric ($\epsilon=0$) case.
In the weakly interacting limit ($\Delta\ll 1$), the LIOMs $\tau^z_j$ are approximately the occupation numbers of single-particle localized orbitals, from which the effective Hamiltonian of the $\eta$ chain is derived:
\begin{equation}
    H_\text{eff}=\sum_{\langle ij\rangle}J^\text{eff}_{ij}\eta_i^z\eta_j^z+\sum_j h^\text{eff}_j\eta_j^x
\end{equation}    
where the coefficients scale as
\begin{equation}
J^\text{eff}_{ij}\sim \Delta e^{-|i-j|/\xi},\quad h^\text{eff}_j\sim e^{-2(c-j)/\xi}.
\end{equation}
This is a transverse field Ising model that is in the paramagnetic (ferromagnetic) phase near (far from) the mirror center.
That is, the IOMs in the $\eta$ chain are approximately
\begin{equation}\label{eq:etaIOM}
    \{\eta^x_j\}_{j\geq j_0},\quad \{\eta^z_i\eta^z_j\}_{\langle ij\rangle,j<j_0},\quad \prod_{j<j_0}\eta^x_j,
\end{equation}
with $j_0$ separating the paramagnetic and ferromagnetic phases.
Note that the IOMs are not all local, as the last one is a string of Pauli operators.
This set of IOMs indicates that the mirror oscillations between $j_0$ and $\widetilde{j_0}$ are unsynchronized, while those outside this interval are synchronized.
Moreover, with a fixed $O(1)$ ratio of active sites near $j_0$, the width of this unsynchronized region scales as
\begin{equation}
    \widetilde{j_0}-j_0\sim-\xi\ln\Delta,
\end{equation}
and the frequency of the synchronized oscillation scales as
\begin{equation}\label{eq:omega_sync}
    \omega_\text{sync}\sim \frac{1}{\Delta^{|A|}}\prod_{j\in A}e^{-2(c-j)/\xi}
\end{equation}
where $A$ is the set of active sites to the left of $j_0$ and $|A|$ indicates the number of such sites.
This applies to the situation in which there is the leftmost site $b$ beyond which all sites are inactive (otherwise $\omega_\text{sync}=0$), and that there is no large gap of inactive sites between $b$ and $\widetilde{b}$ (otherwise there can be more than one paramagnetic region).

Although these results are derived in the $\Delta\ll1$ limit, and $\omega_\text{sync}$ becomes state-dependent beyond that, the numerical results in Ref.~\cite{Li2025} suggest that Eq.~(\ref{eq:omega_sync}) is the smallest possible frequency among all possible initial states.
On the other hand, the largest possible frequency appears to be
\begin{equation}\label{eq:omegasyncmax}
    \omega^\text{max}_\text{sync}\sim e^{-2(c-b)/\xi},\quad (\Delta\sim 1)
\end{equation}
where $b$ is the leftmost active site.
Also, from the synchronization pattern of the numerical results, Eq.~(\ref{eq:etaIOM}) still appears to be a good set of IOMs in the $\Delta\sim 1$ regime, with $j_0$ being very close to the center.
We will assume these behaviors for $\Delta\sim 1$ and that the results below will not be restricted to the weakly interacting regime.

\subsection{Collective freezing}

Now we consider the effect of the symmetry-breaking $H_\epsilon$ term.
Since it leads to an energy difference $\frac{\epsilon}{2}(h'_j-h'_{\widetilde{j}})$ between two spin configurations related by swapping a pair of active sites $j,\widetilde{j}$, the most important contribution in the $\eta$ chain is
\begin{equation}\label{eq:erroreff}
    H^\text{eff}_\epsilon=\sum_j h^{\prime\,\text{eff}}_j\eta_j^z,\quad h^{\prime\,\text{eff}}_j\approx \frac{\epsilon}{2}(h'_j-h'_{\widetilde{j}}),
\end{equation}
which turns the $\eta$ chain into a mixed-field Ising model.
The longitudinal field then competes with other terms, with the tendency to turn the IOMs of the $\eta$ chain into single $\eta^z$ operators.
Since an $\eta^z_j$ eigenstate does not superpose the configuration at site $j$ and $\widetilde{j}$, they become \emph{non-resonant}.

We consider the two regimes of the $\eta$ chain separately.
In the paramagnetic (unsynchronized) regime, the original LIOMs of single $\eta^x$ can be rotated individually to $\eta^z$, with the crossover at 
\begin{equation}
    \epsilon W\sim h^\text{eff}_j\sim \omega_j.
\end{equation}
That is, the unsynchronized resonances become non-resonant one by one, starting from the one furthest from the mirror center as we increase $\epsilon$.
In the ferromagnetic (synchronized) regime, the situation becomes that $H_\epsilon$ competes with the perturbation that lifts the two-fold degeneracy without $H_{\rm cp}$.
As $H_{\rm cp}$ contributes to the off-diagonal element $\sim\omega_\text{sync}$, and $H_\epsilon$ contributes to the diagonal difference, the transition point is estimated to be 
\begin{equation}\label{eq:epsilonfrozen}
    \epsilon W \sqrt{|A|}\sim \omega_\text{sync},
\end{equation}
where the square root comes from the assumption of independent $h'_j$.
For $\epsilon$ larger than this, the entire synchronized oscillation stops.
We say that the particles are \emph{collectively frozen}.
Note that this happens even when some of the sites are resonant at this $\epsilon$ if we turn off the interaction.
This estimation agrees with the numerical result in the supplemental material of Ref.~\cite{Li2025}.

\subsection{The protection layer}\label{sec:protlayer}

\begin{figure*}
    \centering
    \includegraphics[scale=1]{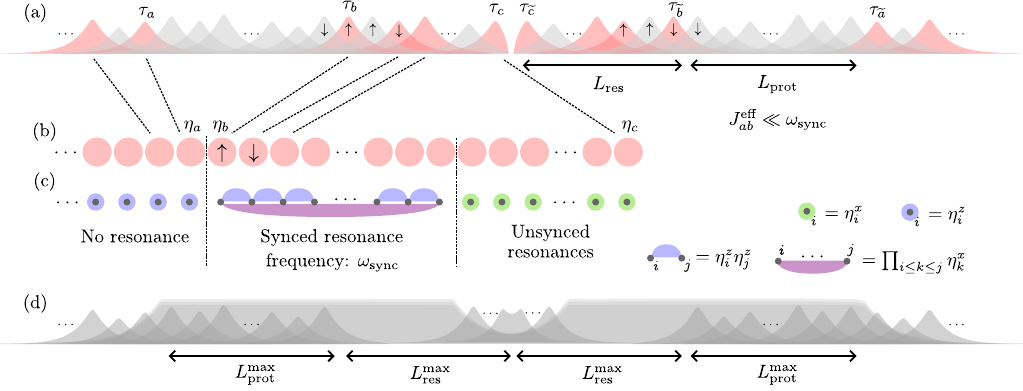}
    \caption{\justifying Illustration of the protection layer and final LIOM structure near an approximate mirror center in a long MBL chain. (a)--(c) has the same visualization setting as Fig.~2 in Ref.~\cite{Li2025}, but now there is a third, non-resonant segment separated from the synchronized segment by the \emph{protection layer}. (d) The final LIOM structure of the coupled chain, where some of the LIOMs become locally extended.}
    \label{fig:protection}
\end{figure*}

From the results above, if we fix $\epsilon$ and the density of active sites, and move $b$ (the first active site) towards the left, eventually the synchronized oscillation will become frozen.
To see this, note that $\omega_\text{sync}$ decays at least exponentially in $c-b$, while $|A|$ only grows linearly, so the crossover point (\ref{eq:epsilonfrozen}) will eventually be reached by increasing $c-b$.
Therefore, if the system size is infinite or large enough, a \emph{typical} eigenstate, which has active sites distributed everywhere in the chain, will not show arbitrarily long many-body mirror resonances around the mirror center.

On the other hand, there are eigenstates where the effective interaction $J^\text{eff}_{ab}$ between two neighboring sites $\langle ab\rangle$ in the $\eta$ chain (which can be far away in the $\tau$ chain) is so weak that the bond essentially cuts the $\eta$ chain into two independent parts.
In this way, the part to the right of it can have an effective size small enough to produce synchronized oscillations, while the part to the left of it is non-resonant.
The smallness of $J^\text{eff}_{ab}$ corresponds to the condition that all the sites in the interval $(a,b)$ in the original chain must be inactive.
We say that the synchronized resonances to the right of $b$ are \emph{protected} by the inactivity of the sites between $(a,b)$, and the segment $(a,b)$ is called the \emph{protection layer}.
This is illustrated in Fig.~\ref{fig:protection}(a)--(c).

The size $L_\text{prot}=b-a+1$ of the protection layer can be estimated by the condition that $J^\text{eff}_{ab}\ll\omega_\text{sync}$ (that is, the influence from site $a$ and to the left of it is not strong enough to influence the resonance in site $b$ and to the right of it).
Since $J^\text{eff}_{ab}\sim \Delta e^{-L_\text{prot}/\xi}$, what we need is to estimate $\omega_\text{sync}$.
We can estimate the range of $\omega_\text{sync}$ among all possible states in $[b,\tilde{b}]$, with the only condition being that $b$ is an active site (this condition ensures that the size $L_\text{res}=c-b+1$ of the resonance region is well defined).
By Eq.~(\ref{eq:omega_sync}) we have, for $\Delta\lesssim1$ and $L_\text{res}\gtrsim -\frac{\xi}{2}\ln\Delta$
\begin{equation}\label{eq:omegasyncminLres}
    \exp\left[-\frac{L_\text{res}^2}{\xi}+\frac{\xi}{4}(\ln\Delta)^2\right]\lesssim\omega_\text{sync}\lesssim\exp\left[-\frac{2L_\text{res}}{\xi}\right].
\end{equation}
Hence, we have the bound of $L_\text{prot}$ in terms of $L_\text{res}$
\begin{equation}\label{eq:LprotBounds}
    2L_\text{res}+\xi\ln\Delta\lesssim L_\text{prot}\lesssim L_\text{res}^2-\left(\frac{\xi}{2}\ln\Delta\right)^2+\xi\ln\Delta
\end{equation}
Note that when $L_\text{res}\sim -\frac{\xi}{2}\ln\Delta$, the upper and lower bounds coincide and become zero.
That is, when the resonance region is just the unsynchronized part, we no longer need the protection layer.

The bounds in Eq.~(\ref{eq:LprotBounds}) are not always achievable; however, $\omega_\text{sync}$ may itself be too small and below the collective freezing bound (\ref{eq:epsilonfrozen}).
That is, $L_\text{res}$ itself has an upper bound depending on what type of states we choose between $[b,\widetilde{b}]$.
We will estimate two bounds, the ``all resonant'' bound $L^\text{all}_\text{res}$, which is the maximum $L_\text{res}$ that \emph{all} states between $[b,\widetilde{b}]$ can lead to a resonance involving site $b$ (as long as there is a suitable protection layer); and the ``maximal resonant'' bound $L^\text{max}_\text{res}$, which is the maximum $L_\text{res}$ that at least \emph{some} state between $[b,\widetilde{b}]$ can lead to a resonance involving site $b$.
To estimate $L^\text{all}_\text{res}$, we again use the minimal possible $\omega_\text{sync}$.
Combining Eqs.~(\ref{eq:omegasyncminLres}) and (\ref{eq:epsilonfrozen}), we have
\begin{equation}
    L^\text{all}_\text{res}\sim\sqrt{\left(\frac{\xi}{2}\ln\Delta\right)^2-\xi\ln\epsilon},
\end{equation}
where we have dropped some subleading terms.
Similarly, to estimate $L^\text{max}_\text{res}$, we use the maximal possible $\omega_\text{sync}$ and combine with (\ref{eq:epsilonfrozen}),
\begin{equation}
    L^\text{max}_\text{res}\sim -\frac{\xi}{2}\ln\epsilon.
\end{equation}
The corresponding $L_\text{prot}$ is denoted by $L^\text{max}_\text{prot}$
\begin{equation}
    L^\text{max}_\text{prot}\sim 2L^\text{max}_\text{res}+\xi\ln\Delta\sim\xi\ln\frac{\Delta}{\epsilon}.
\end{equation}

\subsection{Integrals of motion in the coupled segments}

Finally, we describe the structure of the integrals of motion of the final chain near the mirror center, depicted in Fig.~\ref{fig:protection}(d).
Note that if the total system size is large compared to any scales (e.g., $L^\text{max}_\text{res}+L^\text{max}_\text{prot}$) that are influenced by the mirror center, then we can still call them LIOMs.
This is similar to the picture of a stable rare thermal region in a long random MBL chain.
Each integral of motion within the thermal region must be of the same size as the region, but once we go outside the region, the tail decays exponentially, as it does for other LIOMs, making it local in a length scale larger than the size of the thermal region.
We will see that the mirror center we have been considering also ``stretches out'' the LIOMs, making them non-local up to some scale.

We will go from the mirror center outwards, discussing the structure of each region, measured using the approximate distance from the center:
\begin{itemize}
    \item Less than $-\frac{\xi}{2}\ln\Delta$\\
    In this region, there is no synchronization, so the LIOMs are essentially the same as in the non-interacting case.
    That is, they are the occupation numbers of the orbitals, which are even and odd superpositions of the original half-chain orbitals (which may be distorted near the center due to non-perturbative effects).
    Each LIOM consists of two peaks here, as depicted in the center part of Fig.~\ref{fig:protection}(d).
    \item Between $-\frac{\xi}{2}\ln\Delta$ and $L^\text{all}_\text{res}$\\
    In this region, each site may or may not be involved in a resonance, depending on the sites further away from the center.
    Therefore, the LIOMs must be ``stretched out''.
    Moreover, since every configuration can lead to synchronized resonance as long as there is a protection layer outside it, any LIOM operator must involve the entire region on both sides of the mirror centers, as depicted by extended gray regions in Fig.~\ref{fig:protection}(d).
    \item Between $L^\text{all}_\text{res}$ and $L^\text{max}_\text{res}$\\
    This region is similar to the previous case, except that it may be possible to have certain LIOMs that are only peaked on one side of the mirror.
    This is due to the possibility that by observing the configuration in this region on one side of the mirror, it may be enough to imply that no resonance can occur here.
    Such LIOMs must involve many sites, but the exact form depends on how to combine the original LIOMs $\tau^z$, and there is no general way to determine them without knowing the details between $\omega_\text{sync}$ and the configurations.
    They are not depicted in Fig.~\ref{fig:protection}(d).
    \item Between $L^\text{max}_\text{res}$ and $L^\text{max}_\text{res}+L^\text{max}_\text{prot}$\\
    In this region, the LIOMs $\tau_j$ of the original half-chain remain valid LIOMs (except for being slightly distorted by the perturbation), as they definitely do not involve a mirror resonance.
    They are shown in Fig.~\ref{fig:protection}(d) as single-peaked LIOMs.
    On the other hand, they do control whether the states within $L^\text{max}_\text{res}$ can resonate, so the stretched-out LIOM does stretch to this region.
    \item Beyond $L^\text{max}_\text{res}+L^\text{max}_\text{prot}$\\
    In this region, again, the $\tau_j$ remain LIOMs. Moreover, no central LIOMs are stretched to here, except for their exponential tails.
    This is because the configuration here is never involved in determining whether a configuration can participate in a resonance\footnote{Note that $L^\text{max}_\text{res}+L^\text{max}_\text{prot}$ is indeed the maximum possible $L_\text{res}+L_\text{prot}$, as smaller $L_\text{res}$ involving higher-order perturbation of $H_\text{cp}$ will reach the freezing bound more quickly, so will not be able to make $L_\text{prot}$ larger.}.
\end{itemize}

\subsection{Extension to HMS}

The results above have been for a single mirror center, that is, a single level of the HMS.
A simple way to extend to multiple levels is to assume that the LIOMs in the $\tau$ chain are not entirely random, but themselves contain some stretched-out LIOMs resulting from the mirror resonances in the lower levels of the HMS.
Intuitively, instead of taking two random LIOM chains and coupling them together, as visualized in Fig.~\ref{fig:protection}(a), we are coupling two LIOM chains where the internal structure of each already looks like that of Fig.~\ref{fig:protection}(d).
To distinguish different levels, we will now add a subscript $n$ to all of the symbols introduced above.

Now, although all of the asymptotic estimates above are for random LIOMs, the tails of the stretched-out LIOMs are also expected to decay exponentially at a distance of $L^\text{max}_{\text{res},n}+L^\text{max}_{\text{prot},n}$ away from the mirror center $c_{n}$.
Hence, we expect that if this radius $L^\text{max}_{\text{res},n}+L^\text{max}_{\text{prot},n}$ around $c_{n}$ is much smaller than the distance between $c_{n}$ and $c_{n+1}$, it makes sense to keep using the above formalism and ask which region of the $(n+1)$th level contains the stretched-out region of the $n$th level.
For example, if we have
\begin{multline}\label{eq:full-prot}
    \left[c_{n}-L^\text{max}_{\text{res},n}-L^\text{max}_{\text{prot},n},c_{n}+L^\text{max}_{\text{res},n}+L^\text{max}_{\text{prot},n}\right]\\
    \subset \left[c_{n+1}-L^\text{all}_{\text{res},n+1}+O(\xi),c_{n+1}+\frac{\xi}{2}\ln\Delta-O(\xi)\right],
\end{multline}
that is, the stretched-out region of the $n$th level is within the all-resonance region of the $(n+1)$th level (with an $O(\xi)$ padding to avoid accidental resonances), then we expect the HMS to be stable.
Moreover, since a local configuration in the all-resonant region of level $n+1$ can participate in the mirror resonance of at least one full configuration of level $n+1$, and is frozen in at least one of such configurations, the resonance probability $p_{n+1}$ is nontrivial (neither 0 nor 1), leading to the MBC phenomenology.
Note that as the $L_{\text{prot},n+1}$ scales at least linearly in $L_{\text{res},n+1}$, which itself scales at least exponentially in $n$, $p_n$ decays at least doubly exponentially in $n$.
Thus, this model is always deep in the MBC-L phase (we will discuss generalization below, where it is otherwise).

On the other hand, there are many more possible scenarios beyond Eq.~(\ref{eq:full-prot}), which includes the transition from MBC-L to the single particle critical phase (if the LHS goes into the unsynchronized region), to the MBL phase (if it goes outside $L^\text{max}_{\text{prot},n+1}$.
A detailed study of each of them is outside the scope of this paper.
Also, if the $O(\xi)$ padding is not enough (due to a lack of general theory on accidental resonances, we cannot give an exact condition on what is ``enough''), the effect of accidental resonances may interfere with the mirror resonances, or even take over the dynamics so that the system becomes ETH, a scenario that we will discuss in Appendix~\ref{sec:rareConstruction} in the context of avalanche instability.

\subsection{Generalizations}

Here, we consider the generalization beyond the random-field XXZ model symmetric around a bond with global independent symmetric breaking.
To simplify the discussion (as a first approximation), we do not distinguish between random models and the quasiperiodic model in the absence of distinctive features.
In particular, in Appendix~\ref{sec:rareConstruction} below, we will assume that the above results for the random-field XXZ model directly work for the AA model without generalizations.

\subsubsection{Existence of weak bonds}

The above theory assumes a constant localization length $\xi$ across the system, which is not true if the system contains weak bonds (e.g.\ the EAAH model).
One simple approximation to incorporate weak bonds is to replace the exponential tails $\exp(-|i-j|/\xi)$ in all the discussion above with $\exp(-\mathrm{d}(i,j))$, where $\mathrm{d}(i,j)$ is some effective distance function that jumps a lot when a weak bond is reached.
Another way to think about it is to treat weak bonds as equivalent to a long distance, in which all ``sites'' are inactive (or simply have no sites between them).

Either way, the protection condition $J^\text{eff}_{ab}\ll\omega_\text{sync}$ can become much easier to satisfy if there is a weak bond between $a$ and $b$, as requiring $J^\text{eff}_{ab}$ to be small no longer necessitates the condition that many (dynamical) sites within it are inactive.
Therefore, depending on the scaling of the weakness of the bonds, it becomes possible to realize MBC-E, unlike the constant-$\xi$ case, where it is always MBC-L.
This corresponds to the argument in the main text that the interacting EAAH model can be in the MBC-E phase.

\subsubsection{Sharp symmetry breaking}

We have assumed $H_\epsilon$ to be uniformly random across the entire chain.
Although this is a good model for AA-type quasiperiodic systems (as the error term when we shift the phase is itself a cosine function in $j$), it is not good for the Fibonacci-type quasiperiodic model, where the potential is exactly mirror symmetric within the palindrome, but becomes an $O(1)$ error outside it.
Instead of modeling the error as having a strength $\epsilon$ across all sites, it is better to model it as an exponential decay from the edge of the palindrome. That is, (\ref{eq:erroreff}) is replaced by
\begin{equation}
    H^\text{eff}_\epsilon=\sum_j h^{\prime\,\text{eff}}_j\eta_j^z,\quad h^{\prime\,\text{eff}}_j\sim e^{(j-j_e)/\xi},
\end{equation}
within $j>j_e$, and model the chain as having no symmetry for $j\leq j_e$.
The results remain mostly the same.

\subsubsection{Particle non-conserving systems and long-range hopping}

Now we consider the scenario if there is no particle conservation (as in our solvable model) or if there are long-range hoppings.
These two modifications have a similar effect in the sense that now $H_{\rm cp}$ can move more than one particle at a time.
Suppose the amplitude still follows the decay rate of $\xi$; one can estimate that $\omega_\text{sync}$ now always follows Eq.~(\ref{eq:omegasyncmax}) (as there is no higher-order-perturbation bottleneck).
This implies that $L^\text{max}_\text{res}$ and $L^\text{all}_\text{res}$ no longer have asymptotic separations.

In this case, one can see that our solvable model can be viewed as modeling a subset of LIOMs (with others assumed to be inactive).
To see this, consider a subset of sites near the boundary of the all-resonance region (which plays the role of the resonance region of our solvable model).
Suppose all other sites are inactive, and we run over all possible configurations within the subset.
Due to the lack of asymptotic separation of $L^\text{max}_\text{res}$ and $L^\text{all}_\text{res}$, we expect the corresponding protection layer to intersect with the part outside $L^\text{max}_\text{res}$ and therefore be non-resonant. Now we take the interaction of all these non-resonant protection layers over the resonance configuration.
The resulting non-resonant degrees of freedom become the non-resonant region of our solvable model.
If we additionally replace those enforced-inactive sites with weak bonds, it reduces exactly to the freezing/protection conditions of our solvable model.

\section{Construction of rare initial phases in the interacting AA model}\label{sec:rareConstruction}

In this appendix, we present a construction of the rare initial phase in the interacting AA model mentioned in Sec.~\ref{sec:rareHMS}, at which the system either becomes MBC-L or thermalizes due to accidental resonances.
The derivation is based on the theory of Appendix~\ref{sec:symmetricRandom}, and we will use the approximation that the interacting AA model behaves like a random MBL phase in the absence of HMS.

\begin{figure*}
    \centering
    \includegraphics[scale=1]{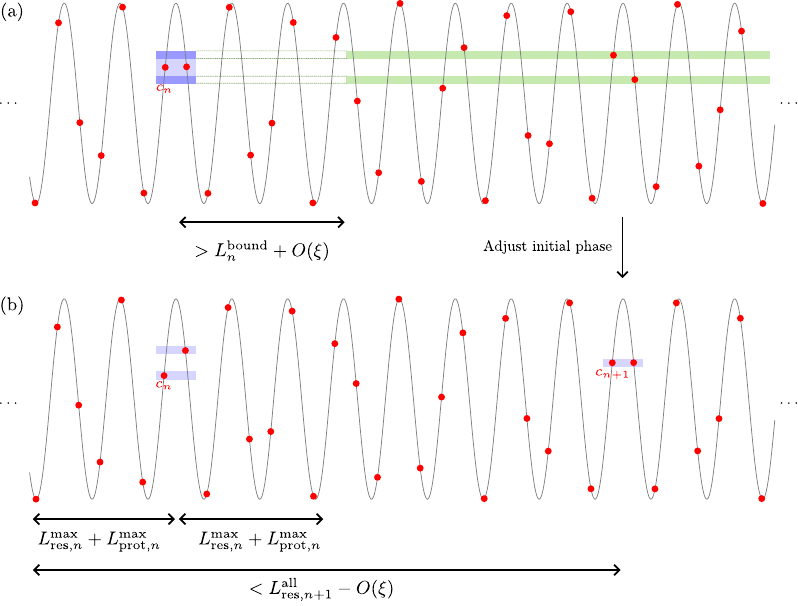}
    \caption{\justifying Construction of the rare initial phase that leads to MBC in the AA model. The gray curve is the cosine function from which the AA potentials $h_j$ are obtained, and the red dots represent $h_j$. (a) The spatial profile of $h_j$ after we have already constructed the approximate mirror point $c_n$ for the $n$th level of the hierarchy, with blue region corresponding to $[\phi_n-\Delta\phi_n,\phi_n+\Delta\phi_n]$.
     The green ribbons contain the candidates for $c_{n+1}$. (b) After adjusting the initial phase, the $(n+1)$th level of the hierarchy is constructed with $c_{n+1}$ being the approximate mirror point. All lengths and sizes are only for illustration and are not taken from real data.}
    \label{fig:aa}
\end{figure*}

The construction is done by iteration on the index $n$ of the level in the hierarchy.
For each $n$, we construct a site $c_n$ and an interval $[\phi_n-\Delta\phi_n,\phi_n+\Delta\phi_n]$ so that if $\phi$ lies between them, we have a finite HMS up to level $n$, with the approximate mirror center of the $n$th level being the bond $c_n,c_n+1$, with the mirror center being exact if $\phi=\phi_n$.
Additionally, we will have $[\phi_{n+1}-\Delta\phi_{n+1},\phi_{n+1}+\Delta\phi_{n+1}]\subseteq[\phi_n-\Delta\phi_n,\phi_n+\Delta\phi_n]$ for each $n$, leading to the existence of $\phi_\infty$ that lies in all of these intervals, producing a true HMS.
This construction is a generalization of the proof in Ref.~\cite{Jitomirskaya1994} and works for all quasiperiodic models coming from an even function (such as the cosine in the AA model), not just for the AA.
However, in our case, it is not a rigorous proof due to the possible interplay with accidental rare resonances, over which we do not have full control.

Although there are four local phases of the cosine that can produce a mirror center (and also four that produce an anti-mirror center and also lead to many-body resonances), for simplicity of demonstration, we only use a particular one of them (as depicted in Fig.~\ref{fig:aa}).

The choice for the first level can be quite arbitrary.
Picking any site $c_1$, we have a $\phi_1$ at which $c_1+1/2$ is an exact mirror center.
Then we can pick an arbitrary $\Delta\phi_1<\pi$ to form the first interval $[\phi_1-\Delta\phi_1,\phi_1+\Delta\phi_1]$.
Note that for any $\phi$ in this interval, we have the corresponding length scales $L^\text{all}_{\text{res},1}$, $L^\text{max}_{\text{res},1}$, and $L^\text{max}_{\text{prot},1}$ as defined in Sec.~\ref{sec:protlayer}, which all go to infinity at $\phi_1$, so picking a phase tolerance $\Delta\phi_1$ corresponds to fixing the set of minimal length scales at which resonances can occur in the first level.

Next, we construct $c_{n+1}$ given $c_n$ and $[\phi_n-\Delta\phi_n,\phi_n+\Delta\phi_n]$.
This is depicted in Fig.~\ref{fig:aa}, where the entire blue region in Fig.~\ref{fig:aa}(a) indicates the area where $h_{c_n}$ and $h_{c_n+1}$ lie when $\phi\in[\phi_n-\Delta\phi_n,\phi_n+\Delta\phi_n]$.
Note that we again have the minimum length scales $L^\text{all}_{\text{res},n}$, $L^\text{max}_{\text{res},n}$, and $L^\text{max}_{\text{prot},n}$ over which resonances in the $n$th level can occur.
However, in order to construct the next level, we also need to fix a maximum resonance length scale; otherwise, there is no bound on where the next mirror center can be placed.
To do this, we carve out the middle half of the interval so that when $\phi\in[\phi_n-\Delta\phi_n,\phi_n-\frac{3}{4}\Delta\phi_n]\cup[\phi_n+\frac{3}{4}\Delta\phi_n,\phi_n+\Delta\phi_n]$, the length scale $L^\text{max}_{\text{res},n}+L^\text{max}_{\text{prot},n}$ has an upper bound $L^\text{bound}_n$.
The corresponding area where $h_{c_n}$ and $h_{c_n+1}$ lie is shown in dark blue in Fig.~\ref{fig:aa}(a).
Now in the region of $j>c_n+1+L^\text{bound}_n$ or $j<c_n-L^\text{bound}_n$ (we additionally need to skip over a distance of $O(\xi)$ to avoid accidental resonances, but we do not know how to rigorously control it), we search for a candidate of $c_{n+1}$.
The case for $j>c_n+1+L^\text{bound}_n+O(\xi)$ is shown in Fig.~\ref{fig:aa}(a), where the candidate region is indicated as green ribbons aligned with the dark blue region.
The key is that there is definitely going to be a pair of $h_{j}$ and $h_{j+1}$ lying in the interior of the green ribbons due to the irrationality of the sampled points.
By shifting the phase slightly, we can make that pair become an exact mirror center, as shown in Fig.~\ref{fig:aa}(b), which becomes the mirror center $c_{n+1},c_{n+1}+1$ of the next level.
Note that this also makes $\phi_{n+1}\in(\phi_n-\Delta\phi_n,\phi_n-\frac{3}{4}\Delta\phi_n)\cup(\phi_n+\frac{3}{4}\Delta\phi_n,\phi_n+\Delta\phi_n)$, so all of the previous bounds are satisfied.

To construct $\Delta\phi_{n+1}$, we note that there is always a closed interval containing $\phi_{n+1}$ in which $\phi$ is still in $(\phi_n-\Delta\phi_n,\phi_n-\frac{3}{4}\Delta\phi_n)\cup(\phi_n+\frac{3}{4}\Delta\phi_n,\phi_n+\Delta\phi_n)$.
In addition, there is also one containing $\phi_{n+1}$ in which Eq.~(\ref{eq:full-prot}) is satisfied.
That is, the entire region near $c_n$ where the LIOM structure is stretched out is within the all-resonance region near $c_{n+1}$.
This means that all resonances in the $n$th level can participate in the $(n+1)$th level as long as the configuration outside it satisfies the protection layer condition (again, we may need to add additional $O(\xi)$ padding to avoid accidental resonances).

Without the effect of accidental resonances, the system at the rare initial phase is expected to be in MBC-L, as explained in Appendix~\ref{sec:symmetricRandom}.
However, it remains possible that accidental resonances drive the system towards ETH.
In particular, in each iteration of the construction above, there is some freedom to choose how large the $O(\xi)$ distance we need to skip over when finding the next mirror center, and how large the additional padding we add in Eq.~(\ref{eq:full-prot}); inadequate padding will lead to accidental resonances between non-mirror configurations.
That is, more combinations of configurations are allowed to resonate, not just the exact mirrored one, so the effective probability of resonance $p_n$ becomes larger, driving the system towards ETH.

Another way to think of the possibility of an avalanche is that ``rare initial phases become common'' due to accidental resonances.
During the construction of the rare initial phase, after we choose $\phi_{n+1}$, the possible choice of $\Delta\phi_{n+1}$ may become larger if the ``all resonance'' requirement also includes accidental non-mirror resonances.
Due to this effect, it may happen that the intervals $[\phi_n-\Delta\phi_n,\phi_n+\Delta\phi_n]$ do not converge to a single point as $n\to\infty$, but to an interval full of ``rare'' initial phases, which would mean that all initial phases give resonances up to infinity, thus delocalizing the entire chain.

\bibliographystyle{apsrev4-1}
\bibliography{references}

\end{document}